\documentclass[a4paper]{article}%
\usepackage[utf8]{inputenc}
\usepackage[T1]{fontenc}
\usepackage{lmodern}
\usepackage[ngerman,english]{babel}
\usepackage{a4wide}
\usepackage[affil-it]{authblk}
\usepackage[automark]{scrpage2}
\usepackage{graphicx}
\usepackage{cancel}
\usepackage{amsmath}
\usepackage{amssymb}
\usepackage{dsfont}
\usepackage{tensor}
\usepackage{slashed}
\usepackage{multirow}
\usepackage[colorlinks,pdfpagelabels, pdfstartview = FitH, bookmarksopen =
true, bookmarksnumbered = true, linkcolor = blue, plainpages = false,
hypertexnames = false, citecolor = red]{hyperref}
\usepackage[toc,page]{appendix}
\usepackage{cite}
\usepackage{amsfonts}%
\begin{document}

\title{Classical and quantum theory of the massive spin-two field}
\author{Adrian Koenigstein\textsuperscript{1,2}, Francesco Giacosa\textsuperscript{1,3},
Dirk H.\ Rischke\textsuperscript{1}}
\affil{\textsuperscript{1}Institut f\"ur Theoretische Physik, Johann Wolfgang 
Goethe-Universit\"at\\ Max-von-Laue-Str.\ 1, 60438 Frankfurt am Main, Germany}
\affil{\textsuperscript{2}Frankfurt Institute for Advanced Studies\\ Ruth-Moufang-Str.\ 1, 
60438 Frankfurt am Main, Germany}
\affil{\textsuperscript{3}Institute of Physics, Jan Kochanowski University\\ 25-406 Kielce,
Poland}
\maketitle

\begin{abstract}
In this paper, we review classical and quantum field theory of massive
non-interacting spin-two fields. We derive the equations of motion and
Fierz-Pauli constraints via three different methods: the eigenvalue equations
for the Casimir invariants of the Poincar\'{e} group, a Lagrangian approach, and a
covariant Hamilton formalism. We also present the conserved
quantities, the solution of the equations of motion in terms of polarization tensors,
and the tree-level propagator. We then discuss canonical quantization by
postulating commutation relations for creation and annihilation operators.
We express the energy, momentum, and spin operators in terms
of the former. As an application, quark-antiquark currents for tensor mesons are
presented. In particular, the current for tensor mesons with quantum numbers 
$J^{PC}=2^{-+}$ is, to our knowledge, given here for the first time.
\end{abstract}
\tableofcontents


\affil{\textsuperscript{1}Institut fuer Theoretische Physik, Johann Wolfgang
Goethe Universitaet\\ Max-von-Laue-Str. 1, 60438 Frankfurt am Main, Germany}
\affil{\textsuperscript{2}Institute of Physics, Jan Kochanowski University\\
25-406 Kielce, Poland}

\setcounter{tocdepth}{2}

\section{Introduction}

\label{sec:motivation}

A first important step toward the understanding of elementary and composite
particles in a relativistic context was made by O.\ Klein and W.\ Gordon: the
so-called Klein-Gordon (KG) equation gives the correct relation between mass,
energy, and momentum of all relativistic particles and is capable of describing
the dynamics of scalar fields in the non-interacting limit. It can be used to
study pions and other (pseudo-)scalar mesons as well as the recently
discovered Higgs particle.

A fundamental property which naturally emerges when special relativity is
applied to classical and quantum field theories is the spin, 
which is semi-integer for fermions and integer for bosons. P.~Dirac introduced the 
famous Dirac equation for fermions with spin $1/2$, which was
able to describe relativistic electrons and leads to the correct energy levels
of the hydrogen atom. The Dirac equation forms the basis for the description 
of all fundamental matter particles in the Standard Model, i.e., the
quarks and leptons. It can also be used to describe 
composites of quarks, e.g.\ baryons with spin $1/2.$

Later on, A.\ Proca \cite{proca} developed an equation which describes massive
particles with spin one. Nowadays the Proca equation finds an application in
effective theories for hadrons, in order to describe composite vector and axial-vector 
mesons, such as e.g.\ $\rho$ and $a_1$ mesons. In the limit of zero masses, 
the Proca equation correctly reproduces the (inhomogeneous) 
Maxwell equation for the photon field.

At present, no fundamental particle with spin larger than one appears in the Standard
Model. However, in an extension of the latter which encompasses the gravitational force,
gravitons as spin-two particles might enter. On the other hand, composite
particles with high spin exist: for instance, in the baryonic sector the
famous $\Delta$ particle has spin $J=3/2$, while in the mesonic sector there
is a whole class of tensor mesons ($J=2$ with different parity and charge
conjugation quantum numbers). At the same time, the quantum field-theoretical description of
particles with arbitrary spin is an interesting subject on its own. This field
was initiated by Fierz \cite{Fierz} and Fierz and Pauli in 1939 \cite{Fierzalt}, where
they postulated the so-called Fierz-Pauli constraints.

In the present work, we focus on the classical and quantum field theory of
massive non-interacting spin-two fields. 
Although these have been subject of many works 
\cite{Aurilia,Baaklini,Bhargava,Bouatta,Chang1,Chang2,Dalmazi,Dalmazi2,Deser,Fierz,
Fierzalt,Huang,Macfarlane,Rivers,Wagenaar},
usually only certain aspects of the theory of spin-two fields have been
considered and a comprehensive review is still missing. With
this work, we aim to fill this gap. In addition, we emphasize the
importance of deriving the Fierz-Pauli constraints from fundamental principles. 
We also present the operators for energy, momentum, and spin of spin-two fields 
in terms of number operators, which to our knowledge has not been shown before.

This paper is organized as follows. In Sec.\ \ref{sec:Sec2} we
discuss the classical theory of spin-two fields. We derive the equations of motion
and the Fierz-Pauli constraints via three different methods. The first is 
based on the observation that fields with a given spin form irreducible representations
of the Poincar\'{e} group. Thus, they must fulfill eigenvalue equations for the two
Casimir operators of this group. From these, we demonstrate how to
extract the equations of motion and the Fierz-Pauli constraints. The second method
is the conventional Lagrangian approach: one postulates a Lagrangian from which one
derives the equations of motion and the Fierz-Pauli constraints 
via the Euler-Lagrange equations. The third method
is a novel approach based on a Lorentz-invariant Hamilton density which is obtained
from the Lagrangian via a covariant Legendre transformation, i.e., by also replacing
spatial derivatives of the field by canonically conjugate fields. Equations of motion
and Fierz-Pauli constraints emerge from the covariant canonical equations and agree
with the results obtained with the other two methods. We also
derive the energy-momentum tensor and conserved quantities and we present the
solution of the equations of motion in terms of a basis of polarization tensors
for spin-two fields. We conclude this section by presenting the tree-level propagator.

In Sec.\ \ref{sec:Sec3} we quantize the classical spin-two field via postulating
commutation relations for the creation and annihiliation operators. We also 
present the Hamilton, momentum, and spin operators.
In particular, the $z$-component of the spin operator assumes a simple and
physically intuitive form.
We conclude this work with a summary and an application
to quark-antiquark currents for tensor mesons with quantum numbers
$J^{PC}=2^{++}$, $J^{PC}=2^{--}$, and $J^{PC}=2^{-+}$. In particular, the
latter is -- to our knowledge -- presented here for the first time. Many technical
details are relegated to various appendices.
Our units are $\hbar = c = 1$ and the metric tensor is $g_{\mu \nu} = {\rm diag}(+,-,-,-)$.

\section{Classical spin-two fields}
\label{sec:Sec2}

In this section we first discuss the classical equations of motion for non-interacting massive
spin-two fields. Here, we follow a novel approach which is based on the observation that 
the representations of any group can be classified according to the eigenvalues 
of the respective Casimir operators. Therefore, a field in a given
representation of a symmetry group has to fulfill eigenvalue equations for the Casimir 
operators of this group. The Casimir operators are functions
of the generators of the group. In our case, the symmetry group is the Poincar\'{e}
group. As a first step, we therefore construct the generators
of this group in the representation for spin-two fields.
We then compute the Casimir operators in this representation and show that the respective 
eigenvalue equations for these operators comprise 
the equations of motions as well as the Fierz-Pauli constraints for the fields.
In this section, we focus on spin-two fields. However, for the sake of the completeness,
we demonstrate the validity of this approach also for spin-1/2, spin-one, and spin-3/2 fields
in App.\ \ref{app:fierzpauliforspin}.  Next, we shall also discuss the 
more conventional derivation of the equations of motion and the constraints for spin-two
fields from a Lagrangian. In the construction of this Lagrangian we put special emphasis on 
the fact that the equations of motion as well as the constraints must follow from the 
Euler-Lagrange equations. Subsequently, we construct a
covariant extension of the conventional Hamilton density and show that
the equations of motion and constraints follow also from the canonical equations.
At the end of this section, we list the conserved quantities, solve the classical
equations of motion for non-interacting massive spin-two fields, and quote the
tree-level propagator for spin-two fields.

\subsection{Generators of the Poincar\'{e} group}
\label{generators}

The fundamental symmetry that each relativistic classical and quantum
field theory has to fulfill is form invariance under Poincar\'{e}
transformations which include  space-time
translations and Lorentz transformations, see e.g.\ Ref.\ \cite[p.55-64]{Ryder}. 
The four-vector $x^\mu$
of space-time coordinates transforms under Poincar\'{e} transformations as
\begin{equation}
x^{\mu}\longrightarrow x^{\prime}{}^{\mu}=
\tensor{\Lambda}{^\mu_\nu}x^{\nu}+a^{\mu}\,, \label{eq:poincaretrans}
\end{equation}
where $a^{\mu}$ is a real four-vector and $\tensor{\Lambda}{^\mu_\nu}$ is
the $(4\times4)$-matrix of Lorentz transformations fulfilling $g_{\mu\nu}
\tensor{\Lambda}{^\mu_\alpha}\tensor{\Lambda}{^\nu_\beta}=g_{\alpha\beta
}$. The corresponding infinitesimal transformation reads
\begin{align}
x^{\mu}\longrightarrow x^{\prime}{}^{\mu}=x^{\mu}+\delta
\tensor{\omega}{^\mu_\nu}x^{\nu}+\epsilon^{\mu}\,, \label{eq:LT_inf}
\end{align}
where $\tensor{\delta\omega}{^\mu^\nu} = - \tensor{\delta\omega}{^\nu^\mu}$ is 
an infinitesimal rank-two tensor and $\epsilon^\mu$
an infinitesimal four-vector.

The algebra of the Poincar\'{e} group involves the six generators $I^{\alpha\beta}$ 
of the Lorentz group and the four generators $P^{\alpha}$ of space-time translations,
\begin{align}
\left[  I^{\alpha\beta},I^{\gamma\delta}\right]  _{-}  &  =i \left( -g^{\alpha\gamma
}I^{\beta\delta}+g^{\alpha\delta}I^{\beta\gamma}+g^{\beta\gamma}%
I^{\alpha\delta}-g^{\beta\delta}I^{\alpha\gamma}\right) \,,\label{eq:lorentzalgebra}
\\
\left[  P^{\alpha},P^{\beta}\right]  _{-}  &
=0\vphantom{\left[ I^{\alpha\beta} , I^{\gamma\delta} \right]_{-}}\,, \label{eq:kommutatorPP}
\\
\left[  P^{\gamma},I^{\alpha\beta}\right]  _{-}  &  =i \left( g^{\gamma\alpha}P^{\beta
}-g^{\gamma\beta}P^{\alpha} \right)%
\vphantom{\left[ I^{\alpha\beta} , I^{\gamma\delta} \right]_{-}}\,. \label{eq:kommutatorPI}
\end{align}
We now determine the generators in the representation for spin-two fields.

\subsubsection{Generators for translations}
In order to derive the generators $P^\alpha$ for translations we require, as usual 
\cite{Reinhardt}, that the new field $T^{\prime}{}^{\mu\nu}$
at the translated coordinate $x^{\prime}{}^\tau$
is equal to  the old field $T^{\mu\nu}$ at the original coordinate $x^\tau$:
\begin{align}
T^{\prime}{}^{\mu\nu} (x^{\prime}{}^\tau) & = T^{\mu\nu}(x^\tau) \,.
\end{align}
Adding $T^{\prime}{}^{\mu\nu} (x^\tau)$ to both sides of this equation, 
this can be arranged into
\begin{align} \label{eq:addzero}
T^{\prime}{}^{\mu\nu} (x^\tau) & = T^{\mu\nu} (x^\tau) 
- \left[ T^{\prime}{}^{\mu\nu} (x^{\prime}{}^\tau) - T^{\prime}{}^{\mu\nu} (x^\tau) \right] \,.
\end{align}
It is sufficient to consider infinitesimal translations, cf.\ Eq.\ (\ref{eq:LT_inf}). Thus
we may expand the term in brackets to first order in  $\epsilon^\tau$,
\begin{align}
T^{\prime}{}^{\mu\nu} (x^\tau) & = T^{\mu\nu} (x^\tau) 
- \left[ T^{\prime}{}^{\mu\nu} (x^\tau) + \epsilon_\alpha \partial^\alpha T^{\prime}{}^{\mu\nu} 
(x^\tau) + O(\epsilon^2) - T^{\prime}{}^{\mu\nu} (x^\tau) \right] \nonumber
\\
& = T^{\mu\nu} (x^\tau) - g^\mu_\sigma g^\nu_\rho \epsilon_\alpha \partial^\alpha 
T^{\sigma\rho}(x^\tau) + O(\epsilon^2) \,, \label{eq:tensortransgen1} 
\end{align}
where to order $O(\epsilon^2)$ we were allowed to replace 
$\partial^\alpha T^{\prime}{}^{\mu\nu} (x^\tau) \equiv \partial^\alpha 
T^{\mu \nu}(x^\tau)$.
On the other hand, a space-time translation of a field can be expressed in terms of a group
element acting on the original field:
\begin{align}
T^{\prime}{}^{\mu\nu} (x^\tau) & = \exp \left( -i\, a_\alpha P^\alpha \right)
\tensor{}{^\mu_\sigma^\nu_\rho}\, T^{\sigma\rho}(x^\tau) \,, \label{eq:exptransgen}
\end{align}
where $a_\alpha$ is the translation vector, cf.\ Eq.\ (\ref{eq:poincaretrans}), and 
$\tensor{(P^\alpha)}{^\mu_\sigma^\nu_\rho}$ is the generator for translations in 
spin-two field representation. For infinitesimal translations 
$a_\alpha \longrightarrow \epsilon_\alpha$ and we may expand the exponential in Eq.\ 
(\ref{eq:exptransgen}) to obtain
\begin{align}
T^{\prime}{}^{\mu\nu} (x^\tau) & = \left[ g^\mu_\sigma g^\nu_\sigma 
- i\, \epsilon_\alpha \tensor{(P^\alpha)}{^\mu_\sigma^\nu_\rho} 
+ O(\epsilon^2) \right] T^{\sigma\rho}(x^\tau) \nonumber
\\
& = T^{\mu\nu}(x^\tau) - i\, \epsilon_\alpha \tensor{(P^\alpha)}{^\mu_\sigma^\nu_\rho} 
T^{\sigma\rho}(x^\tau) + O(\epsilon^2) \,.
\end{align}
Comparing this result with Eq.\ (\ref{eq:tensortransgen1}) one extracts the 
generators for translations as
\begin{align}
\tensor{(P^\alpha)}{^\mu_\sigma^\nu_\rho} & = -i\, g^\mu_\sigma g^\nu_\rho \partial^\alpha \,. 
\label{eq:generatortensortrans}
\end{align}

\subsubsection{Generators for Lorentz transformations}
We apply the same method as in the preceding subsection to derive the generators
$I^{\alpha \beta}$ for Lorentz transformations. We start 
with the Lorentz transformation of a rank-two tensor,
\begin{align}
T^{\prime}{}^{\mu\nu}(x^{\prime}{}^\tau) & = \tensor{X}{^\mu_\sigma^\nu_\rho} 
T^{\sigma\rho}(x^\tau) =\tensor{\Lambda}{^\mu_\sigma}\tensor{\Lambda}{^\nu_\rho}
T^{\sigma\rho}(x^\tau)\, . \label{eq:LT_tensor}
\end{align}
This equation describes the relation between the transformed field $T^{\prime\mu\nu}$ at 
the new position $x^{\prime}{}^\tau$ and the original field $T^{\mu\nu}$ at the old 
position $x^\tau$. Note that both field and coordinate are transformed. 
In consequence, this relation only yields the part of the generator which is related to the
spin of the field, whereas the part that is related to angular momentum will not appear. 
In order to extract the complete expression for the generators of Lorentz transformations 
in spin-two field representation, one has to compare the transformed field
$T^{\prime\mu\nu}$ with the original one $T^{\mu\nu}$ at the same coordinate $x^\tau$. 
As in Eq.\ (\ref{eq:addzero}), this is done by adding 
$T^{\prime}{}^{\mu\nu}(x^\tau)$ to both sides of Eq.\ (\ref{eq:LT_tensor}) 
and rearranging the terms into the form
\begin{align}
T^{\prime}{}^{\mu\nu}(x^\tau) &  = \tensor{\Lambda}{^\mu_\sigma} 
\tensor{\Lambda}{^\nu_\rho} T^{\sigma\rho}(x^\tau) 
- \left[ T^{\prime}{}^{\mu\nu}(x^{\prime}{}^\tau) - T^{\prime}{}^{\mu\nu}(x^\tau) \right] \,.
\end{align}
For infinitesimal transformations, cf.\ Eq.\ (\ref{eq:LT_inf}), we replace
$\tensor{\Lambda}{^\mu_\sigma} \longrightarrow g^\mu_\sigma
+ \delta \tensor{\omega}{^\mu_\sigma}+ O(\delta \omega^2)$ and expand
the term in brackets to first order in 
$\delta \tensor{\omega}{^\tau_\sigma} x^\sigma$:
\begin{align}
T^{\prime}{}^{\mu\nu} (x^\tau)
& = \left[ g_{\sigma}^{\mu} + \delta \tensor{\omega}{^\mu_\sigma} + O(\delta \omega^{2}) \right]
 \left[ g_{\rho}^{\nu} + \delta \tensor{\omega}{^\nu_\rho} + O(\delta\omega^{2}) \right] 
 T^{\sigma\rho} (x^\tau) \vphantom{\frac{1}{2}} \nonumber
\\
& - \left[ T^{\prime}{}^{\mu\nu} (x^\tau) + \delta\omega_{\alpha\beta} x^\beta \partial^\alpha 
T^{\prime}{}^{\mu\nu} (x^\tau) + O(\delta\omega^2) - T^{\prime}{}^{\mu\nu} (x^\tau) \right] \,. 
\label{eq:LTcomplete}
\end{align}
To first order in $\delta \omega$, we may replace $\partial^\alpha T^{\prime}{}^{\mu\nu}(x^\tau)
\rightarrow \partial^\alpha T^{\mu\nu}(x^\tau)$. We now make explicit use of the
antisymmetry of $\delta \omega_{\alpha \beta}$. This is
important, since otherwise we will not obtain the correct expression for the
generators $I^{\alpha\beta}$. Equation (\ref{eq:LTcomplete}) can then be rewritten as
\begin{align}
T^{\prime}{}^{\mu\nu} (x^\tau)
& = T^{\mu\nu} (x^\tau) + \frac{\delta\omega_{\alpha\beta}}{2} \left[ g^\mu_\sigma 
( g^{\nu\alpha} g^\beta_\rho - g^{\nu\beta} g^\alpha_\rho) + g^\nu_\rho 
(g^{\mu\alpha} g^\beta_\sigma - g^{\mu\beta} g^{\alpha\sigma}) \right. 
\vphantom{\frac{\delta}{2}} \nonumber
\\
& \hphantom{ \quad = T^{\mu\nu} (x^\tau) + \frac{\delta\omega_{\alpha\beta}}{2} }  
+ \left. g^\mu_\sigma g^\nu_\rho (x^\alpha\partial^\beta - x^\beta \partial^\alpha) \right] 
T^{\sigma\rho} (x^\tau) + O(\delta\omega^2) \,. \vphantom{\frac{\delta}{2}} \label{eq:LTgen1}
\end{align}
A Lorentz transformation of the rank-two
tensor field can also be written in terms of an element of the Lorentz group
acting on this field,
\begin{equation}
T^{\prime}{}^{\mu\nu}(x^\tau) = \exp \tensor{ \left( - i \,\frac{\omega_{\alpha \beta}}{2} 
I^{\alpha \beta} \right) }{^\mu_\sigma^\nu_\rho} \, T^{\sigma\rho}(x^\tau) \,,
\end{equation}
where $\tensor{( I^{\alpha \beta} )}{^\mu_\sigma^\nu_\rho}$ are the generators
for Lorentz transformation in the rank-two tensor field representation.
For an infinitesimal transformation, $\omega_{\alpha \beta} \longrightarrow \delta
\omega _{\alpha \beta}$, the exponential can be expanded,
\begin{align}
T^{\prime}{}^{\mu\nu}(x^\tau) &  = \left[ g^\mu_\sigma g^\nu_\rho 
- i \, \frac{\delta \omega_{\alpha \beta}}{2} \tensor{( I^{\alpha \beta})}{^\mu_\sigma^\nu_\rho} 
+ O(\delta\omega^2) \right] T^{\sigma\rho} (x^\tau) \nonumber
\\
& = T^{\mu\nu}(x^\tau) -i \,\frac{\delta\omega_{\alpha\beta}}{2}
\tensor{(I^{\alpha\beta})}{^\mu_\sigma^\nu_\rho} T^{\sigma\rho} (x^\tau)+ O(\delta\omega^2)
\vphantom{\left( \delta^\mu_\sigma + \frac{\delta\omega_{\alpha\beta}}{2} 
\tensor{(M^{\alpha\beta})}{^\mu_\sigma} + O(\delta\omega^2) \right)}\,. 
\vphantom{\left[\frac{\delta \omega_{\alpha \beta}}{2}\right]}
\label{eq:LT_inf_exp2}
\end{align}
By comparing Eqs.\ (\ref{eq:LTgen1}) and (\ref{eq:LT_inf_exp2}), we read off
the generators as
\begin{align}
\tensor{(I^{\alpha\beta})}{^\mu_\sigma^\nu_\rho} & = i \left[ g_{\sigma}^{\mu} 
( g^{\nu\alpha} g_{\rho}^{\beta} - g^{\nu\beta} g_{\rho}^{\alpha} ) 
+ g_{\rho}^{\nu} ( g^{\mu\alpha} g_{\sigma}^{\beta} - g^{\mu\beta} g_{\sigma}^{\alpha} ) 
+ g^\mu_\sigma g^\nu_\rho (x^\alpha\partial^\beta - x^\beta \partial^\alpha) \right]  \,. 
\label{eq:generators}
\end{align}
This procedure to obtain the explicit form of the group generators has also been
employed e.g.\ in Reinhardt \& Greiner
\cite[p.119,148,149]{Reinhardt} and Greiner \cite[p.391-392]{Greiner}. For an application to
spin-1/2, spin-one, and spin-3/2 fields, see App.\ \ref{app:fierzpauliforspin}.

The generators (\ref{eq:generators}) can be decomposed into a spin part
$\tensor{(S^{\alpha\beta})}{^\mu_\sigma^\nu_\rho}$ and an angular momentum part 
$\tensor{(L^{\alpha\beta})}{^\mu_\sigma^\nu_\rho}$:
\begin{align}
\tensor{(S^{\alpha\beta})}{^\mu_\sigma^\nu_\rho} & = i \left[ g_{\sigma}^{\mu} 
( g^{\nu\alpha} g_{\rho}^{\beta} - g^{\nu\beta} g_{\rho}^{\alpha} ) 
+ g_{\rho}^{\nu} ( g^{\mu\alpha} g_{\sigma}^{\beta} - g^{\mu\beta} g_{\sigma}^{\alpha} ) 
\right] \, ,
\label{eq:tensorspingen}
\\
\tensor{(L^{\alpha\beta})}{^\mu_\sigma^\nu_\rho} & = i\, g^\mu_\sigma g^\nu_\rho 
(x^\alpha\partial^\beta - x^\beta \partial^\alpha) \, .\label{eq:tensordrehgen}
\end{align}
The result for the spin part $\tensor{(S^{\alpha\beta})}{^\mu_\sigma^\nu_\rho}$ 
agrees with that given in Ref.\ \cite[p.213]{Macfarlane}, where it was termed `spin operator'.
This denotation is, however, not quite correct, as we shall see in 
Sec.\ \ref{sec:conservedquantities}, because it is only part of the spin operator. 

In order to check whether Eqs.\ (\ref{eq:generatortensortrans}) and (\ref{eq:generators}) 
are the correct expressions for the generators of the Poincar\'{e} group in the representation
appropriate for rank-two tensor fields, it is necessary to prove that they fulfill
the group algebra (\ref{eq:lorentzalgebra}) -- (\ref{eq:kommutatorPI}). This is 
explicitly demonstrated in App.\ \ref{app:generatoralgebra}. As is shown there,
$\tensor{(S^{\alpha\beta})}{^\mu_\sigma^\nu_\rho}$ and 
$\tensor{(L^{\alpha\beta})}{^\mu_\sigma^\nu_\rho}$ fulfill the Lorentz algebra 
(\ref{eq:lorentzalgebra}) separately. 
On the other hand, relation (\ref{eq:kommutatorPI}) is only fulfilled, if the complete 
generator $\tensor{(I^{\alpha\beta})}{^\mu_\sigma^\nu_\rho} 
= \tensor{(S^{\alpha\beta})}{^\mu_\sigma^\nu_\rho} 
+ \tensor{(L^{\alpha\beta})}{^\mu_\sigma^\nu_\rho}$ is used. This is due to the fact that 
the spin part $\tensor{(S^{\alpha\beta})}{^\mu_\sigma^\nu_\rho}$, which is closely linked 
to Lorentz boosts, commutes with $\tensor{(P^\alpha)}{^\mu_\sigma^\nu_\rho}$,
whereas rotations, which are generated by 
$\tensor{(L^{\alpha\beta})}{^\mu_\sigma^\nu_\rho}$, do not commute with translations.

\subsection{Eigenvalue equations for the Casimir operators}
\label{sec:fundamentalproperties}
The Casimir operators $C_{i}$ of a group are the operators which commute with all 
generators of the group. This means that these operators and
their eigenvalues are invariant under the 
symmetry transformations of the group. In consequence, the eigenvalues of
these operators can be used to classify the representations of the group.
The Poincar\'{e} group has two Casimir operators,
\begin{align}
C_{1}  &  \equiv P^{2}=P_{\alpha}P^{\alpha}\,, \label{eq:momentumcasimir}\\
C_{2}  &  \equiv W^{2}=W_{\alpha}W^{\alpha} = - \frac{1}{2} I_{\alpha\beta}I^{\alpha\beta} 
P^2 + I_{\gamma\beta}I^{\alpha\beta} P_{\alpha}P^{\gamma}\,, \label{eq:paulicasimir}%
\end{align}
where
\begin{equation} \label{eq:PauliLubanski}
W^{\alpha}=- \frac{1}{2}\epsilon^{\alpha\beta\gamma\delta} I_{\beta\gamma}P_{\delta} 
= - \frac{1}{2}\epsilon^{\alpha\beta\gamma\delta} P_{\beta} I_{\gamma\delta}
\end{equation}
is the so-called Pauli-Lubanski pseudovector. 
The expression on the right-hand side of Eq.\ (\ref{eq:paulicasimir}) is proven in App.\
\ref{app:secondcasimiroperator}. The last equality in Eq.\ (\ref{eq:PauliLubanski}) 
follows from the commutation relation (\ref{eq:kommutatorPI}) 
and the antisymmetry of the Levi-Civita tensor. The eigenvalue of $C_{1}$ is the 
squared mass $m^{2}$ of the particle, while the
eigenvalue of $C_{2}$ is $-m^{2}s(s+1)$, where $s$ is the spin of the particle,
see e.g.\ Ref.\ \cite[p.55-64]{Ryder}.

The next step is to find the equations of motions for a field of a given spin. The 
standard approach is the Bargmann-Wigner method \cite{Bargmann:1948ck} (see also Refs.\ 
\cite[chap.15]{Greiner} and \cite[p.63]{Huang}). Another way is to use
projection operators as discussed in Ref.\ \cite[p.1684ff]{Aurilia} and further
elaborated in Ref.\ \cite[p.213-217]{Macfarlane} for the spin-two case.
In the following we present yet another, and to our knowledge novel, 
way to derive the equations of motion
as well as the Fierz-Pauli constraints. This will be done via eigenvalue equations for the 
Casimir operators. In this section, we apply this approach to 
rank-two tensor fields $T^{\alpha\beta}$. In App.\ \ref{app:fierzpauliforspin}
we demonstrate its validity for spin-1/2, spin-one, and spin-3/2 fields. Therefore, we 
conjecture that it is valid also for fields with arbitrary spin.

The eigenvalue equation for the first Casimir operator $C_1$ (\ref{eq:momentumcasimir})
reads:
\begin{align}
\tensor{(P^2)}{^\mu_\gamma^\nu_\delta} T^{\gamma\delta} 
= \tensor{(P_\alpha)}{^\mu_\sigma^\nu_\rho} 
\tensor{(P^\alpha)}{^\sigma_\gamma^\rho_\delta} T^{\gamma\delta} = m^{2} T^{\mu\nu}\,. 
\label{eq:C_1}
\end{align}
Using the generator (\ref{eq:generatortensortrans}) for translations in spin-two field 
representation, this equation becomes the well-known Klein-Gordon (KG) equation:
\begin{align}
-\Box \,T^{\mu\nu}=m^{2}T^{\mu\nu}\,. \label{eq:kleingordon}
\end{align}
We now proceed by computing $W^{2}$ from Eq.\ (\ref{eq:PauliLubanski}) in rank-two
tensor field representation, i.e., using Eqs.\  
(\ref{eq:generatortensortrans}) and (\ref{eq:generators}). 
As shown in App.\ \ref{app:secondcasimirspin2}, the result is
\begin{align}
\tensor{(W^2)}{^\mu_\gamma^\nu_\delta}  &  = 
2 \left[ \left( 2 g_{\gamma}^{\mu} g_{\delta}^{\nu} 
- g^{\mu\nu} g_{\gamma\delta}  + g_{\delta}^{\mu} g_{\gamma}^{\nu}  \right)\right.
\Box \nonumber
\\
& \qquad + \left.  g_{\gamma\delta} \partial^{\mu} \partial^{\nu} + g^{\mu\nu} \partial_{\gamma} 
\partial_{\delta}  -  g_{\delta}^{\nu} 
\partial^{\mu} \partial_{\gamma} -  g_{\gamma}^{\mu} \partial^{\nu} \partial_{\delta} 
-  g_{\gamma}^{\nu} \partial^{\mu} \partial_{\delta} - g_{\delta}^{\mu} \partial^{\nu} 
\partial_{\gamma}\right]  \vphantom{\tensor{(W^2)}{^\mu_\gamma^\nu_\delta}} \,.
\label{eq:casimirtwo}
\end{align}
In analogy to Eq.\ (\ref{eq:C_1}) one obtains the eigenvalue
equation for $C_2=W^2$, acting on a rank-two tensor field, as
\begin{equation} \label{eq:C_2}
\tensor{(W^2)}{^\mu_\gamma^\nu_\delta}T^{\gamma\delta}
=-m^{2}s(s+1)T^{\mu\nu}\,,
\end{equation}
where we have to set $s=2$. Using the explicit expression (\ref{eq:casimirtwo}) for $W^2$,
one obtains with the abbreviation $\tensor{T}{^\mu_\mu} \equiv T$:
\begin{align}
-6m^{2} T^{\mu\nu}  & = 
2 \left( 2\, \Box \,T^{\mu \nu} - g^{\mu\nu}\, \Box\, T + \Box\, T^{\nu \mu}  \right. \nonumber
\\
& \quad + \left. \partial^{\mu}\partial^{\nu}T+g^{\mu\nu}\partial_{\gamma}
\partial_{\delta}T^{\gamma\delta} -\partial^{\mu}\partial_{\gamma}T^{\gamma\nu}
-\partial^{\nu}\partial_{\delta}T^{\mu\delta}-\partial^{\mu}
\partial_{\delta}T^{\nu\delta}-\partial^{\nu}\partial_{\gamma}T^{\gamma\mu} \right) 
 \,.
\end{align}
We now employ Eq.\ (\ref{eq:kleingordon}) to obtain
\begin{align}
0  & = m^2 \left( T^{\mu\nu} + g^{\mu\nu}\, T-  T^{\nu\mu} \right) \nonumber \\
& \quad +\partial^{\mu}\partial^{\nu}T+g^{\mu\nu}
\partial_{\gamma}\partial_{\delta}T^{\gamma\delta} -
\partial^{\mu}\partial_{\gamma}T^{\gamma\nu}-\partial^{\nu}\partial_{\delta}T^{\mu\delta}
-\partial^{\mu}\partial_{\delta}T^{\nu\delta}-\partial^{\nu}\partial_{\gamma}T^{\gamma\mu}\,. 
\label{eq:equofmot}
\end{align}
A rank-two tensor can always be decomposed into a part which is symmetric,
$T_{s}^{\mu \nu} = (T^{\mu \nu}+ T^{\nu \mu})/2$, and one which is
antisymmetric, $T_{a}^{\mu \nu} = (T^{\mu \nu}- T^{\nu \mu})/2$. For the latter,
Eq.\ (\ref{eq:equofmot}) reduces to
\begin{equation} \label{eq:C_2T_a}
0=2 m^{2}T_{a}^{\mu\nu}\; \; \longrightarrow \; \; T_{a}^{\mu\nu}=0 \,,
\end{equation}
if $m^2 \neq 0$. This means that a spin-two field can be described by a symmetric
rank-two tensor. 

For the remainder of the calculation, we may now take $T^{\mu \nu}$ to be symmetric.
Then, Eq.\ (\ref{eq:equofmot}) becomes
\begin{equation}
0= g^{\mu\nu}\, m^2 T+\partial^{\mu}\partial^{\nu}T+g^{\mu\nu}\partial_{\gamma
}\partial_{\delta}T^{\gamma\delta}-2\, \partial^{\mu}\partial_{\gamma}%
T^{\gamma\nu}-2\, \partial^{\nu}\partial_{\gamma}T^{\gamma\mu}\,.
\label{eq:equofmot2}
\end{equation}
Taking the trace, we obtain using the KG equation (\ref{eq:kleingordon})
\begin{equation} \label{eq:tracelessness}
0 = \left( 4 m^2 + \Box\right)  T = 3\, m^2 T \; \; \longrightarrow \; \; T=0 \,,
\end{equation}
i.e., the symmetric rank-two tensor field describing spin-two particles is traceless.
Using this fact and taking the four-divergence of Eq.\ (\ref{eq:equofmot2}) leads to
\begin{equation} \label{eq:18}
0=-\partial^{\nu}\partial_{\gamma}\partial_{\delta}T^{\gamma\delta}%
-2\,\Box\,\partial_{\gamma}T^{\gamma\nu} \,.
\end{equation}
Taking the four-divergence again gives (yet again employing the KG equation)
\begin{equation}
0 = - 3\, \Box \, \partial_{\gamma}\partial_{\delta}T^{\gamma\delta}
= 3\, m^2 \partial_{\gamma}\partial_{\delta}T^{\gamma\delta} \;\; \longrightarrow \;\;
\partial_{\gamma}\partial_{\delta}T^{\gamma\delta} =0 \,.
\end{equation}
Inserting this back into Eq.\ (\ref{eq:18}) we obtain (again using the KG equation)
\begin{equation} \label{eq:Lorentz}
0 = -2\, \Box\, \partial_{\gamma}T^{\gamma\nu} = 2\, m^2
\partial_{\gamma}T^{\gamma\nu} \;\; \longrightarrow \;\;
\partial_{\gamma}T^{\gamma\nu} =0 \,.
\end{equation}

In summary, the eigenvalue equations (\ref{eq:C_1}), (\ref{eq:C_2})
for the two Casimir operators have been shown to be equivalent to the following 
equations for a spin-two field:
\begin{align}
0  &  =(\Box+m^{2})T^{\mu\nu} \,,\label{eq:kleingordoneq}\\
0  &  =T^{\mu\nu}-T^{\nu\mu} \,,\label{eq:symmetrie}\\
0  &  = g_{\mu\nu}T^{\mu\nu} \,,\label{eq:traceless}\\
0  &  =\partial_{\mu}T^{\mu\nu} \,, \label{eq:lorentzcond}%
\end{align}
cf.\ Eqs.\ (\ref{eq:kleingordon}), (\ref{eq:C_2T_a}), (\ref{eq:tracelessness}), 
and (\ref{eq:Lorentz}).
Equation (\ref{eq:kleingordoneq}) is the equation of motion for the rank-two tensor
field, i.e., each component obeys the KG equation. Equations (\ref{eq:symmetrie}),
(\ref{eq:traceless}), and (\ref{eq:lorentzcond}) are additional equations
known as Fierz-Pauli constraints. These constraints constitute $6+1+4=11$ additional
conditions which reduce the number of  degrees of freedom of
a rank-two tensor field from 16 to $16-11=5$. These are just the $2s+1=5$ 
spin projections or polarization directions for a spin-two particle. 

The generalization of Eqs.\ (\ref{eq:kleingordoneq}) -- (\ref{eq:lorentzcond}) 
for arbitrary spin was already provided
by Fierz \cite[p.5-8,18-20]{Fierzalt} and in a more modern notation by Bouatta
et al. \cite[p.1]{Bouatta}, Huang et al. \cite[p.65,66]{Huang}, and Chang
\cite[p.2]{Chang2}. For arbitrary integer spin $N$ (bosons) they read
\begin{align*}
0  &  =(\Box+m^{2})\phi^{\mu_1 \cdots \mu_N}\,,\\
0  &  =\phi^{\mu_1 \cdots \mu_i \cdots \mu_j  \cdots \mu_N}
-\phi^{\mu_1 \cdots \mu_j \cdots \mu_i  \cdots \mu_N} \;\; \forall \; 1 \leq i\neq j \leq N
\,,\\
0  &  =g_{\mu_i\mu_j}\phi^{\mu_1 \cdots \mu_i \cdots \mu_j  \cdots \mu_N}
\;\; \forall \; 1 \leq i\neq j \leq N\,,\\
0  &  =\partial_{\mu_i}\phi^{\mu_1 \cdots \mu_i \cdots \mu_N}\;\; \forall \; 1 \leq i \leq N\,, 
\end{align*}
and for arbitrary semi-integer spin (fermions) they are
\begin{align*}
0  &  =(\Box+m^{2})\psi^{\mu_1 \cdots \mu_N}\,,\\
0  &  =(i\slashed{\partial}-m)\psi^{\mu_1 \cdots \mu_N}\,,\\
0  &  =\psi^{\mu_1 \cdots \mu_i \cdots \mu_j  \cdots \mu_N}
-\psi^{\mu_1 \cdots \mu_j \cdots \mu_i  \cdots \mu_N} \;\; \forall \; 1 \leq i\neq j \leq N
\,,\\
0  &  =g_{\mu_i\mu_j}\psi^{\mu_1 \cdots \mu_i \cdots \mu_j  \cdots \mu_N}
\;\; \forall \; 1 \leq i\neq j \leq N\,,\\
0  &  =\partial_{\mu_i}\psi^{\mu_1 \cdots \mu_i \cdots \mu_N}\;\; \forall \; 1 \leq i \leq N\,,  \\
0 & =\gamma_{\mu_i}\psi^{\mu_1 \cdots \mu_i \cdots \mu_N}\;\; \forall \; 1 \leq i \leq N\,.
\end{align*}
For the general determination of the number of degrees of freedom 
see Ref.\ \cite{Fierzalt} and
the works of Barut \& Raczka \cite[p.635]{Barut} and Huang \cite[p.63]{Huang}.

\subsection{Lagrangian}

\label{sec:lagrangianhamiltonian}

In this section, we discuss the derivation of the equations of motion and the
Fierz-Pauli constraints from a suitably defined Lagrangian.
This Lagrangian was already given in Ref.\ \cite[p.3]{Dalmazi} and reads
\begin{align}
\mathcal{L}  &  = \;\;\, \,\frac{1}{8} \,  \left[ \partial_{\mu} \left(T_{\alpha\beta}
+T_{\beta \alpha}\right)\right] \partial^{\mu} \left(T^{\alpha\beta}
+T^{\beta \alpha}\right)  -\frac{1}{2} \, (\partial_{\mu} T) \partial^{\mu}T 
-  \frac{1}{4} \, \left[\partial_{\alpha}\left( T^{\alpha\beta} +T^{\beta \alpha} \right)\right]
\partial^{\mu} \left( T_{\mu\beta} + T_{\beta \mu} \right) \nonumber\\
&\quad + \vphantom{\partial_\mu T^\beta}\frac{1}{2} \, \left[\partial_{\mu}
\left( T^{\mu\nu}+ T^{\nu \mu} \right)\right] \partial_{\nu} T
-\frac{m^{2}}{8}  \vphantom{\partial_\mu T^\beta}
\left(T_{\mu\nu} + T_{\nu\mu}\right) \left(T^{\mu\nu}+T^{\nu\mu}\right) 
+\frac{m^{2}}{2} T^{2}\,.
\label{eq:lagrangedichte}
\end{align}
In the following we shall explicitly demonstrate that this Lagrangian
leads to Eqs.\ (\ref{eq:kleingordoneq}), (\ref{eq:symmetrie}), (\ref{eq:traceless}), and
(\ref{eq:lorentzcond}) via the Euler-Lagrange equations
\begin{align}
\partial_{\mu} \frac{\partial\mathcal{L}}{\partial(\partial_{\mu}T_{\alpha\beta})} 
- \frac{\partial\mathcal{L}}{\partial T_{\alpha\beta}}=0\, .
\end{align}
Inserting the Lagrangian (\ref{eq:lagrangedichte}) into the Euler-Lagrange equations
we obtain
\begin{align}
0 &  =\frac{1}{2} \, \Box \left(T_{\mu\nu}+T_{\nu\mu} \right)  
-g_{\mu\nu}\, \Box \, T
-\frac{1}{2}  \partial_{\mu}\partial^{\alpha}\left(T_{\alpha\nu}+T_{\nu\alpha}
\right) - \frac{1}{2} \partial_{\nu}\partial^{\alpha} \left( T_{\alpha\mu}+T_{\mu\alpha}\right)
\nonumber\\
&  \quad
+  \partial_{\mu}\partial_{\nu}T+g_{\mu\nu}\, \frac{1}{2}\, \partial_{\alpha
}\partial_{\beta}\left(T^{\alpha\beta} +T^{\beta \alpha} \right) \vphantom{\frac{1}{2}}
+\frac{m^{2}}{2}\left(  T_{\mu\nu}+T_{\nu\mu}\right)  -g_{\mu\nu}
m^{2}T\vphantom{\frac{1}{2}}\ . \label{eq:zwangsbeweg}
\end{align}
Note that only the symmetric combination $T^{\alpha \beta}+T^{\beta \alpha}$ enters 
these equations, i.e., only the symmetric part of $T^{\alpha \beta}$ is
a dynamical quantity. The antisymmetric part does not play any role. 
Of course, this is a consequence
of the way the Lagrangian (\ref{eq:lagrangedichte}) is constructed.
Nevertheless, one should not assume from the very beginning, i.e., in the Lagrangian
itself, that $T^{\alpha \beta}$ is symmetric, otherwise one will 
not obtain the correct form of the covariant Hamilton density and the
tree-level propagator, see Secs.\ \ref{sec:canonical} and \ref{sec:propagator}.

In the following, we shall therefore assume that $T^{\alpha \beta} = T^{\beta \alpha}$,
i.e., that Eq.\ (\ref{eq:symmetrie}) holds. Then, Eq.\ (\ref{eq:zwangsbeweg}) simplifies to
\begin{align}
0 &  = \Box \, T_{\mu\nu}- g_{\mu\nu}\, \Box\, T
- \vphantom{\partial_\beta T^{\alpha\beta}}  \partial_{\mu}\partial^{\alpha}
T_{\alpha\nu}- \partial_{\nu}\partial^{\alpha}T_{\mu\alpha}
+  \partial_{\mu}\partial_{\nu}T+g_{\mu\nu}\partial_{\alpha}\partial_{\beta}
T^{\alpha\beta}  +m^{2}T_{\mu\nu}-g_{\mu\nu}
m^{2}T\,.\label{eq:symzwangsbew}
\end{align}
This equation is known as the Fierz-Pauli equation, cf.\
Hinterbichler \cite[p.16]{Hinterbichler}. We now take the four-divergence, in analogy
to the discussion in Sec.\ \ref{sec:fundamentalproperties} and in Ref.\
\cite[p.16]{Hinterbichler}. This is similar
to the derivation of the Proca equation (cf.\ Reinhardt \& Greiner \cite[p.152,153]{Reinhardt}). 
After cancelling identical terms, we obtain
\begin{align} \label{eq:LC_2}
\partial_{\mu}T^{\mu\nu}=\partial^{\nu}T\,,
\end{align}
which can be reinserted into Eq.\ (\ref{eq:symzwangsbew}), resulting in
\[
0 = \Box \, T_{\mu\nu} -   \partial_{\mu}\partial_{\nu} T  +m^{2}T_{\mu\nu}-g_{\mu\nu}%
m^{2}T\,.
\] 
Taking the trace, we see that
\[
0=-3\, m^{2}T\,,
\]
i.e., the rank-two tensor field is traceless, which is Eq.\ (\ref{eq:traceless}). Using this fact
in Eq.\ (\ref{eq:LC_2}) yields the Lorentz condition (\ref{eq:lorentzcond}). Finally,
using $T=0$ and $\partial_\mu T^{\mu \nu} =0$ in Eq.\ (\ref{eq:symzwangsbew}), we obtain
the KG equation (\ref{eq:kleingordoneq}). This completes the derivation of the
equations of motion and the Fierz-Pauli constraints starting from the Lagrangian
(\ref{eq:lagrangedichte}).

The Lagrangian (\ref{eq:lagrangedichte}) is different from the ones used in
other works. Usually some of the Fierz-Pauli constraints are
introduced from the beginning and are not extracted via the Euler-Lagrange
equations, see for instance Baaklini \& Tuite
\cite[p.13]{Baaklini}, Wagenaar \& Rijken \cite[p.3]{Wagenaar}, Rivers
\cite[p.395]{Rivers}, and Hinterbichler \cite[p.13]{Hinterbichler}. Indeed,
even in the very first discussion of spin-two fields by Fierz \& Pauli 
the assumption is made that the tensor field $T^{\mu\nu}$ is symmetric and 
traceless \cite[p.216]{Fierz}. Alternatively, some authors introduced additional auxiliary fields
(Lagrange multipliers) in
the Lagrangian in order to derive the equations of motion and the constraints, 
see Fierz \& Pauli \cite[p.216]{Fierz} and Chang \cite[p.1310]{Chang2}, 
while other authors derive their Lagrangian from projection operators \cite[p.219-223]{Macfarlane}.
 In this context we would also like to mention the seminal work
by Bhargava \& Watanabe\cite{Bhargava}, who quoted a Lagrangian for the
complex spin-two field. Finally, we also refer to the works of Dalmazi
\cite{Dalmazi2} and Bouatta et al.\ \cite{Bouatta}, who include several arbitrary 
parameters in the Lagrangian, which are then fixed by physical requirements.

\subsection{Covariant Hamilton density}
\label{sec:canonical}

In this section, we discuss the derivation of the equations of motion and the
Fierz-Pauli constraints from a covariant Hamilton density and the pertaining
canonical equations. This procedure is described e.g.\ in 
Refs.\ \cite[p.3,4]{StruckmeierUn} and \cite{StruckmeierUn2}, to which we
refer for details. To this end, one has to perform a so-called complete Legendre
transformation of the Lagrangian, i.e., one not only replaces the temporal derivatives
$\partial_0 \phi_I$ of a field $\phi_I$ but also its spatial derivatives $\vec{\nabla} \phi_I$ 
by associated canonically conjugate fields. Consequently, 
if the field $\phi_I$ is a Lorentz tensor of rank $n$, then these form a Lorentz tensor 
of rank $n+1$ defined as
\begin{align} \label{eq:can_conj_field}
\pi_{I}^{\mu}=\frac{\partial\mathcal{L}}{\partial(\partial_{\mu}\phi_{I})} \, .%
\end{align}
Then, the so-called covariant Hamilton density
$\mathcal{H}_{\text{cov}}$ reads
\begin{align} 
\mathcal{H}_{\text{cov}}(\phi_{I},\pi_{I}^{\mu})=\pi_{I}^{\mu}%
\partial_{\mu}\phi_{I}-\mathcal{L}(\phi_{I},\partial_{\mu}\phi_{I}) \,. \label{eq:covhamdens}%
\end{align}
Note that this quantity is manifestly Lorentz-invariant.
The canonical equations can derived from a variational principle as \cite{StruckmeierUn}
\begin{align}
\frac{\partial\mathcal{H}_{\text{cov}}}{\partial\pi_{I}^{\mu}}  &
=\partial_\mu \phi_{I} \,,\label{eq:canonicalone}\\
\frac{\partial\mathcal{H}_{\text{cov}}}{\partial\phi_{I}}  &  = -
\partial_\mu \pi_{I}^{\mu} \, . \label{eq:canonicaltwo}%
\end{align}
The difference between the covariant Hamilton density $\mathcal{H}_{\rm cov}$ and the 
usual Hamilton density $\mathcal{H}$ will be clarified when we discuss the energy-momentum 
tensor in Sec.\ \ref{sec:energymomentumtensor}.

In the case of the spin-two field the canonically conjugate field is given by
\begin{align}
\Pi^{\sigma\rho\gamma}  &  =\frac{\partial\mathcal{L}}{\partial\left(
\partial_{\gamma}T_{\sigma\rho}\right)  } \vphantom{\frac{1}{2}} \nonumber\\
&  =+\frac{1}{2}\, \partial^{\gamma} \left(  T^{\sigma\rho}+T^{\rho\sigma}\right)  
-g^{\rho\sigma}\partial^{\gamma}T+\frac{1}{2}\left(  g^{\gamma\sigma}\partial^{\rho}T
+g^{\rho\gamma}\partial^{\sigma}T\right) \nonumber\\
&  \quad-\frac{1}{2}\, g^{\gamma\sigma}\partial_{\mu}\left(  T^{\mu\rho}+T^{\rho\mu}
\right) -\frac{1}{2}\, g^{\gamma\rho}\partial_{\mu}\left( T^{\sigma\mu}+T^{\mu\sigma}\right)
+\frac{1}{2}\,g^{\rho\sigma }\partial_{\mu}\left( T^{\mu\gamma}+T^{\gamma\mu}\right) 
  \,. \label{eq:momentumfield}
\end{align}
Note that at this stage no assumption about $T^{\sigma \rho}$ is made. In particular, the
Fierz-Pauli constraints are not used at this stage but, as we shall see, follow from
the canonical equations. This means that also the unphysical degrees of freedom 
of $T^{\sigma \rho}$ enter in Eq.\ (\ref{eq:momentumfield}).
We now derive the covariant Hamilton density via performing the complete
Legendre transformation for the spin-two field (\ref{eq:covhamdens}), for details
see App.\ \ref{app:legendretransformation}. The result is
\begin{align}
\mathcal{H}_{\text{cov}}  &  = +\frac{1}{2}\,\Pi_{\sigma\rho\gamma}\Pi^{\sigma\rho\gamma}
-\frac{1}{4}\,\tensor{\Pi}{_\alpha^\alpha_\gamma}\tensor{\Pi}{_\beta^\beta^\gamma}
-\frac{1}{6}\,\tensor{\Pi}{_\sigma_\alpha^\alpha}\tensor{\Pi}{^\sigma_\beta^\beta}
-\frac{1}{6}\,\tensor{\Pi}{_\alpha_\rho^\alpha}\tensor{\Pi}{_\beta^\rho^\beta}\nonumber\\
&  \quad+\frac{m^{2}}{8}\left(  T_{\mu\nu}+T_{\nu\mu} \right) \left(T^{\mu\nu}+T^{\nu\mu}\right)  
-\frac{m^{2}}{2}T^{2}\,. \label{eq:hamiltonian}%
\end{align}
In App.\ \ref{app:canonicalequations} it is shown that Eqs.\
(\ref{eq:kleingordoneq}), (\ref{eq:symmetrie}), (\ref{eq:traceless}), and
(\ref{eq:lorentzcond}) follow from the canonical equations (\ref{eq:canonicalone}),
(\ref{eq:canonicaltwo}) with the Hamilton density (\ref{eq:hamiltonian}). 
This proves that the covariant Hamilton formalism is an alternative approach which
is completely equivalent to the Lagrangian description. 
For the standard non-covariant Hamilton density for the
spin-two field, see for example Baaklini \& Tuite \cite[p.13]{Baaklini},
which takes into account all constraints of the spin-two field. 
This standard Hamilton density is the $00$--component
of the energy-momentum tensor and related to the energy of the
field, cf.\ Eq.\ (\ref{eq:energiefunktion}). However, from this Hamilton density one cannot
derive all constraints for spin-two fields. Wagenaar \& Rijken circumvented this problem
by adding Lagrange multipliers for the constraints to a
non-covariant Hamilton density \cite[p.5]{Wagenaar}. Then, they were able
to derive the equations of motion and the constraints from this Hamilton density.

\subsection{Energy-momentum tensor}

\label{sec:energymomentumtensor} The energy-momentum tensor of the spin-two
field is given by:
\begin{align}
\Theta_{\mu\nu}  &  =\frac{\partial\mathcal{L}}{\partial\left(  \partial^{\mu
}\tensor{T}{^\alpha^\beta}\right)  }\partial_{\nu}%
\tensor{T}{^\alpha^\beta}-g_{\mu\nu}\mathcal{L}\nonumber\\
&  =\;\;\;  \frac{1}{2}\left(  \partial_{\mu}T_{\alpha\beta}+\partial_{\mu}T_{\beta\alpha}
\right)\partial_{\nu}T^{\alpha\beta}  -\left(\partial_{\mu}T\right) \partial_{\nu}T\nonumber\\
&  \quad-\frac{1}{2}\left(  \partial_{\nu}T_{\mu\alpha}
+\partial_{\nu}T_{\alpha\mu}\right) \left(\partial_{\beta}T^{\alpha\beta}
+ \partial_{\beta}T^{\beta\alpha}\right) \nonumber\\
&  \quad+\frac{1}{2}\left(  \partial_{\nu}T_{\mu\alpha}+\partial_{\nu}T_{\alpha\mu}
\right) \partial^{\alpha}T+\frac{1}{2}\left( \partial^{\alpha}T_{\alpha\mu}
+\partial^{\alpha}T_{\mu\alpha}\right)   \partial_{\nu}T 
- g_{\mu\nu}\mathcal{L}\,. \label{eq:ENIM}%
\end{align}
Alternatively it is also possible to express the energy-momentum tensor in terms of 
the covariant Hamilton density (\ref{eq:covhamdens}). Using the definition 
(\ref{eq:can_conj_field}) of the canonically conjugate field, the
Legendre transformation (\ref{eq:covhamdens}), and the canonical equation 
(\ref{eq:canonicalone}), one obtains
\begin{align}
\Theta_{\mu\nu}  &  = \tensor{\Pi}{_\alpha_\beta_\mu} 
\frac{\partial \mathcal{H}_{\rm cov}}{\partial \tensor{\Pi}{_\alpha_\beta^\nu}} 
- g_{\mu\nu} \left( \tensor{\Pi}{^\alpha^\beta^\gamma} 
\frac{\partial \mathcal{H}_{\rm cov}}{\partial \tensor{\Pi}{^\alpha^\beta^\gamma}} 
- \mathcal{H}_{\rm cov} \right)\,.
\end{align}
Both expressions lead to equivalent results. In addition, it is now obvious that the 
standard Hamilton density $\mathcal{H} \equiv \Theta_{00}$ does not coincide 
with the covariant Hamilton density $\mathcal{H}_{\rm cov}$.

In general one is interested in the energy-momentum tensor for physical solutions. 
To this end, one uses the Fierz-Pauli equations (\ref{eq:kleingordoneq}), (\ref{eq:symmetrie}),
(\ref{eq:traceless}), and (\ref{eq:lorentzcond}) to obtain:
\begin{equation}
\Theta_{\mu\nu}=\left( \partial_{\mu}T_{\alpha\beta}\right) \partial_{\nu}T^{\alpha\beta
}-g_{\mu\nu}\frac{1}{2}\left[ \left(  \partial_{\gamma}T_{\alpha\beta}\right) 
\partial^{\gamma}T^{\alpha\beta}-m^{2}T_{\alpha\beta}T^{\alpha\beta}\right]  \,,
\label{eq:energieimpulstensor}%
\end{equation}
which coincides with the expression of Rivers \cite[p.397]{Rivers}. (Also
Bhargava \& Watanabe mention the importance of this tensor and claim that it
is independent of all parameters they have in their Lagrangian, but do not
derive it explicitly \cite[p.273]{Bhargava}.)
In general, an energy-momentum tensor can always be symmetrised, see 
Reinhardt \& Greiner \cite{Reinhardt}. Here this is not necessary because 
the tensor (\ref{eq:energieimpulstensor}) is already symmetric in the indices 
$\mu$ and $\nu$.
Finally, the divergence of the energy-momentum tensor vanishes, as expected
for a system that is invariant under space-time translations:
\begin{align}
\partial^{\mu}\Theta_{\mu\nu}  &  =\left(\Box \, T_{\alpha\beta}\right)\partial_{\nu}%
T^{\alpha\beta}+\left(\partial_{\mu}T_{\alpha\beta}\right) \partial^{\mu}\partial_{\nu
}T^{\alpha\beta}-\left(\partial_{\gamma}T_{\alpha\beta}\right)\partial^{\gamma}%
\partial_{\nu}T^{\alpha\beta}+m^{2}T^{\alpha\beta}
\partial_{\nu}T_{\alpha\beta}\vphantom{\left( \Box + m^2 \right)}\nonumber\\
&  =\left(\partial_{\nu}T^{\alpha\beta}\right) \left(  \Box+m^{2}\right)  T_{\alpha\beta}
=0\vphantom{\left( \Box + m^2 \right)}\,,%
\end{align}
where we used Eq.\ (\ref{eq:kleingordoneq}).

\subsection{Conserved quantities}

\label{sec:conservedquantities} The aim of this section is to find the
conserved quantities following from the Noether theorem.
The Lagrangian (\ref{eq:lagrangedichte}) is Poincar\'{e}-invariant. As is well known, 
space-time translation invariance is responsible for the
conservation of the four-momentum $P_{\nu}=\int\mathrm{d}^{3}x\Theta_{0\nu}$.
The time-like part $H=P_{0}=\int\mathrm{d}^{3}x\Theta_{00}$ is usually referred 
to as the Hamilton function and coincides with the energy of the system. 
Recall that it is not possible to derive all constraints
from $H$, if one uses $\Theta_{00}$ from Eq.\ (\ref{eq:energieimpulstensor}), since 
the constraints have already been taken into account to derive this expression. 
We now define
\begin{equation}
\Pi^{\alpha \beta} \equiv \Pi^{\alpha \beta 0} = \partial^0 T^{\alpha \beta}\,, \label{eq:impulsfeld}
\end{equation}
where we used Eq.\ (\ref{eq:momentumfield}) and applied the constraints 
(\ref{eq:symmetrie}), (\ref{eq:traceless}), and (\ref{eq:lorentzcond}).
Then, using Eq.\ (\ref{eq:energieimpulstensor})
the explicit forms for $H$ and $\vec{P}$ for spin-two fields are:
\begin{align}
H & = \frac{1}{2} \int \mathrm{d}^{3} x \left[ \Pi_{\alpha\beta} \Pi^{\alpha\beta} 
+ \left(\vec{\nabla} T_{\alpha\beta}\right) \cdot \vec{\nabla} T^{\alpha\beta} 
+ m^{2} T_{\alpha\beta} T^{\alpha\beta} \right] \,  \vphantom{\int} \,, \label{eq:energiefunktion}
\\
\vec{P} & = - \int \mathrm{d}^{3} x \, \Pi_{\alpha\beta} \,\vec{\nabla}T^{\alpha\beta} 
\, \vphantom{\int} \,. \label{eq:impulsfunktion}%
\end{align}
A similar result for the Hamilton function $H$ of the spin-two field was already
given in Refs.\ \cite[p.1314]{Chang2} and  \cite[p.1262]{Chang1}.

The remaining symmetry transformations are the Lorentz transformations, which imply the
conservation of angular momentum $\vec{L}$ and spin $\vec{S}$, see e.g.\ Ref.\
\cite[p.44-46]{Reinhardt}. In general the angular-momentum density reads:
\begin{align}
\mathcal{L}^{\varepsilon\eta} & = \Theta^{0\eta} x^\varepsilon - \Theta^{0\varepsilon} x^\eta
\,.
\end{align}
Then the $m$-component of the angular-momentum vector is
\begin{align}
L_{m} & = \frac{1}{2}\varepsilon_{mnl}\int\mathrm{d}^{3}x \, \mathcal{L}_{nl} \nonumber
\\
&  =\frac{1}{2}\varepsilon_{mnl}\int\mathrm{d}^{3}x \, \frac{\partial\mathcal{L}}%
{\partial\left(  \partial^{0}T_{\alpha\beta}\right)  }\left[ \left(  \partial_l
T_{\alpha\beta}\right) x_{n}-\left(\partial_n T_{\alpha\beta}\right)x_{l}\right]  \ \nonumber
\\
&  =-\varepsilon_{mnl}\int\mathrm{d}^{3}x \,\Pi_{\alpha\beta}\left(  \partial_n
T^{\alpha\beta}\right)x_{l}  \ \,.%
\end{align}
The covariant spin density is defined as
\begin{align}
\mathcal{S}^{\varepsilon\eta}  &  
= -i \,\frac{\partial\mathcal{L}}{\partial \left(\partial^{0}T^{\mu\nu}\right)} 
\left(  S^{\varepsilon\eta}\right)
^{\mu\lambda\nu\rho}T_{\lambda\rho}\nonumber
\\
&  =-i \, \Pi_{\mu\nu}\left(  S^{\varepsilon\eta}\right)^{\mu\lambda\nu\rho}T_{\lambda\rho}%
\vphantom{\frac{\partial\mathcal{L}}{\partial\left(\partial^{0}T^{\mu\nu}\right)}}\nonumber
\\
&  =\Pi_{\mu\nu}\left[  \left(  g^{\varepsilon\mu}g^{\eta\lambda}-g^{\eta\mu
}g^{\varepsilon\lambda}\right)  g^{\nu\rho}+\left(  g^{\varepsilon\nu}%
g^{\eta\rho}-g^{\eta\nu}g^{\varepsilon\rho}\right)  g^{\mu\lambda}\right]
T_{\lambda\rho} \vphantom{\frac{\partial\mathcal{L}}{\partial\left(\partial^{0}T^{\mu\nu}\right)}} 
\nonumber\\
&  =2\left(  \Pi^{\varepsilon\rho}\tensor{T}{^\eta_\rho}-\Pi^{\eta\rho
}\tensor{T}{^\varepsilon_\rho}\right)
\vphantom{\frac{\partial\mathcal{L}}{\partial\left(\partial^{0}T^{\mu\nu}\right)}}\,,%
\label{eq:cov_spin_dens}
\end{align}
where we used the symmetry of $\Pi^{\alpha \beta}$ and $T^{\alpha \beta}$, as well as
the definition of the spin part (\ref{eq:tensorspingen}) of the generator for 
Lorentz transformations in spin-two field representation.
The $m$-component of the spin vector is defined as
\begin{align}
S_{m}  &  =\frac{1}{2}\varepsilon_{mnl}\int\mathrm{d}^{3}x\,
\mathcal{S}_{nl} \nonumber
\\
&  =2\, \varepsilon_{mnl}\int\mathrm{d}^{3}x\,
\tensor{\Pi}{_n_\rho}\tensor{T}{_l^\rho}\,,
\label{eq:spinfunktion}
\end{align}
where we used Eq.\ (\ref{eq:cov_spin_dens}) and the symmetry of 
$\Pi^{\alpha \beta}$ and $T^{\alpha \beta}$.
Finally, in vector notation $\vec{L}$ and $\vec{S}$ read:
\begin{align}
\vec{L}  &  =\int\mathrm{d}^{3}x\, \Pi_{\alpha\beta}\left(  \vec{x}\times\vec{\nabla}%
T^{\alpha\beta}\right)  \,,\label{eq:drehimpulsfunktion}\\
\vec{S}  &  =2\int\mathrm{d}^{3}x\, \vec{\Pi}_{\rho}\times\vec{T}^{\rho}\,.
\label{eq:spincompact}%
\end{align}
Here we defined $\vec{\Pi}_\rho = \Pi_{n\rho} \vec{e}_n,\, \vec{T}^{\rho}
= \tensor{T}{_n^\rho} \vec{e}_n$, where $\vec{e}_n$ is the Cartesian
unit vector in spatial $n$-direction.
Note that the factor 2 in front of expression (\ref{eq:spincompact}) corresponds to
the fact that our fields are rank-two Lorentz tensors that obey the constraints
(\ref{eq:symmetrie}), (\ref{eq:traceless}), and (\ref{eq:lorentzcond}).
As a consequence, if the integral in Eq.\ (\ref{eq:spincompact}) is properly normalized,
the correct value for the spin arises already from classical
relativistic field theory, without any need of quantization. 
To our knowledge, expressions (\ref{eq:spinfunktion}) and (\ref{eq:spincompact}) 
are novel results (for angular momentum, see Chang \cite[p.1262]{Chang2}).

\subsection{Solution of the equations of motion and polarization tensors}
\label{sec:solutionpolarisation}

The solution of the free KG equation (\ref{eq:kleingordoneq}) for real-valued
scalar \cite[p.76f]{Reinhardt}
and vector fields \cite[p.160f]{Reinhardt} is well known. This is
easily generalized for spin-two fields,
\begin{equation}
\tensor{T}{_\mu_\nu}(x)=\int\frac{\mathrm{d}^{3}k}{\sqrt{2\pi}^{3}}\frac
{1}{2\omega_{k}}\sum_{\lambda=1}^{5}\epsilon_{\mu\nu}(\vec{k},\lambda)\left[
a(\vec{k},\lambda)e^{-ikx}+a^{\ast}(\vec{k},\lambda)e^{+ikx}\right]
_{k^{0}=\omega_{k}}\,, \label{kgtensor}%
\end{equation}
where $\omega_k=\sqrt{\vec{k}^{\,2} + m^2}$ and where
we introduced polarization tensors $\epsilon_{\mu\nu}(\vec{k},\lambda)$ in
generalization of the polarization vectors $\epsilon_\mu(\vec{k},\lambda)$ of the
spin-one case.

The polarization tensors can be explicitly determined using the constraints
(\ref{eq:symmetrie}), (\ref{eq:traceless}), and (\ref{eq:lorentzcond}). 
When applied to Eq.\ (\ref{kgtensor}) one
obtains\footnote{Let us remark that, in the case of 
massless fields, one has additional constraints from the local (gauge) symmetry,
which further reduce the number of degrees of freedom. Consequently, 
this leads to a reduction in the number of polarization
vectors or tensors, respectively.}:
\begin{align}
\tensor{\epsilon}{^\mu^\nu}(\vec{k},\lambda) 
- \tensor{\epsilon}{^\nu^\mu}(\vec{k},\lambda) & = 0\,, \label{eq:symm}\\
k_{\mu}\tensor{\epsilon}{^\mu^\nu}(\vec{k},\lambda)  &  =0\,,
\label{eq:spatial}\\
g_{\mu\nu}\tensor{\epsilon}{^\mu^\nu}(\vec{k},\lambda)  &
=0\label{eq:trace}\, .
\end{align}
We choose a specific solution fulfilling the orthonormality condition
\begin{equation}
\tensor{\epsilon}{_\mu_\nu}(\vec{k},\lambda)\tensor{\epsilon}{^\mu^\nu}(\vec{k},\lambda
^{\prime})=\delta_{\lambda\lambda^{\prime}}\,. \label{eq:ortho}%
\end{equation}
In the rest frame of the field, where $k^{\mu}=(m,\vec{0})^{T}$, a possible
choice is:
\begin{align}
\tensor{\epsilon}{^\mu^\nu}(\vec{0},1)& =\frac{1}{\sqrt{2}}%
\begin{pmatrix}
0 & 0 & 0 & 0\\
0 & 1 & 0 & 0\\
0 & 0 & -1 & 0\\
0 & 0 & 0 & 0
\end{pmatrix}
\,, \label{pol1} \\[0.1cm]
\tensor{\epsilon}{^\mu^\nu}(\vec{0},2)& =\frac{1}{\sqrt{2}}%
\begin{pmatrix}
0 & 0 & 0 & 0\\
0 & 0 & 1 & 0\\
0 & 1 & 0 & 0\\
0 & 0 & 0 & 0
\end{pmatrix}
\,, \label{pol2} \\[0.1cm]
\tensor{\epsilon}{^\mu^\nu}(\vec{0},3)& =\frac{1}{\sqrt{2}}%
\begin{pmatrix}
0 & 0 & 0 & 0\\
0 & 0 & 0 & 1\\
0 & 0 & 0 & 0\\
0 & 1 & 0 & 0
\end{pmatrix}
\,, \label{pol3}\\[0.1cm]
\tensor{\epsilon}{^\mu^\nu}(\vec{0},4)& =\frac{1}{\sqrt{2}}%
\begin{pmatrix}
0 & 0 & 0 & 0\\
0 & 0 & 0 & 0\\
0 & 0 & 0 & 1\\
0 & 0 & 1 & 0
\end{pmatrix}
\,, \label{pol4} \\[0.1cm]
\tensor{\epsilon}{^\mu^\nu}(\vec{0},5)& =\frac{1}{\sqrt{6}}%
\begin{pmatrix}
0 & 0 & 0 & 0\\
0 & 1 & 0 & 0\\
0 & 0 & 1 & 0\\
0 & 0 & 0 & -2
\end{pmatrix}
\,. \label{pol5}
\end{align}
In order to find a representation of these polarization tensors
in another inertial frame, one has to perform a Lorentz boost, see App.\
\ref{app:polarisationboost}. 

The polarization tensors appear in several forms in different papers. Chang
\cite[p.1263]{Chang1} only mentions the results without discussing the details
and Hinterbichler \cite[p.17-19]{Hinterbichler} and Huang et al.\
\cite[p.67,68]{Huang} give a short presentation of this topic. Note that, in the
massless case (such as in the case of the graviton), only two of these five
tensors are needed. For example, Carroll uses $\lambda=1,2$ in his treatment
of gravitational waves \cite[p.151]{Carroll}. He also discusses interesting
visualizations of oscillations along these polarization directions
\cite[p.153,154]{Carroll} which are helpful to get a better idea of the
behavior of such fields.

An important property of the polarization tensors
is the completeness relation, see App.\ \ref{app:completenessrelation}:
\begin{equation}
\sum_{\lambda=1}^{5}\epsilon_{\mu\nu}(\vec{k},\lambda)\epsilon_{\alpha\beta}%
(\vec{k},\lambda)=\frac{1}{2}(G_{\mu\alpha}G_{\nu\beta}+G_{\mu\beta}G_{\nu\alpha})
-\frac{1}{3}G_{\mu\nu}G_{\alpha\beta} \,, \label{eq:vollrel}%
\end{equation}
where the tensor $G_{\mu\nu}$ is defined as
\begin{align}
G_{\mu \nu} & \equiv
g_{\mu\nu}-\frac{k_{\mu}k_{\nu}}{m^{2}} \;. \label{Gproj}
\end{align}
This tensor will appear again in the discussion of the tree-level
propagator, Sec.\ \ref{sec:propagator},
and when quantizing the fields.

\subsection{The tree-level propagator}

\label{sec:propagator} 

The tree-level propagator for the spin-two field is well known,
see e.g.\ Zee \cite[p.35,36]{Zee}, and was already described by Hinterbichler
\cite[p.21]{Hinterbichler}. Here we derive it from the Lagrangian 
(\ref{eq:lagrangedichte}), which we write in the form
\[
\mathcal{L}=\frac{1}{2}T^{\nu\rho}D_{\nu\rho\alpha\beta}^{-1}T^{\alpha\beta}\, .
\]
Fourier-transforming the differential operator $D_{\nu\rho\alpha\beta}^{-1}$ 
into momentum space (and for the sake of convenience denoting
the result with the same symbol), we obtain\footnote{
Hinterbichler \cite[p.21]{Hinterbichler}
obtains (aside from different sign conventions) the same result. Nevertheless, 
his Lagrangian already accounts for the symmetry of the tensor fields. This requires 
that the analogue of the differential operator
(\ref{eq:differentialoperator}) in his treatment must be symmetrized by hand.}
\begin{align}
D_{\nu\rho\alpha\beta}^{-1}  &  = + \frac{1}{2}\left(  k^{2}-m^{2}\right)
\left(  g_{\nu\alpha}g_{\rho\beta}+g_{\nu\beta}g_{\rho\alpha}
 -2\,   g_{\nu\rho}g_{\alpha\beta}\right)  \vphantom{\frac{1}{2}}\nonumber\\
&  \quad-\frac{1}{2}\left(  g_{\rho\beta}k_{\nu}k_{\alpha}+g_{\rho\alpha
}k_{\nu}k_{\beta}+g_{\nu\alpha}k_{\rho}k_{\beta}+g_{\nu\beta}k_{\rho}%
k_{\alpha} -2\,  g_{\nu\rho}k_{\alpha}k_{\beta} -2\, g_{\alpha\beta}k_{\nu}k_{\rho
} \right) \vphantom{\frac{1}{2}}\,. \label{eq:differentialoperator}%
\end{align}
The tree-level propagator $P^{\alpha\beta\sigma\lambda}$ of the spin-two field is 
the inverse of this differential operator,
\begin{equation}
D_{\nu\rho\alpha\beta}^{-1}P^{\alpha\beta\sigma\lambda}=\frac{1}{2}\left(
g_{\nu}^{\sigma}g_{\rho}^{\lambda}+g_{\nu}^{\lambda}g_{\rho}^{\sigma}\right)\, .
\label{eq:inversionprop}%
\end{equation}
Note that the rank-four identity tensor is symmetric in all indices. 
In order to determine $P_{\alpha\beta\sigma\lambda}$
we write it as a sum of all possible rank-four tensors formed from $k^\mu$ and
the metric tensor\footnote{A similar approach is proposed in exercise 1.5.1
of Zee \cite[p.39]{Zee} to find the completeness relation (\ref{eq:completenessrel}),
which is the numerator of the tree-level propagator.}. Choosing this expansion to be
symmetric under the exchanges $\sigma \leftrightarrow \lambda$,
$\alpha \leftrightarrow \beta$, $(\sigma \lambda) \leftrightarrow (\alpha \beta)$, we
obtain:
\begin{align}
P_{\alpha\beta\sigma\lambda}  &  =A\left(  g_{\alpha\sigma}g_{\beta\lambda
}+g_{\alpha\lambda}g_{\beta\sigma}\right)+B\,  g_{\alpha\beta}g_{\sigma\lambda} \nonumber\\
&  \quad+C\left(  g_{\beta\lambda}k_{\alpha}k_{\sigma}+g_{\beta\sigma
}k_{\alpha}k_{\lambda}+g_{\alpha\sigma}k_{\beta}k_{\lambda}+g_{\alpha\lambda
}k_{\beta}k_{\sigma}\right) \nonumber\\
&  \quad+D\left(  g_{\sigma\lambda}k_{\alpha}k_{\beta}+g_{\alpha\beta
}k_{\sigma}k_{\lambda}\right) +E\,  k_{\alpha}k_{\beta}k_{\sigma}k_{\lambda}\,,
\label{eq:approachprop}
\end{align}
with coefficients $A,B,C,D,E$ that are determined in App.\ \ref{app:propagatorcalc}. 
The result is
\begin{equation} \label{eq:treelevelprop}
P_{\alpha\beta\sigma\lambda}=\frac{G_{\alpha\sigma}G_{\beta\lambda}%
+G_{\alpha\lambda}G_{\beta\sigma}-\frac{2}{3}\,  G_{\alpha\beta}%
G_{\sigma\lambda}  }{2\left(  k^{2}-m^{2}\right)  } \,.
\end{equation}
Note that our derivation of the tree-level propagator is based on the
Lagrangian (\ref{eq:lagrangedichte}), which a priori does not assume that the
spin-two field is symmetric and traceless. In comparison,
Wagenaar \& Rijken derive a non-covariant version of the
tree-level propagator \cite[p.10]{Wagenaar}. For other works which discuss the 
tree-level propagator of spin-two fields, see Bhargava \& Watanabe
\cite[p.276]{Bhargava}, Macfarlane \& Tait \cite[p.223]{Macfarlane}, Chang
\cite[p.1263]{Chang2}, and Rivers \cite[p.394]{Rivers}.

\section{Quantized spin-two fields}
\label{sec:Sec3}

\subsection{Canonical quantization}
\label{sec:canonicalquantisation} 

Various works show how to quantize a
spin-two field by making use of the canonical commutation relations.
Here, we first provide a short description of this procedure and then apply it
to the conserved quantities.
The first step in the procedure of second quantization is to promote the fields to operators:
\begin{equation}
T_{\mu\nu}(x)\longrightarrow\hat{T}_{\mu\nu}(x) \text{ , } 
\Pi_{\mu\nu}(x)\longrightarrow\hat{\Pi}_{\mu\nu}(x)\,.%
\end{equation}
Then, the Fourier coefficients $a(\vec{k},\lambda)$ and $a^{\ast}(\vec{k},\lambda)$
become annihilation and creation operators:
\begin{align}
a(\vec{k},\lambda)\longrightarrow\hat{a}(\vec{k},\lambda)\text{ , }a^{\ast}%
(\vec{k},\lambda)\longrightarrow\hat{a}^{\dagger}(\vec{k},\lambda)\,.
\label{eq:creationannihilation}%
\end{align}
This leads to the following operators for the field and its canonically
conjugate field, which can be expressed in terms of the time derivative of the tensor field
according to Eq.\ (\ref{eq:impulsfeld}),
\begin{align}
\tensor{\hat{T}}{_\mu_\nu}(x) & = \int \frac{\mathrm{d}^{3}k}{\sqrt{2\pi}^{3}} 
\frac{1}{2\omega_{k}} \sum_{\lambda=1}^{5} \epsilon_{\mu\nu}(\vec{k},\lambda) 
\left[ \hat{a}(\vec{k},\lambda) e^{-ikx} + \hat{a}^{\dagger}(\vec{k},\lambda) e^{+ikx} 
\right]_{k^{0}=\omega_{k}} \,, \label{eq:loesung}
\\
\tensor{\hat{\Pi}}{_\mu_\nu}(x) & = \int \frac{\mathrm{d}^{3}k}{\sqrt{2\pi}^{3}} 
\frac{1}{2\omega_{k}} \sum_{\lambda=1}^{5} \epsilon_{\mu\nu}(\vec{k},\lambda) 
\left[ -i\, \omega_{k}\, \hat{a}(\vec{k},\lambda) e^{-ikx} + i\, \omega_{k}\, 
\hat{a}^{\dagger}(\vec{k},\lambda) e^{+ikx} \right]_{k^{0}=\omega_{k}} \,. 
\label{eq:kankonjlosung}
\end{align}
Following the discussion in Sec.\ \ref{sec:solutionpolarisation}, 
these operators fulfill Eqs.\ (\ref{eq:kleingordoneq}),
(\ref{eq:symmetrie}), (\ref{eq:traceless}), and (\ref{eq:lorentzcond}). 

There are two possibilities to proceed with quantization: (i) one postulates commutation
relations for the fields (\ref{eq:loesung}), (\ref{eq:kankonjlosung}) and
derives corresponding relations for the creation and annihilation operators
(\ref{eq:creationannihilation}).
This approach is rather similar to the discussion of the
Poisson brackets in classical mechanics, see e.g.\ Baklini \& Tuite \cite[p.14]{Baaklini}
for the case of vector fields. Equivalently, (ii) one may
postulate commutation relations for the creation and annihilation operators and derives
corresponding relations for the fields, see e.g.\ the discussion in
Reinhardt \& Greiner \cite[p.163]{Reinhardt} for the Proca field. 
Here, we proceed with option (ii) and postulate
\begin{align}
\left[  \hat{a}(\vec{k},\lambda),\hat{a}^{\dagger}(\vec{k}^{\prime},\lambda^{\prime})\right] _{-}  
&  =2\, \omega_{k}\, \delta_{\lambda\lambda^{\prime}}\, 
\delta^{3}(\vec{k}-\vec{k}^{\prime})\,,\label{eq:antikomm}\\
\left[  \hat{a}(\vec{k},\lambda),\hat{a}(\vec{k}^{\prime},\lambda^{\prime})\right]  _{-}  
&  =0\,,\\
\left[  \hat{a}^{\dagger}(\vec{k},\lambda),
\hat{a}^{\dagger}(\vec{k}^{\prime},\lambda^{\prime})\right]  _{-}  &  =0\,.
\end{align}
It is useful to define the so-called number operator $\hat{N}(\vec{k},\lambda)$ via:
\begin{equation} \label{eq:numberop}
\hat{a}^{\dagger}(\vec{k},\lambda)\hat{a}(\vec{k},\lambda)=2\, \omega_{k}\, %
\delta^{3}(0)\, \hat{N}(\vec{k},\lambda)\,,
\end{equation}
where $\delta^{3}(0) \rightarrow V/(2 \pi)^3$ in a finite volume $V < \infty$.

Now it is straightforward to calculate the commutators of the fields. The explicit
calculation is relegated to App.\ \ref{app:commutators}. 
The equal-time commutation relations for 
the spatial components of the spin-two field and its canonically conjugate field then read:
\begin{align}
\left[ \tensor{\hat{T}}{_i_j}(t,\vec{x}) , \tensor{\hat{T}}{_l_k}(t,\vec
{y}) \right]_{-} & = 0 \vphantom{\left(\frac{\partial^2}{3}\right)} \,, \label{eq:commutatorTT}
\\
\left[ \tensor{\hat{\Pi}}{_i_j}(t,\vec{x}) , \tensor{\hat{\Pi}}{_l_k}(t,\vec
{y}) \right]_{-} & = 0 \vphantom{\left(\frac{\partial^2}{3}\right)} \,, \label{eq:commutatorPP}
\\
\left[ \tensor{\hat{T}}{_i_j}(t,\vec{x}) , \tensor{\hat{\Pi}}{_l_k}(t,\vec{y})\right]_{-}
& = \left[ \frac{1}{2} \left( \tilde{G}_{ik}\tilde{G}_{jl} +\tilde{G}_{il}\tilde{G}_{jk} \right)
- \frac{1}{3}  \tilde{G}_{ij}\tilde{G}_{kl} \right] i \, \delta^{3}(\vec{x}-\vec{y}) \,,
\label{eq:commutatorTP}
\end{align}
where 
\begin{align} \label{eq:tildeGproj}
\tilde{G}_{ij} \equiv - \delta_{ij} + \frac{\partial_{i}\partial_{j}}{m^{2}}
\end{align}
is the Fourier-transform of the on-shell projection operator (\ref{Gproj}).

It is remarkable that the expression in brackets in the last commutator
(\ref{eq:commutatorTP}) is (up to a factor of two) the Fourier-transform of
the numerator of the propagator (\ref{eq:treelevelprop}). It would certainly have been difficult to
guess this result and thus quantize the spin-two field following option (i) discussed
in the beginning of this subsection.

Similar results for the equal-time commutation relations were also given by
Wagenaar \& Rijken \cite[p.9]{Wagenaar} where, as already mentioned, Lagrange
multipliers were used. Also, Bhargava \& Watanabe presented an expression for
the covariant commutator of the charged tensor field \cite[p.277]{Bhargava}. 
Chang derived a
covariant commutation relation for the field \cite[p.1261]{Chang1}, which is
in accordance with Eq.\ (\ref{eq:nichtvertasuch}). More details can be found 
in Ref.\ \cite[p.1312]{Chang2}. Another discussion which leads to similar results is
given by Rivers \cite[p.400,403]{Rivers}.

\subsection{Operators for conserved quantities}
\label{sec:specialoperators} 

The conserved quantities presented in Sec.\ \ref{sec:conservedquantities}
become operators once the fields are quantized. Here we express the Hamilton
operator, the momentum operator, and the spin operator in $z$-direction as functions 
of the creation and annihilation operators (for more details see 
App.\ \ref{app:operators}).

The Hamilton operator is given by the quantized version of Eq.\
(\ref{eq:energiefunktion}):
\begin{align}
\hat{H}  &  =\frac{1}{2}\int\mathrm{d}^{3}x\left(  \hat{\Pi}_{\mu\nu}\hat{\Pi
}^{\mu\nu}+\vec{\nabla}\hat{T}_{\mu\nu}\cdot \vec{\nabla}\hat{T}^{\mu\nu}+m^{2}%
\hat{T}_{\mu\nu}\hat{T}^{\mu\nu}\right)
\vphantom{\sum_{\lambda,\lambda'=1}^{5}}\nonumber\\
&  =\delta^{3}(0)\int\mathrm{d}^{3}k\sum_{\lambda=1}^{5}\omega_{k}\left[
\hat{N}(\vec{k},\lambda)+\frac{1}{2}\right]\,, \label{eq:HOp}
\end{align}
while the momentum of the field, Eq.\ (\ref{eq:impulsfunktion}), takes the form:
\begin{align}
\hat{\vec{P}}  &  =-\int \mathrm{d}^{3}x \,  \hat{\Pi}_{\alpha\beta}\vec{\nabla}\hat
{T}^{\alpha\beta} \vphantom{\sum_{\lambda,\lambda'=1}^{5}}\nonumber
\\
&  = \delta^{3}(0)\int\mathrm{d}^{3}k\sum_{\lambda=1}^{5}\vec{k}\, \hat{N}(\vec{k},\lambda)  \,.
\label{eq:POp}
\end{align}
These results are familiar and consistent with the results for fields with other spin values.
The energy of the field is given by a vacuum term and the number of all
excitations with all possible polarisations and momenta multiplied by the respective
energy. The momentum operator can be interpreted analogously, except that
there is no momentum associated with the vacuum.

The derivation of the $z$-component of the spin operator is presented 
in App.\ \ref{app:operators}, with the result:
\begin{align}
\hat{S}_{z}  &  =-2\int \mathrm{d}^{3}x 
\left(  \hat{\Pi}_{1\rho}\hat{T}^{2\rho}-\hat{\Pi}_{2\rho}\hat{T}^{1\rho}\right) 
\vphantom{\sum_{\lambda,\lambda'=1}^{5}}\nonumber
\\
&  =\delta^{3}(0)\int\mathrm{d}^{3}k\left[  2\, \hat{N}(\vec{k},+)-2\, \hat{N}%
(\vec{k},-)+\hat{N}(\vec{k},\Delta)-\hat{N}(\vec{k},\Box)\right]  \,,
\label{eq:spinoperator}%
\end{align}
where we used number operators in a circularly polarized bases,
denoted by $\lambda = +, - , \Delta, \Box,$ and $0$, see Eq.\ (\ref{eq:circpol}).
This is a physically very intuitive result, since it expresses
the total spin as the sum over all particles with definite momentum and
(circular) polarization direction, weighted
with the modulus of the spin projection in $z$-direction.
The result (\ref{eq:spinoperator}) is analogous to that for vector fields, see e.g. Reinhardt \&
Greiner \cite[p.166]{Reinhardt}. 
Note that only in the circularly polarized basis the $z$-component of the spin operator
$\hat{S}_z$ is diagonal in the polarization directions, such that
the corresponding number operators can be employed.
Had we chosen a different quantization axis, e.g.\ the $x$-direction, the diagonalization
of the spin operator is not as straightforward, but can be achieved by a rotation of the basis
of polarization tensors, or by choosing a basis for the annihilation and
creation operators which is different from Eq.\ (\ref{eq:circpol}).

\section{Conclusions and outlook}

\label{sec:outlook}

Spin-two fields play an important role in gravity \cite{Carroll,Deser} and in its various
extensions where even massive gravitons appear, see e.g.\ Refs.\
\cite{arkani,rubakov} and refs.\ therein, as well as in extensions of the
standard model \cite{colladay}. In the framework of QCD, tensor mesons
naturally emerge as composite quark-antiquark bound states
\cite{Suzuki,Ye,Anisovich,Burakovsky,Cotanch,Giacosa}. It is therefore
important to have a solid theoretical basis for the field-theoretical description
of spin-two fields. To this
end, in this work we have reviewed known features and presented new aspects of
massive spin-two fields in both classical and quantum field theory. 

For classical tensor fields, we have
determined the equations of motion as well as the Fierz-Pauli constraints 
for tensor fields via three different methods: (i) the eigenvalue equations
for the Casimir operators, (ii) a Lagrangian, and (iii) a covariant Hamilton approach. 
We have also
computed the energy-momentum tensor, conserved quantities, and the solution
of the equations of motion in terms of a basis of polarization tensors. 
To conclude the discussion of classical fields, we have also presented the
corresponding tree-level propagator.

Quantization required us to postulate commutation relations for the annihilation and 
creation operators. From these and the solution of the equations of motion we
have derived equal-time commutation relations for the fields and their canonically
conjugate counterparts. We have also computed the Hamilton, momentum, and
spin operators.

As an application, let us consider tensor mesons in QCD. Conventionally,
mesons are bound states of a quark and an antiquark, but other, more exotic
possibilities, such as glueballs, hybrids, tetraquarks, etc.\ exist, too (see the
general discussion in Ref.\ \cite{olive} and for the
phenomenology of tensor mesons see e.g.\ Refs.\
\cite{Suzuki,Ye,Anisovich,Burakovsky,Cotanch,Giacosa} and refs.\ therein).
The theoretical description of such objects, be it via lattice QCD or
via QCD sum rules, requires us to construct composite operators with 
the appropriate quantum numbers. 

In the conventional picture of a meson, i.e., a quark-antiquark bound state, 
the spin can either be $S=0$ or $S=1$, while the orbital angular momentum
$L$ can assume all integer values, $L=0,1,2,\ldots$. The total angular momentum 
$\vec{J}=\vec{L}+\vec{S}$ is an integer limited by $\left\vert L-S\right\vert \leq J\leq
L+S$. In order to obtain tensor mesons with $J=2$, we can therefore
either combine $S=0$ with $L=2$, or we can combine
$S=1$ with $L=1,2,$ or $3$.

In general one classifies different particle states in quantum field theory by
their transformation behavior under parity and charge conjugation. For neutral
unflavored mesons these values are given by:
\[
P=\left(  -1\right)  ^{L+1}\ ,\text{ }C=\left(  -1\right)  ^{L+S}\ .
\]
Together with the total spin $J$, the quantum numbers are summarized by $J^{PC}$.
For the specific case of $J=2$, there are three possibilities, which
we study in the following together with their microscopic quark-antiquark
currents. These are objects constructed from fermionic quark fields
$q_{i}(x)$ (where $i$ stands for flavor $u,d,s,c,b,t$) and should have the
correct behavior under Poincar\'{e}, parity, and flavor transformations, as well as
charge conjugation and time reversal. Moreover, they should fulfill
the KG equation (\ref{eq:kleingordoneq}) and the constraints
(\ref{eq:symmetrie}), (\ref{eq:traceless}), and (\ref{eq:lorentzcond}). 
Thus, the following spin-two quark-antiquark currents emerge:

(1) \textbf{Tensor mesons }$J^{PC}=$\textbf{\ $2^{++}.$ }They arise from
$S=1$ and $L=1$ or $3$, which couple to $J^{PC}=2^{++}$. 
The corresponding current was previously
introduced e.g.\ in Ref.\ \cite{Wang}:
\begin{equation}
(X_{\mu\nu})_{ji}=i\bar{q}_{i}\left[  \gamma_{\mu}\overleftrightarrow
{\partial}_{\nu}+\gamma_{\nu}\overleftrightarrow{\partial}_{\mu}-\frac{2}{3}
\left(  g_{\mu\nu}-\frac{k_{\mu}k_{\nu}}{k^2}\right)  \gamma_{\delta}
\overleftrightarrow{\partial}^{\delta}\right]  q_{j}\, ,
\end{equation}
where $\overleftrightarrow{\partial}_{\mu} \equiv \overrightarrow{\partial}_\mu
- \overleftarrow{\partial}_\mu$ and $k^\mu$ is the total momentum of the meson.

(2) \textbf{Axial-tensor mesons }$J^{PC}=$ \textbf{$2^{--}.$} They arise from
$S=1$ and $L=2.$ The respective current was introduced in Ref.\ \cite{chenzhu}
and reads explicitly:
\begin{equation}
(B_{\mu\nu})_{ji}=i\bar{q}_{i}\left[  \gamma_{\mu}\gamma^{5}
\overleftrightarrow{\partial}_{\nu}+\gamma_{\nu}\gamma^{5}
\overleftrightarrow{\partial}_{\mu}-\frac{2}{3}\left(  g_{\mu\nu}
-\frac{k_{\mu}k_{\nu}}{k^2}\right)  \gamma_{\delta}
\gamma^{5}\overleftrightarrow{\partial}^{\delta}\right]  q_{j}\,.
\end{equation}

(3) \textbf{Pseudotensor mesons }$J^{PC}=$\textbf{$2^{-+}.$ }They arise from
$S=0$ and $L=2$. The respective current, to our knowledge
presented here for the first time, is:
\begin{equation}
(T_{\mu\nu})_{ji}=i\bar{q}_{i}\left[  \overleftrightarrow{\partial}_{\mu
}\gamma^{5}\overleftrightarrow{\partial}_{\nu}-\frac{2}{3}\left(  g_{\mu\nu
}-\frac{k_{\mu}k_{\nu}}{k^2}\right)  \left(  \overleftrightarrow{\partial
}_{\alpha}\gamma^{5}\overleftrightarrow{\partial}^{\alpha}\right)  \right]
q_{j}\,.%
\end{equation}
The decays of such pseudotensor mesons will be discussed in a forthcoming publication.

\bigskip

\textbf{Acknowledgements:} A.K.\ thanks J.\ Struckmeier for introducing him to the
covariant Hamilton formalism. A.K.\ also acknowledges valuable discussions with 
P.\ Kovacs, J.\ Reinhardt, J.\ Struckmeier, and M.\ Zetenyi.
F.G.\ thanks K.\ Shekhter for useful discussions. 

\bigskip

\appendix
\addcontentsline{toc}{section}{Appendices}

\section{Poincar\'{e} algebra for the generators in spin-two field representation}
\label{app:generatoralgebra} 

In this appendix we prove that the generators (\ref{eq:generatortensortrans}) and 
(\ref{eq:generators}) in spin-two field representation fulfil the Poincar\'{e} algebra
(\ref{eq:lorentzalgebra}) -- (\ref{eq:kommutatorPI}). 
We start with Eq.\ (\ref{eq:lorentzalgebra}), which reads for spin-two fields
\begin{align}
\tensor{ { \left[ {I_{\alpha\beta}} , {I_{\eta\epsilon}} \right]_{-}} }{^\mu_\gamma^\nu_\delta}
& \equiv \tensor{(I_{\alpha\beta})}{^\mu_\sigma^\nu_\rho} 
\tensor{(I_{\eta\epsilon})}{^\sigma_\gamma^\rho_\delta} - 
\tensor{(I_{\eta\epsilon})}{^\mu_\sigma^\nu_\rho}
\tensor{(I_{\alpha\beta})}{^\sigma_\gamma^\rho_\delta}\nonumber \\
& = i \left[ -g_{\alpha\eta} \tensor{(I_{\beta\epsilon})}{^\mu_\gamma^\nu_\delta} 
+ g_{\alpha\epsilon} \tensor{(I_{\beta\eta})}{^\mu_\gamma^\nu_\delta} 
+ g_{\beta\eta} \tensor{(I_{\alpha\epsilon})}{^\mu_\gamma^\nu_\delta} 
- g_{\beta\epsilon} \tensor{(I_{\alpha\eta})}{^\mu_\gamma^\nu_\delta} \right] \,. 
\label{eq:lorentzalgebra_tensor}
\end{align}
Now we use the decomposition into a spin part 
$\tensor{(S_{\alpha\beta})}{^\mu_\sigma^\nu_\rho}$, cf.\ Eq.\ (\ref{eq:tensorspingen}), 
and an angular momentum part $\tensor{(L_{\alpha\beta})}{^\mu_\sigma^\nu_\rho}$,
cf.\ Eq.\ (\ref{eq:tensordrehgen}), and insert this into the left-hand side of 
Eq.\ (\ref{eq:lorentzalgebra_tensor}):
\begin{align}
\tensor{ { \left[ {I_{\alpha\beta}} , {I_{\eta\epsilon}} \right]_{-}} }{^\mu_\gamma^\nu_\delta}
& =  \quad
\tensor{ { \left[ {S_{\alpha\beta}} , {S_{\eta\epsilon}} \right]_{-}} }{^\mu_\gamma^\nu_\delta}
+ \tensor{ { \left[ {S_{\alpha\beta}} , {L_{\eta\epsilon}} \right]_{-}} }{^\mu_\gamma^\nu_\delta}
 \nonumber
\\
& \quad \, +
\tensor{ { \left[ {L_{\alpha\beta}} , {S_{\eta\epsilon}} \right]_{-}} }{^\mu_\gamma^\nu_\delta}
+ \tensor{ { \left[ {L_{\alpha\beta}} , {L_{\eta\epsilon}} \right]_{-}} }{^\mu_\gamma^\nu_\delta}
\,.
\label{eq:LScommutators}
\end{align}
Now we compute the four commutators on the right-hand side separately. Inserting the 
explicit expression (\ref{eq:tensorspingen}) for the generators in the first commutator, 
we obtain
\begin{align}
\tensor{ { \left[ {S_{\alpha\beta}} , {S_{\eta\epsilon}} \right]_{-}} }{^\mu_\gamma^\nu_\delta}
& =\tensor{(S_{\alpha\beta})}{^\mu_\sigma^\nu_\rho}
\tensor{(S_{\eta\epsilon})}{^\sigma_\gamma^\rho_\delta}
-\tensor{(S_{\eta\epsilon})}{^\mu_\sigma^\nu_\rho}
\tensor{(S_{\alpha\beta})}{^\sigma_\gamma^\rho_\delta}\nonumber\\
&  = - [g_{\sigma}^{\mu}(  g_{\alpha}^{\nu}g_{\beta\rho}-g_{\beta}^{\nu}g_{\alpha\rho}) 
+g_{\rho}^{\nu} ( g_{\alpha}^{\mu}g_{\beta\sigma}-g_{\beta}^{\mu}g_{\alpha\sigma}) ] 
\vphantom{\tensor{ { \left[ {S_{\alpha\beta}} , {S_{\eta\epsilon}} \right]_{-}}
 }{^\mu_\gamma^\nu_\delta}} 
\nonumber\\
&  \quad \times [g_{\gamma}^{\sigma} ( g_{\eta}^{\rho}g_{\epsilon \delta}
-g_{\epsilon}^{\rho}g_{\eta\delta})  +   g_{\delta}^{\rho} (g_{\eta}^{\sigma}g_{\epsilon\gamma}
-g_{\epsilon}^{\sigma}g_{\eta\gamma})]  
\vphantom{\tensor{ { \left[ {S_{\alpha\beta}} , {S_{\eta\epsilon}} \right]_{-}}
 }{^\mu_\gamma^\nu_\delta}} 
\nonumber\\
&  \quad +  [g_{\sigma}^{\mu}(  g_{\eta}^{\nu}g_{\epsilon\rho} -g_{\epsilon}^{\nu}g_{\eta\rho})
+ g_{\rho}^{\nu}(  g_{\eta}^{\mu}g_{\epsilon\sigma}-g_{\epsilon}^{\mu}g_{\eta\sigma})  ] 
\vphantom{\tensor{ { \left[ {S_{\alpha\beta}} , {S_{\eta\epsilon}} \right]_{-}}
 }{^\mu_\gamma^\nu_\delta}} 
\nonumber\\
&  \quad \times  [  g_{\gamma}^{\sigma} (  g_{\alpha}^{\rho
}g_{\beta\delta}-g_{\beta}^{\rho}g_{\alpha\delta})+  g_{\delta}^{\rho} 
(  g_{\alpha}^{\sigma}g_{\beta\gamma}-g_{\beta}^{\sigma} g_{\alpha\gamma}) ] 
\vphantom{\tensor{ { \left[ {S_{\alpha\beta}} , {S_{\eta\epsilon}} \right]_{-}}
 }{^\mu_\gamma^\nu_\delta}} 
\nonumber\\
& = -g_{\gamma}^{\mu} \left[ g_{\alpha}^{\nu} \left( g_{\beta\eta}g_{\epsilon\delta}
- g_{\beta\epsilon}g_{\eta\delta} \right) 
- g_{\beta}^{\nu} \left( g_{\alpha\eta}g_{\epsilon\delta}
- g_{\alpha\epsilon}g_{\eta\delta}\right) \vphantom{g_{\beta}^{\mu}} \right]
\vphantom{\tensor{ { \left[ {S_{\alpha\beta}} , {S_{\eta\epsilon}} \right]_{-}}
 }{^\mu_\gamma^\nu_\delta}} 
\nonumber \\
& \quad - ( g_{\alpha}^{\nu}g_{\beta\delta}- g_{\beta}^{\nu}g_{\alpha\delta})
( g_{\eta}^{\mu}  g_{\epsilon\gamma} -g_{\epsilon}^{\mu} g_{\eta\gamma})
\vphantom{\tensor{ { \left[ {S_{\alpha\beta}} , {S_{\eta\epsilon}} \right]_{-}}
 }{^\mu_\gamma^\nu_\delta}} 
 \nonumber \\
& \quad -( g_{\alpha}^{\mu}g_{\beta\gamma} - g_{\beta}^{\mu}g_{\alpha\gamma})
(g_{\eta}^{\nu} g_{\epsilon\delta} -  g_{\epsilon}^{\nu}g_{\eta\delta}) 
\vphantom{\tensor{ { \left[ {S_{\alpha\beta}} , {S_{\eta\epsilon}} \right]_{-}}
 }{^\mu_\gamma^\nu_\delta}} 
\nonumber \\
&  \quad - g_{\delta}^{\nu}\left[ g_{\alpha}^{\mu} \left( g_{\beta\eta}  g_{\epsilon\gamma}
 - g_{\beta\epsilon}g_{\eta\gamma}\right)
-g_{\beta}^{\mu} \left( g_{\alpha\eta}g_{\epsilon\gamma}
-  g_{\alpha\epsilon}g_{\eta\gamma}\right)\right]
\vphantom{\tensor{ { \left[ {S_{\alpha\beta}} , {S_{\eta\epsilon}} \right]_{-}}
 }{^\mu_\gamma^\nu_\delta}} 
\nonumber\\
& \quad +g_{\gamma}^{\mu} \left[ g_{\eta}^{\nu} \left( g_{\epsilon\alpha}g_{\beta\delta}
- g_{\epsilon\beta}g_{\alpha\delta}\right) 
- g_{\epsilon}^{\nu} \left( g_{\eta\alpha}g_{\beta\delta}
- g_{\eta\beta}g_{\alpha\delta}\right) \vphantom{g_{\beta}^{\mu}} \right]
\vphantom{\tensor{ { \left[ {S_{\alpha\beta}} , {S_{\eta\epsilon}} \right]_{-}}
 }{^\mu_\gamma^\nu_\delta}} 
\nonumber \\
& \quad + ( g_{\eta}^{\nu}g_{\epsilon\delta} - g_{\epsilon}^{\nu}g_{\eta\delta})
( g_{\alpha}^{\mu} g_{\beta\gamma} -  g_{\beta}^{\mu} g_{\alpha\gamma})  
\vphantom{\tensor{ { \left[ {S_{\alpha\beta}} , {S_{\eta\epsilon}} \right]_{-}}
 }{^\mu_\gamma^\nu_\delta}} 
\nonumber \\
& \quad +  ( g_{\eta}^{\mu}g_{\epsilon\gamma} - g_{\epsilon}^{\mu}g_{\eta\gamma})
( g_{\alpha}^{\nu} g_{\beta\delta}- g_{\beta}^{\nu}g_{\alpha\delta} )  
\vphantom{\tensor{ { \left[ {S_{\alpha\beta}} , {S_{\eta\epsilon}} \right]_{-}}
 }{^\mu_\gamma^\nu_\delta}} 
\nonumber \\ 
&  \quad+g_{\delta}^{\nu} \left[ g_{\eta}^{\mu}
\left( g_{\epsilon\alpha}g_{\beta\gamma}-g_{\epsilon\beta}g_{\alpha\gamma}\right) 
-g_{\epsilon}^{\mu}\left( g_{\eta\alpha}g_{\beta\gamma}
-g_{\eta\beta}g_{\alpha\gamma} \right) \vphantom{g_{\beta}^{\mu}} \right]
\vphantom{\tensor{ { \left[ {S_{\alpha\beta}} , {S_{\eta\epsilon}} \right]_{-}}
 }{^\mu_\gamma^\nu_\delta}} 
\,.
\end{align}
The terms in the second and third line cancel against those in the sixth and seventh
line. The remaining ones can be rearranged as
\begin{align}
\tensor{ { \left[ {S_{\alpha\beta}} , {S_{\eta\epsilon}} \right]_{-}}
 }{^\mu_\gamma^\nu_\delta} 
 &  = + g_{\alpha\eta}\left[g_{\gamma}^{\mu} (g_{\beta}^{\nu}g_{\epsilon\delta}
-g_{\epsilon}^{\nu}g_{\beta\delta})+ g_{\delta}^{\nu} (g_{\beta}^{\mu}g_{\epsilon\gamma}
- g_{\epsilon}^{\mu}g_{\beta\gamma}) \right]
\nonumber\\
&  \quad-g_{\alpha\epsilon}\left[ g_{\gamma}^{\mu}(g_{\beta}^{\nu}g_{\eta\delta}
-g_{\eta}^{\nu}g_{\beta\delta})+g_{\delta}^{\nu} (g_{\beta}^{\mu}g_{\eta\gamma}
-g_{\eta}^{\mu}g_{\beta\gamma})\right]
\vphantom{\left[\tensor{(I_{\eta\epsilon})}{^\sigma_\gamma^\rho_\delta} \right]_{-}}\nonumber\\
& \quad-g_{\beta\eta}\left[ g_{\gamma}^{\mu} (g_{\alpha}^{\nu}g_{\epsilon\delta}
-g_{\epsilon}^{\nu}g_{\alpha\delta})+ g_{\delta}^{\nu} (g_{\alpha}^{\mu}g_{\epsilon\gamma}
-g_{\epsilon}^{\mu}g_{\alpha\gamma}) \vphantom{g_{\beta}^{\mu}}  \right]
\vphantom{\left[\tensor{(I_{\eta\epsilon})}{^\sigma_\gamma^\rho_\delta} \right]_{-}}\nonumber\\
&  \quad+ g_{\beta\epsilon}\left[ g_{\gamma}^{\mu} (g_{\alpha}^{\nu}g_{\eta\delta
}-g_{\eta}^{\nu}g_{\alpha\delta}) +g_{\delta}^{\nu}  (g_{\alpha}^{\mu}g_{\eta\gamma}-g_{\eta
}^{\mu}g_{\alpha\gamma}) \vphantom{g_{\beta}^{\mu}} \right]
\vphantom{\left[\tensor{(I_{\eta\epsilon})}{^\sigma_\gamma^\rho_\delta} \right]_{-}}\nonumber\\
&  =i \left[ -g_{\alpha\eta} \tensor{(S_{\beta\epsilon})}{^\mu_\gamma^\nu_\delta}
+g_{\alpha\epsilon} \tensor{(S_{\beta\eta})}{^\mu_\gamma^\nu_\delta}
+g_{\beta\eta} \tensor{(S_{\alpha\epsilon})}{^\mu_\gamma^\nu_\delta}
-g_{\beta\epsilon} \tensor{(S_{\alpha\eta})}{^\mu_\gamma^\nu_\delta}\right]
\vphantom{\left[\tensor{(S_{\eta\epsilon})}{^\sigma_\gamma^\rho_\delta} \right]_{-}}
\,, \label{eq:provelorentz1}
\end{align}
where we used the definition (\ref{eq:tensorspingen}) of the spin generators. 
We see that the spin part $\tensor{(S^{\alpha\beta})}{^\mu_\sigma^\nu_\rho}$ fulfills 
the Lorentz algebra (\ref{eq:lorentzalgebra}) by itself.
Next we show that the mixed commutators in Eq.\ (\ref{eq:LScommutators}) 
vanish identically:
\begin{align}
\tensor{ { \left[ {S_{\alpha\beta}} , {L_{\eta\epsilon}} \right]_{-}}
 }{^\mu_\gamma^\nu_\delta} 
& = - \left[g^\mu_\sigma (g^\nu_\alpha g_{\beta\rho} - g^\nu_\beta g_{\alpha\rho}) 
+ g^\nu_\rho (g^\mu_\alpha g_{\beta\sigma} - g^\mu_\beta g_{\alpha\sigma})\right] 
g^\sigma_\gamma g^\rho_\delta (x_\eta \partial_\epsilon - x_\epsilon \partial_\eta) \nonumber
\\
& \quad + g^\mu_\sigma g^\nu_\rho (x_\eta \partial_\epsilon - x_\epsilon \partial_\eta) 
\left[ g^\sigma_\gamma (g^\rho_\alpha g_{\beta\delta} - g^\rho_\beta g_{\alpha\delta}) 
+ g^\rho_\delta (g^\sigma_\alpha g_{\beta\gamma} - g^\sigma_\beta g_{\alpha\gamma}) 
\right] \nonumber
\\
& = 0 \vphantom{\left[\tensor{(I_{\eta\epsilon})}{^\sigma_\gamma^\rho_\delta} \right]_{-}} \,.
\end{align}
Finally, we need to compute
\begin{align}
\tensor{ { \left[ {L_{\alpha\beta}} , {L_{\eta\epsilon}} \right]_{-}}
 }{^\mu_\gamma^\nu_\delta} 
 & = i^2 \left[ g^\mu_\sigma g^\nu_\rho (x_\alpha \partial_\beta - x_\beta \partial_\alpha) 
g^\sigma_\gamma g^\rho_\delta (x_\eta \partial_\epsilon - x_\epsilon \partial_\eta) \right. 
\vphantom{\left[\tensor{(I_{\eta\epsilon})}{^\sigma_\gamma^\rho_\delta} \right]_{-}} \nonumber
\\
& \qquad \left. - g^\mu_\sigma g^\nu_\rho (x_\eta \partial_\epsilon - x_\epsilon \partial_\eta) 
g^\sigma_\gamma g^\rho_\delta (x_\alpha \partial_\beta - x_\beta \partial_\alpha) \right] 
\vphantom{\left[\tensor{(I_{\eta\epsilon})}{^\sigma_\gamma^\rho_\delta} \right]_{-}} \nonumber
\\
& = i^2 g^\mu_\gamma g^\nu_\delta ( x_\alpha x_\eta \partial_\beta \partial_\epsilon 
 - x_\alpha x_\epsilon \partial_\beta \partial_\eta  
 - x_\beta x_\eta \partial_\alpha \partial_\epsilon
+ x_\beta x_\epsilon \partial_\alpha \partial_\eta 
\vphantom{\left[\tensor{(I_{\eta\epsilon})}{^\sigma_\gamma^\rho_\delta} \right]_{-}} \nonumber
\\
& \qquad \quad \quad + g_{\beta\eta} x_\alpha \partial_\epsilon 
 - g_{\beta\epsilon} x_\alpha \partial_\eta 
- g_{\alpha\eta} x_\beta \partial_\epsilon
+ g_{\alpha\epsilon} x_\beta \partial_\eta 
\vphantom{\left[\tensor{(I_{\eta\epsilon})}{^\sigma_\gamma^\rho_\delta} \right]_{-}} \nonumber
\\
& \qquad \quad \quad - x_\alpha x_\eta \partial_\beta \partial_\epsilon 
+ x_\beta x_\eta \partial_\alpha \partial_\epsilon 
+ x_\alpha x_\epsilon \partial_\beta \partial_\eta 
- x_\beta x_\epsilon \partial_\alpha \partial_\eta 
\vphantom{\left[\tensor{(I_{\eta\epsilon})}{^\sigma_\gamma^\rho_\delta} \right]_{-}} \nonumber
\\
& \qquad \quad \quad - g_{\alpha\epsilon} x_\eta \partial_\beta 
+ g_{\beta\epsilon} x_\eta \partial_\alpha
+ g_{\alpha\eta} x_\epsilon \partial_\beta 
- g_{\beta\eta} x_\epsilon \partial_\alpha) 
\vphantom{\left[\tensor{(I_{\eta\epsilon})}{^\sigma_\gamma^\rho_\delta} \right]_{-}} \nonumber
\\
& = i \left[ -g_{\alpha\eta} \tensor{(L_{\beta\epsilon})}{^\mu_\gamma^\nu_\delta} 
+ g_{\alpha\epsilon} \tensor{(L_{\beta\eta})}{^\mu_\gamma^\nu_\delta} 
+ g_{\beta\eta} \tensor{(L_{\alpha\epsilon})}{^\mu_\gamma^\nu_\delta} 
- g_{\beta\epsilon} \tensor{(L_{\alpha\eta})}{^\mu_\gamma^\nu_\delta} \right] \,, 
\label{eq:provelorentz2}
\end{align}
which means that also the angular momentum part 
$\tensor{(L_{\alpha\beta})}{^\mu_\sigma^\nu_\rho}$ fulfills the Lorentz algebra 
(\ref{eq:lorentzalgebra}) by itself. Furthermore, combining 
Eqs.\ (\ref{eq:provelorentz1}) and (\ref{eq:provelorentz2}) proves 
Eq.\ (\ref{eq:lorentzalgebra_tensor}).

The second commutation relation of the Poincar\'{e} algebra (\ref{eq:kommutatorPP}) can also 
be calculated explicitly. One inserts Eq.\ (\ref{eq:generatortensortrans}) and finds:
\begin{align}
\tensor{ { \left[ {P_{\alpha}} , {P_{\beta}} \right]_{-}}
 }{^\mu_\gamma^\nu_\delta} 
& = i^2 \left( g^\mu_\sigma g^\nu_\rho \partial_\alpha g^\sigma_\gamma g^\rho_\delta 
\partial_\beta - g^\mu_\sigma g^\nu_\rho \partial_\beta g^\sigma_\gamma 
g^\rho_\delta \partial_\alpha \right) \nonumber
\\
& = 0 \vphantom{\left[ \tensor{(P_\alpha)}{^\mu_\sigma^\nu_\rho} \right]} \,.
\end{align}
In order to prove the third commutation relation (\ref{eq:kommutatorPI}) one uses again the 
decomposition $\tensor{(I_{\alpha\beta})}{^\mu_\gamma^\nu_\delta} = 
\tensor{(S_{\alpha\beta})}{^\mu_\gamma^\nu_\delta} + 
\tensor{(L_{\alpha\beta})}{^\mu_\gamma^\nu_\delta}$:
\begin{align}
\tensor{ { \left[ {P_{\alpha}} , {I_{\eta\epsilon}} \right]_{-}} }{^\mu_\gamma^\nu_\delta} 
& = \tensor{ { \left[ {P_{\alpha}} , {S_{\eta\epsilon}} \right]_{-}} }{^\mu_\gamma^\nu_\delta} 
+ \tensor{ { \left[ {P_{\alpha}} , {L_{\eta\epsilon}} \right]_{-}} }{^\mu_\gamma^\nu_\delta} 
\,. \label{eq:commutatorPI}
\end{align}
The first commutator is given by
\begin{align}
\tensor{ { \left[ {P_{\alpha}} , {S_{\eta\epsilon}} \right]_{-}} }{^\mu_\gamma^\nu_\delta} 
& = g^\mu_\sigma g^\nu_\rho \partial_\alpha \left[ g^\sigma_\gamma (g^\rho_\eta 
g_{\epsilon\delta} - g^\rho_\epsilon g_{\eta\delta}) 
+ g^\rho_\delta (g^\sigma_\eta g_{\epsilon\gamma} 
- g^\sigma_\epsilon g_{\eta\gamma}) \right] \nonumber
\\
& \quad - \left[ g^\mu_\sigma (g^\nu_\eta g_{\epsilon\rho} - g^\nu_\epsilon g_{\eta\rho}) 
+ g^\nu_\rho (g^\mu_\eta g_{\epsilon\sigma} - g^\mu_\epsilon g_{\eta\sigma})
 \right] g^\sigma_\gamma g^\rho_\delta \partial_\alpha
\vphantom{\left[\tensor{(P_\alpha)}{^\mu_\sigma^\nu_\rho}\right]} \nonumber
\\
& = 0 \vphantom{\left[\tensor{(P_\alpha)}{^\mu_\sigma^\nu_\rho}\right]} \,,
\end{align}
whereas the second term leads to
\begin{align}
\tensor{ { \left[ {P_{\alpha}} , {L_{\eta\epsilon}} \right]_{-}} }{^\mu_\gamma^\nu_\delta} 
& = -i^2 \left[ g^\mu_\sigma g^\nu_\rho \partial_\alpha g^\sigma_\gamma 
g^\rho_\delta (x_\eta \partial_\epsilon - x_\epsilon \partial_\eta) 
- g^\mu_\sigma g^\nu_\rho (x_\eta \partial_\epsilon - x_\epsilon \partial_\eta) 
g^\sigma_\gamma g^\rho_\delta \partial_\alpha \right] 
\vphantom{\left[\tensor{(P_\alpha)}{^\delta_\gamma^\nu_\rho}\right]} \nonumber
\\
& = -i^2 g^\mu_\gamma g^\nu_\delta (g_{\alpha\eta}\partial_\epsilon 
- g_{\alpha\epsilon}\partial_\eta) 
\vphantom{\left[\tensor{(P_\alpha)}{^\delta_\gamma^\nu_\rho}\right]} \nonumber
\\
& = i \left[ g_{\alpha\eta} \tensor{(P_\epsilon)}{^\mu_\gamma^\nu_\delta} 
- g_{\alpha\epsilon} \tensor{(P_\eta)}{^\mu_\gamma^\nu_\delta} \right] 
\vphantom{\left[\tensor{(P_\alpha)}{^\delta_\gamma^\nu_\rho}\right]} \,.
\end{align}
On the one hand this shows that the third commutation relation (\ref{eq:kommutatorPI}) is 
fulfilled, while on the other hand this proves that translations do not commute with rotations, 
whereas the spin part $\tensor{(S_{\alpha\beta})}{^\mu_\gamma^\nu_\delta}$ of the generator 
commutes with the generator of translations. In summary, we have shown that 
the generators of the Poincar\'{e} group in spin-two field representation fulfill the Poincar\'{e}
algebra (\ref{eq:lorentzalgebra}), (\ref{eq:kommutatorPP}), and (\ref{eq:kommutatorPI}).

\section{Second Casimir operator}
\label{app:secondcasimiroperator}
In this appendix we explicitly calculate the square of the Pauli-Lubanski
pseudovector (\ref{eq:paulicasimir}).
Using the identity 
\begin{equation}
\epsilon_{\mu \alpha \beta \gamma} \, \epsilon^{\mu\nu\rho\sigma} 
= g^\nu_\alpha (g^\sigma_\beta g^\rho_\gamma - g^\rho_\beta g^\sigma_\gamma) 
+ g^\rho_\alpha (g^\nu_\beta g^\sigma_\gamma - g^\sigma_\beta g^\nu_\gamma) 
+ g^\sigma_\alpha (g^\rho_\beta g^\nu_\gamma - g^\nu_\beta g^\rho_\gamma)\,,
\end{equation}
we obtain
\begin{align}
W^{2}& = \frac{1}{4} \epsilon^{\mu\nu\rho\sigma} \, \epsilon_{\mu\alpha\beta\gamma} 
I_{\nu\rho} P_\sigma I^{\alpha\beta} P^\gamma \vphantom{\left[\frac{1}{2}\right]} 
\nonumber
\\
& = \frac{1}{4} \left[g^\nu_\alpha (g^\sigma_\beta g^\rho_\gamma 
- g^\rho_\beta g^\sigma_\gamma) + g^\rho_\alpha (g^\nu_\beta g^\sigma_\gamma 
- g^\sigma_\beta g^\nu_\gamma) + g^\sigma_\alpha (g^\rho_\beta g^\nu_\gamma 
- g^\nu_\beta g^\rho_\gamma)\right] I_{\nu\rho} P_\sigma I^{\alpha\beta} 
P^\gamma \vphantom{\left[\frac{1}{2}\right]} \nonumber
\\
& = \frac{1}{4} \left( I_{\alpha\gamma} P_\beta -I_{\alpha\beta} P_\gamma 
+ I_{\beta\alpha} P_\gamma - I_{\gamma\alpha} P_\beta 
+ I_{\gamma\beta} P_\alpha -I_{\beta\gamma} P_\alpha  \right) I^{\alpha\beta} P^\gamma 
\vphantom{\left[\frac{1}{2}\right]} \nonumber
\\
& = - \frac{1}{2} I_{\alpha\beta} g_{\gamma\delta} P^\delta I^{\alpha\beta} P^\gamma 
+ I_{\gamma\beta} g_{\alpha\delta} P^\delta I^{\alpha\beta} P^\gamma 
\vphantom{\left[\frac{1}{2}\right]} \nonumber
\\
& = \left( - \frac{1}{2} I_{\alpha\beta} g_{\gamma\delta} + I_{\gamma\beta} 
g_{\alpha\delta} \right) \left[ i \left( g^{\delta\alpha} P^\beta - g^{\delta\beta} P^\alpha \right) 
+ I^{\alpha\beta}P^\delta \vphantom{\frac{1}{2}}\right] P^\gamma 
\vphantom{\left[\frac{1}{2}\right]} \nonumber
\\
& = - \frac{1}{2} I_{\alpha\beta}I^{\alpha\beta} P^2 + I_{\gamma\beta}
I^{\alpha\beta} P_{\alpha}P^{\gamma} 
\vphantom{\left[\frac{1}{2}\right]} \,, \label{eq:W2}
\end{align}
where we used the antisymmetry of $I^{\alpha \beta}$ in the fourth and sixth lines and
the commutation relation (\ref{eq:kommutatorPI}) in the fifth line.

\section{Second Casimir operator in spin-two field representation}
\label{app:secondcasimirspin2} 
Just like the generators, for the spin-two case $W^2$, Eq.\ (\ref{eq:W2}), is a rank-four tensor, 
\begin{align}
\tensor{(W^2)}{^\mu_\gamma^\nu_\delta} 
& = -\frac{1}{2} \tensor{(I_{\alpha\beta})}{^\mu_\sigma^\nu_\rho} 
\tensor{(I^{\alpha\beta})}{^\sigma_\eta^\rho_\xi} \tensor{(P_\lambda)}{^\eta_\tau^\xi_\iota} 
\tensor{(P^\lambda)}{^\tau_\gamma^\iota_\delta} \nonumber
\\
& \quad + \tensor{(I_{\lambda\beta})}{^\mu_\sigma^\nu_\rho} 
\tensor{(I^{\alpha\beta})}{^\sigma_\eta^\rho_\xi} \tensor{(P_\alpha)}{^\eta_\tau^\xi_\iota} 
\tensor{(P^\lambda)}{^\tau_\gamma^\iota_\delta} \,.
\end{align}
After inserting the generator for translations (\ref{eq:generatortensortrans}), 
this simplifies to:
\begin{align}
\tensor{(W^2)}{^\mu_\gamma^\nu_\delta} & = \frac{1}{2} 
\tensor{(I^{\alpha\beta})}{^\mu_\sigma^\nu_\rho} 
\tensor{(I_{\alpha\beta})}{^\sigma_\gamma^\rho_\delta} \Box 
-  \tensor{(I_{\lambda\beta})}{^\mu_\sigma^\nu_\rho} 
\tensor{(I^{\alpha\beta})}{^\sigma_\gamma^\rho_\delta}  \partial_\alpha \partial^\lambda\,. 
\label{eq:casimirtensor}
\end{align}
We first compute
\begin{align}
\tensor{(I_{\lambda\beta})}{^\mu_\sigma^\nu_\rho} 
\tensor{(I^{\alpha\beta})}{^\sigma_\gamma^\rho_\delta} & =
i^2 \left[ g^\mu_\sigma (g^\nu_\lambda g_{\beta\rho} - g^\nu_\beta g_{\lambda\rho}) 
+ g^\nu_\rho (g^\mu_\lambda g_{\beta\sigma} - g^\mu_\beta g_{\lambda \sigma}) 
+ g^\mu_\sigma g^\nu_\rho (x_\lambda \partial_\beta - x_\beta \partial_\lambda) \right] 
\nonumber
\\
& \quad \times \left[ g^\sigma_\gamma (g^{\rho\alpha} g^\beta_\delta - g^{\rho\beta}
g^\alpha_\delta) + g^\rho_\delta (g^{\sigma\alpha} g^\beta_\gamma - g^{\sigma\beta} 
g^\alpha_\gamma) + g^\sigma_\gamma g^\rho_\delta (x^\alpha \partial^\beta - x^\beta
\partial^\alpha) \right] 
\vphantom{\left[(x^\beta_\gamma \partial^\beta_\delta)\right]} \nonumber
\\
& = 2\, (g^\mu_\gamma g^\nu_\lambda g^\alpha_\delta + g^\mu_\lambda g^\nu_\delta 
g^\alpha_\gamma + g^\mu_\gamma g^\nu_\delta g^\alpha_\lambda) 
\vphantom{\left[(x^\beta_\gamma \partial^\beta_\delta)\right]} \nonumber
\\
& \quad - g^{\mu\alpha} \left(g^\nu_\lambda g_{\gamma\delta} - g^\nu_\gamma 
g_{\lambda\delta} \right) - g^{\nu\alpha} \left( g^\mu_\lambda g_{\gamma\delta} - 
g^\mu_\delta g_{\lambda\gamma} \right)
\vphantom{\left[(x^\beta_\gamma \partial^\beta_\delta)\right]} \nonumber
\\
& \quad - g^\alpha_\gamma \left( g_{\lambda\delta} g^{\mu\nu} - g^\nu_\lambda
g^\mu_\delta \right) - g^\alpha_\delta \left( g_{\lambda\gamma} g^{\mu\nu}  - 
g^\mu_\lambda g^\nu_\gamma \right)
\vphantom{\left[(x^\beta_\gamma \partial^\beta_\delta)\right]} \nonumber
\\
& \quad + g^\mu_\gamma g^\nu_\lambda (x^\alpha \partial_\delta - x_\delta \partial^\alpha) 
- g^\mu_\gamma g_{\lambda\delta} ( x^\alpha \partial^\nu-x^\nu \partial^\alpha ) 
+ g^\nu_\delta g^\mu_\lambda  (x^\alpha \partial_\gamma - x_\gamma \partial^\alpha)
\vphantom{\left[(x^\beta_\gamma \partial^\beta_\delta)\right]} \nonumber
\\
& \quad - g^\nu_\delta g_{\lambda\gamma} (x^\alpha \partial^\mu-x^\mu \partial^\alpha) 
+ g^\mu_\gamma g^{\nu\alpha} (x_\lambda \partial_\delta - x_\delta \partial_\lambda)
- g^\mu_\gamma g^\alpha_\delta (x_\lambda \partial^\nu-x^\nu \partial_\lambda )
 \vphantom{\left[(x^\beta_\gamma \partial^\beta_\delta)\right]} \nonumber
\\
& \quad +g^\nu_\delta g^{\mu\alpha}(x_\lambda \partial_\gamma - x_\gamma \partial_\lambda)
- g^\nu_\delta g^\alpha_\gamma (x_\lambda \partial^\mu-x^\mu \partial_\lambda ) 
\vphantom{\left[(x^\beta_\gamma \partial^\beta_\delta)\right]} \nonumber
\\
& \quad + g^\mu_\gamma g^\nu_\delta ( x_\lambda x^\alpha \Box - 2\, x_\lambda 
\partial^\alpha - x^\alpha x^\beta \partial_\beta \partial_\lambda - x_\lambda x_\beta 
\partial^\beta \partial^\alpha - g^\alpha_\lambda x_\beta \partial^\beta 
+ x^2 \partial_\lambda \partial^\alpha ) 
\vphantom{\left[(x^\beta_\gamma \partial^\beta_\delta)\right]} \,. \label{eq:cas2tensor}
\end{align}
Now we contract $\lambda$ and $\alpha$,
\begin{align}
\tensor{(I_{\alpha\beta})}{^\mu_\sigma^\nu_\rho} 
\tensor{(I^{\alpha\beta})}{^\sigma_\gamma^\rho_\delta} & =
12\, g^\mu_\gamma g^\nu_\delta - 4\, g^{\mu\nu} g_{\gamma\delta} 
+ 4\, g^\mu_\delta g^\nu_\gamma 
+ 4\, g^\mu_\gamma (x^\nu \partial_\delta - x_\delta \partial^\nu) 
+ 4\, g^\nu_\delta (x^\mu \partial_\gamma - x_\gamma \partial^\mu) \nonumber
\\
& \quad + 2\,g^\mu_\gamma g^\nu_\delta ( x^2 \Box 
- x_\alpha x_\beta \partial^\alpha \partial^\beta - 3\, x_\alpha \partial^\alpha) \,. 
\label{eq:cas2tensor2}
\end{align}
Finally, inserting Eqs.\ (\ref{eq:cas2tensor}) and (\ref{eq:cas2tensor2}) into 
Eq.\ (\ref{eq:casimirtensor}) the second Casimir operator in spin-two field 
representation reads:
\begin{align}
\tensor{(W^2)}{^\mu_\gamma^\nu_\delta}  &  = 2 \left[ \left( 
2\, g_{\gamma}^{\mu} g_{\delta}^{\nu}  - g^{\mu\nu} g_{\gamma\delta} 
+ g_{\delta}^{\mu} g_{\gamma}^{\nu}  \right)\Box  \right.
\nonumber
\\
& \qquad +\left. g_{\gamma\delta} \partial^{\mu} \partial^{\nu} 
+ g^{\mu\nu} \partial_{\gamma} \partial_{\delta} 
-  g_{\delta}^{\nu} \partial^{\mu} \partial_{\gamma} 
- g_{\gamma}^{\mu} \partial^{\nu} \partial_{\delta} 
- g_{\gamma}^{\nu} \partial^{\mu} \partial_{\delta} 
- g_{\delta}^{\mu} \partial^{\nu} \partial_{\gamma}  \right]
\vphantom{\tensor{(W^2)}{^\mu_\gamma^\nu_\delta}} \,.
\end{align}

\section{Fierz-Pauli constraints for spin 1/2, spin one, and spin 3/2}
\label{app:fierzpauliforspin} 
In this appendix, we derive the Fierz-Pauli constraints from the eigenvalue
equations for the Casimir operators for spin-1/2, spin-one, and spin-3/2 fields. 
We first determine the generators $P^\alpha$, $I^{\alpha \beta}$ of the
Poincar\'{e} group for each case, before we embark on the discussion of the
Fierz-Pauli constraints.
\subsection{Generators of the Poincar\'{e} group}
\subsubsection{Spin 1/2:} 
We determine the generators for space-time translations from the requirement 
that the translated field at the translated coordinate is identical to the non-translated field at the 
original coordinate:
\begin{align}
\psi^{\prime}(x^{\prime}{}^\tau) = \psi(x^\tau) \,.
\end{align}
Adding $\psi'(x^\tau)$ on both sides of this equation, rearranging terms, 
and Taylor-expanding
$\psi'(x^{\prime}{}^\tau)$ around $x^\tau$ for an infinitesimal translation
$x^{\prime}{}^\tau = x^\tau + \epsilon^\tau$ leads to:
\begin{align}
\psi^{\prime} (x^\tau) & = \psi(x^\tau) - \left[ \psi^{\prime} (x^\tau) 
+ \epsilon_\alpha \partial^\alpha \psi^{\prime}(x^\tau) + O(\epsilon^2) 
- \psi^{\prime}(x^\tau) \right] \nonumber
\\
& = \psi(x^\tau) - \mathds{1}\, \epsilon_\alpha \partial^\alpha \psi(x^\tau) +O(\epsilon^2) \,,
\label{eq:spinor3}
\end{align}
where, to order $O(\epsilon^2)$, we were able to replace 
$\partial^\alpha \psi^{\prime}(x^\tau)$ by $\partial^\alpha \psi(x^\tau)$.
In order to keep track of Dirac indices, we also explicitly denoted the $(4\times 4)$ unit
matrix in Dirac-spinor space by $\mathds{1}$.
On the other hand, a translation can also be written in terms of
a group element acting on $\psi(x^\tau)$:
\begin{align}
\psi^{\prime}(x^\tau) & = \exp \left( -i\, a_\alpha P^\alpha \right) \psi(x^\tau) \,.
\end{align}
Up to first order in $a_\alpha \equiv \epsilon_\alpha$ this is equal to:
\begin{align}
\psi^{\prime}(x^\tau) 
& = \psi(x^\tau) -i\, \epsilon_\alpha P^\alpha \psi(x^\tau) + O(\epsilon^2) \,. \label{eq:spinor4}
\end{align}
By comparing Eqs.\ (\ref{eq:spinor3}) and (\ref{eq:spinor4}) we read off
the generator for space-time translations in the Dirac-spinor representation as:
\begin{align}
P^\alpha & = -i\, \mathds{1}\, \partial^\alpha \,. \label{eq:spinorgen2}
\end{align}

The next step is the determination of the generators for Lorentz
transformations. The transformation of a spinor reads, see e.g.\
Reinhardt \& Greiner \cite[p.119]{Reinhardt},
\begin{align}
\psi^{\prime} (x^{\prime}{}^\tau) & =  S \psi(x^\tau) 
\, , 
\label{eq:spinor0}
\end{align}
with 
\begin{align}
S \equiv \exp \left( - i\, \frac{\omega_{\alpha\beta}}{4} \sigma^{\alpha\beta} \right)\,,
\end{align}
where $\sigma^{\alpha\beta}=\frac{i}{2}\left[  \gamma^{\alpha},\gamma^{\beta}\right]_{-}$.
Adding $\psi^{\prime} (x^\tau)$ on both sides of Eq.\ (\ref{eq:spinor0}) and
rearranging terms yields
\begin{align}
\psi^{\prime} (x^\tau) 
& = S \psi(x^\tau) - [ \psi^{\prime} (x^{\prime}{}^\tau) - \psi^{\prime} (x^\tau)] \,. 
\label{eq:spinor0a}
\end{align}
For infinitesimal transformations, $\omega_{\alpha\beta} \rightarrow \delta
\omega_{\alpha \beta}$, and
expanding $S$ to first order in $\delta\omega$, as well as Taylor-expanding 
$\psi^{\prime} (x^{\prime}{}^\tau)$ around $x^\tau$ for an
infinitesimal Lorentz transformation $x^{\prime}{}^\tau =x^\tau+
\delta \tensor{\omega}{^\tau_\beta} x^\beta$ to the same order yields:
\begin{align}
\psi^{\prime} (x^\tau)  & = \psi(x^\tau) -i\, \frac{\delta\omega_{\alpha\beta}}{4} 
\left[ \sigma^{\alpha\beta} 
+ 2i\, \mathds{1}\, ( x^\alpha \partial^\beta - x^\beta \partial^\alpha ) \right] \psi(x^\tau) 
+ O(\delta\omega^2) \,, \vphantom{\frac{\delta}{4}} \label{eq:spinor1}
\end{align}
where to order $O(\delta\omega^2)$ we replaced $\partial^\alpha\psi^{\prime} (x^\tau)$ by
$\partial^\alpha\psi (x^\tau)$ and used the antisymmetry of $\delta\omega_{\rho\sigma}$.

On the other hand, the transformation of a spinor under an element of
the Lorentz group reads:
\begin{align}
\psi^{\prime} (x^\tau)  &  = \exp \left( -i\, \frac{\omega_{\alpha\beta}}{2} 
I^{\alpha\beta} \right) \psi(x^\tau) \,. \label{eq:spinor2.1}
\end{align}
For infinitesimal transformations, $\omega_{\alpha\beta} \rightarrow \delta
\omega_{\alpha \beta}$, and expanding the exponential to first order in $\delta\omega$, 
we have
\begin{align}
\psi^{\prime} (x^\tau) & = \psi(x^\tau) -i\, \frac{\delta\omega_{\alpha\beta}}{2} 
I^{\alpha\beta} \psi(x^\tau) + O(\delta\omega^2) \,. \label{eq:spinor2}
\end{align}
Comparing Eqs.\ (\ref{eq:spinor1}) and(\ref{eq:spinor2}) the generator for Lorentz 
transformations in Dirac-spinor representation is
\begin{align}
I^{\alpha\beta} & = \frac{1}{2}\, \sigma^{\alpha\beta} +i\, \mathds{1}\, 
( x^\alpha \partial^\beta - x^\beta \partial^\alpha )\,. \label{eq:spinorgen1}
\end{align}
This expression can be decomposed into a spin part $S^{\alpha\beta}$ and an 
angular momentum part $L^{\alpha\beta}$:
\begin{align}
S^{\alpha\beta} & = \frac{1}{2} \sigma^{\alpha\beta}\,,
\\
L^{\alpha\beta} & = i\, \mathds{1}\,  (x^\alpha \partial^\beta - x^\beta \partial^\alpha)\,.
\end{align}
One can verify that the generator (\ref{eq:spinorgen1}), as
well as the spin and angular momentum parts separately, fulfill the Lorentz algebra 
(\ref{eq:lorentzalgebra}). Moreover, both generators (\ref{eq:spinorgen2}) and
(\ref{eq:spinorgen1}) fulfill the two other commutation relations (\ref{eq:kommutatorPP}) 
and (\ref{eq:kommutatorPI}) of the Poincar\'{e} group. The proof is analogous to that given
in App.\ \ref{app:generatoralgebra} for the rank-two tensor representation of the
generators.

\subsubsection{Spin one:}
The generators for space-time translations in the representation for
spin-one fields are computed analogously to the spin-1/2 case.
We demand that
\begin{align}
A^{\prime}{}^\mu(x^{\prime}{}^\tau) & = A^\mu(x^\tau) \,.
\end{align}
Adding $A^{\prime}{}^\mu(x^\tau)$ to both sides of this equation, rearranging
terms, and expanding $A^{\prime}{}^\mu(x^{\prime}{}^\tau)$ to linear order
in an infinitesimal space-time translation $x^{\prime}{}^\tau = x^\tau +\epsilon^\tau$ leads to:
\begin{align}
A^{\prime}{}^\mu(x^\tau) & = A^\mu(x^\tau) - \left[ A^{\prime}{}^\mu(x^\tau) 
+ \epsilon_\alpha \partial^\alpha A^{\prime}{}^\mu(x^\tau) + O(\epsilon^2) 
- A^{\prime}{}^\mu(x^\tau) \right] \nonumber
\\
& = A^\mu(x^\tau) - g^\mu_\nu \epsilon_\alpha \partial^\alpha A^\nu(x^\tau) + O(\epsilon^2) \,,  
\label{eq:vector3}
\end{align}
where, up to terms of order $O(\epsilon^2)$, we were able to replace
$\partial^\rho A^{\prime}{}^\nu(x^\tau)$ by $\partial^\rho A^\nu(x^\tau)$.
A space-time translation can also be written in terms of a group element
acting on the field:
\begin{align}
A^{\prime}{}^{\mu}(x^\tau) &  =\exp\left( -i\,  a_\alpha P^\alpha\right)  \tensor{}{^\mu_\nu}
A^{\nu}(x^\tau) \,.
\end{align}
For an infinitesimal translation, $a_\alpha \equiv \epsilon_\alpha$, this reads:
\begin{align}
A^{\prime}{}^{\mu}(x^\tau) & = A^\mu(x^\tau) - i\, \epsilon_\alpha \tensor{(P^\alpha)}{^\mu_\nu} 
A^\nu(x^\tau) + O(\epsilon^2) \,. \label{eq:vector4}
\end{align}
Comparing Eqs.\ (\ref{eq:vector3}) and (\ref{eq:vector4}), the generators for
space-time translations can be read off as:
\begin{align}
\tensor{(P^\alpha)}{^\mu_\nu} & = -i\, g^\mu_\nu \, \partial^\alpha \,. 
\label{eq:genvectrans}
\end{align}

A Lorentz transformation of a vector field is defined as
\begin{align}
A^{\prime}{}^\mu(x^{\prime}{}^\tau)  = \tensor{\Lambda}{^\mu_\nu} A^{\nu}(x^\tau) \,.
\end{align}
Adding $A^{\prime}{}^{\mu}(x^\tau)$ on both sides of this equation and 
rearranging terms, we obtain
\begin{align}
A^{\prime}{}^{\mu}(x^\tau) &  = \tensor{\Lambda}{^\mu_\nu} A^{\nu}(x^\tau) 
- \left[ A^{\prime}{}^\mu(x^{\prime}{}^\tau) - A^{\prime}{}^\mu(x^\tau) \right] 
\vphantom{\frac{\delta}{2}} \; .\label{eq:vector0}
\end{align}
In the following step we consider an infinitesimal Lorentz transformation matrix 
$\tensor{\Lambda}{^\mu_\nu} \equiv g^\mu_\nu + \delta \tensor{\omega}{^\mu_\nu}$ 
and expand $A^{\prime}{}^\mu(x^{\prime}{}^\tau) $ around 
$x^\tau$ for an infinitesimal Lorentz transformation $x^{\prime}{}^\tau =
x^\tau + \delta \tensor{\omega}{^\tau_\beta} x^\beta$, 
\begin{align}
A^{\prime}{}^{\mu}(x^\tau) &  =\left[ g_{\nu}^{\mu}+\delta\tensor{\omega}{^\mu_\nu}
+O(\delta\omega^{2})\right] A^{\nu}(x^\tau) \vphantom{\frac{\delta}{2}} \nonumber
\\
& \hphantom{=} - \left[ A^{\prime}{}^\mu(x^\tau) + \delta \tensor{\omega}{^\alpha_\beta} 
x^\beta \partial_\alpha A^{\prime}{}^\mu(x^\tau) + O(\delta\omega^2) 
- A^{\prime}{}^\mu(x^\tau) \right] \vphantom{\frac{\delta}{2}}  \nonumber
\\
& = A^\mu(x^\tau) + \frac{\delta\omega_{\alpha\beta}}{2} \left[ g^{\alpha\mu} 
g^\beta_\nu - g^{\beta\mu} g^\alpha_\nu - g^\mu_\nu (x^\beta \partial^\alpha 
- x^\alpha \partial^\beta) \right] A^\nu(x^\tau) + O(\delta\omega^2) 
\vphantom{\frac{\delta}{2}} \,, \label{eq:vector1}
\end{align}
where we used the antisymmetry of $\delta \omega_{\alpha \beta}$ and, 
to order $O(\delta \omega^2)$, we were able to replace 
$\partial^\alpha A^{\prime}{}^\mu(x^\tau)$ by $\partial^\alpha A^\mu(x^\tau)$.
A Lorentz transformation of a spin-one field can also be expressed via 
a group element acting on the field,
\begin{align}
A^{\prime}{}^{\mu}(x^\tau) &  = \exp\tensor{\left( -i\,  \frac{\omega_{\alpha\beta}}{2} 
I^{\alpha\beta}\right) }{^\mu_\nu} A^{\nu}(x^\tau)\,.
\end{align}
For infinitesimal Lorentz transformations, $\omega_{\alpha\beta} \rightarrow \delta
\omega_{\alpha\beta}$, this reads
\begin{align}
A^{\prime}{}^{\mu}(x^\tau) & = A^{\mu}(x^\tau) 
-i\, \frac{\delta\omega_{\alpha\beta}}{2}\tensor{(I^{\alpha\beta})}{^\mu_\nu} 
A^{\nu}(x^\tau) +O(\delta\omega^{2}) \,. \label{eq:vector2}
\end{align}
Comparing Eqs.\ (\ref{eq:vector1}) and (\ref{eq:vector2}) yields:
\begin{align}
\tensor{(I^{\alpha\beta})}{^\mu_\nu} = i\, \left[g^{\alpha\mu} g^\beta_\nu - g^{\beta\mu} 
g^\alpha_\nu + g^\mu_\nu (x^\alpha \partial^\beta - x^\beta \partial^\alpha) \right] \,. 
\label{eq:genveclor}
\end{align}
It is again possible to identify the spin part, $\tensor{(S^{\alpha\beta})}{^\mu_\nu}$, and 
angular momentum part, $\tensor{(L^{\alpha\beta})}{^\mu_\nu}$, of the generator:
\begin{align}
\tensor{(S^{\alpha\beta})}{^\mu_\nu} & 
= i\, (g^{\alpha\mu} g^\beta_\nu - g^{\beta\mu} g^\alpha_\nu)\,,
\\
\tensor{(L^{\alpha\beta})}{^\mu_\nu} & = i\, g^\mu_\nu (x^\alpha \partial^\beta 
- x^\beta \partial^\alpha)\,.
\end{align}
One can prove that the generators (\ref{eq:genvectrans}) and (\ref{eq:genveclor}) fulfill 
the Poincar\'{e} algebra (\ref{eq:lorentzalgebra}), (\ref{eq:kommutatorPP}), 
and (\ref{eq:kommutatorPI}). 
This calculation is analogous to that in App.\ \ref{app:generatoralgebra} for
the generators in spin-two field presentation.

\subsubsection{Spin 3/2:} 
The transformation of a spin-3/2 field under space-time translations reads
\begin{align}
\psi^{\prime}{}^\mu (x^{\prime}{}^\tau) & = \psi^\mu (x^\tau) \, .
\end{align}
Adding $\psi^{\prime}{}^\mu (x^\tau)$ on both sides of this equation, rearranging terms,
and expanding $\psi^{\prime}{}^\mu (x^{\prime}{}^\tau)$ around $x^\tau$ for an
infinitesimal translation $x^{\prime}{}^\tau =  x^\tau + \epsilon^\tau$, we obtain
\begin{align}
\psi^{\prime}{}^\mu (x^\tau) & = \psi^\mu (x^\tau)- \left[ \psi^{\prime}{}^\mu (x^\tau) 
+ \epsilon_\alpha \partial^\alpha \psi^{\prime}{}^\mu (x^\tau) + O(\epsilon^2) 
- \psi^{\prime}{}^\mu (x^\tau) \right] \nonumber
\\
& = \psi^\mu (x^\tau) - \mathds{1}\, g^\mu_\nu \epsilon_\alpha \partial^\alpha 
\psi^\nu (x^\tau) + O(\epsilon^2) \,, \label{eq:RStransl}
\end{align}
where we replaced $\partial^\alpha \psi^{\prime}{}^\mu (x^\tau)$ to
order $O(\epsilon^2)$ with $\partial^\alpha \psi^\mu (x^\tau)$ and explicitly kept track
of the $(4 \times 4)$-unit matrix $\mathds{1}$ in Dirac space.
A space-time translation can also be written as a group element acting on the field,
\begin{align}
\psi^{\prime}{}^\mu (x^\tau) & = \exp \left( -i\, a_\alpha P^\alpha \right)\tensor{}{^\mu_\nu} 
\psi^\nu(x^\tau) \, .
\end{align}
For infinitesimal translations, $a_\alpha \rightarrow \epsilon_\alpha$, this becomes
\begin{align}
\psi^{\prime}{}^\mu (x^\tau) & = \psi^\mu(x^\tau) - i\, \epsilon_\alpha 
\tensor{(P^\alpha)}{^\mu_\nu} \psi^\nu(x^\tau) + O(\epsilon^2)\,.
\end{align}
Comparison to Eq.\ (\ref{eq:RStransl}) yields 
the generators for space-time translations of Rarita-Schwinger fields:
\begin{align}
\tensor{(P^\alpha)}{^\mu_\nu} & = -i\, \mathds{1}\, g^\mu_\nu \partial^\alpha \,. 
\label{eq:raritagen2}
\end{align}

A Lorentz transformation of a spin-3/2 field reads:
\begin{align}
\psi^{\prime}{}^\mu (x^{\prime}{}^\tau) & = \tensor{\Lambda}{^\mu_\nu} S \psi^\nu (x^\tau)\,.
\end{align}
Adding $\psi^{\prime}{}^\mu (x^\tau)$ on both sides of this equation and rearranging terms,
this becomes
\begin{align}
\psi^{\prime}{}^\mu (x^\tau) & = \tensor{\Lambda}{^\mu_\nu} S \psi^\nu (x^\tau) 
- \left[ \psi^{\prime}{}^\mu (x^{\prime}{}^\tau) - \psi^{\prime}{}^\mu (x^\tau) \right] \,. 
\label{eq:rarita0}
\end{align}
For infinitesimal Lorentz transformations (with parameters $\delta\omega_{\alpha \beta}$) 
this becomes:
\begin{align}
\psi^{\prime}{}^\mu (x^\tau) & = \left[ g^\mu_\nu + \delta \tensor{\omega}{^\mu_\nu} 
+ O(\delta\omega^2) \right] \left[ \mathds{1} - i\, \frac{\delta\omega_{\alpha\beta}}{4} 
\sigma^{\alpha\beta} + O(\delta\omega^2) \right] \psi^\nu(x^\tau) 
\vphantom{\frac{1}{2}} \nonumber
\\
& \quad - \left[ \psi^{\prime}{}^\mu (x^\tau) + \delta\omega_{\alpha\beta} 
x^\beta \partial^\alpha \psi^{\prime}{}^\mu (x^\tau) + O(\delta\omega^2) 
- \psi^{\prime}{}^\mu (x^\tau) \right] \vphantom{\frac{1}{2}} \nonumber
\\
& = \psi^\mu (x^\tau) + \frac{\delta\omega_{\alpha\beta}}{2} \left[ \mathds{1}\, 
( g^{\mu\alpha}g^\beta_\nu - g^{\mu\beta}g^\alpha_\nu ) 
- \frac{i}{2} g^\mu_\nu \sigma^{\alpha\beta} - \mathds{1}\, g^\mu_\nu 
(x^\beta \partial^\alpha - x^\alpha \partial^\beta) \right] \psi^\nu (x^\tau) 
\vphantom{\frac{1}{2}} \nonumber
\\
& \quad  + O(\delta\omega^2) \vphantom{\frac{1}{2}} \,, \label{eq:rarita1}
\end{align}
where we used the antisymmetry of $\delta \omega_{\alpha \beta}$ and, to order
$O(\delta \omega^2)$, were able to replace $\partial^\alpha \psi^{\prime}{}^\mu (x^\tau)$
by $\partial^\alpha \psi^\mu (x^\tau)$.
On the other hand, in terms of a group element acting on the field we have:
\begin{align}
\psi^{\prime}{}^\mu (x^\tau) & = \exp \tensor{ \left( -i\, \frac{\omega_{\alpha\beta}}{2} 
I^{\alpha\beta} \right)}{^\mu_\nu} \psi^\nu(x^\tau) \,,
\end{align}
which for an infinitesimal transformation reads
\begin{align}
\psi^{\prime}{}^\mu (x^\tau) & = \psi^\mu(x^\tau) -i\, \frac{\delta\omega_{\alpha\beta}}{2}
\tensor{(I^{\alpha\beta})}{^\mu_\nu} \psi^\nu(x^\tau) + O(\delta\omega^2) \,. \label{eq:rarita2}
\end{align}
Comparing Eqs.\ (\ref{eq:rarita1}) and (\ref{eq:rarita2}) the generators of Lorentz 
transformations for Rarita-Schwinger fields can be read off as:
\begin{align}
\tensor{(I^{\alpha\beta})}{^\mu_\nu} & = i\, \mathds{1}\, (g^{\mu\alpha}g^\beta_\nu 
- g^{\mu\beta}g^\alpha_\nu) + \frac{1}{2} g^\mu_\nu \sigma^{\alpha\beta} 
+ i\, \mathds{1}\, g^\mu_\nu (x^\alpha \partial^\beta - x^\beta \partial^\alpha) \,.
 \label{eq:raritagen1}
\end{align}
At this point we may again identify the spin part, $\tensor{(S^{\alpha\beta})}{^\mu_\nu}$, 
and the angular momentum part, $\tensor{(L^{\alpha\beta})}{^\mu_\nu}$, 
of the generators:
\begin{align}
\tensor{(S^{\alpha\beta})}{^\mu_\nu} & = i\, \mathds{1}\, (g^{\mu\alpha}g^\beta_\nu 
- g^{\mu\beta}g^\alpha_\nu) + \frac{1}{2} g^\mu_\nu \sigma^{\alpha\beta} \,,
\\
\tensor{(L^{\alpha\beta})}{^\mu_\nu} & = i\, \mathds{1}\, g^\mu_\nu (x^\alpha \partial^\beta
- x^\beta \partial^\alpha)\,.
\end{align}
Again, one can prove that the generators (\ref{eq:raritagen2}) and (\ref{eq:raritagen1}) fulfill 
the Poincar\'{e} algebra (\ref{eq:lorentzalgebra}), (\ref{eq:kommutatorPP}), 
and (\ref{eq:kommutatorPI}). 
This calculation is analogous to that in App.\ \ref{app:generatoralgebra} for
the generators in spin-two field presentation.
\subsection{Eigenvalue equations for the Casimir operators}
\subsubsection{Spin 1/2:}
With the results (\ref{eq:spinorgen2}) and (\ref{eq:spinorgen1}) it is possible to construct the 
Casimir operators $P^2$ and $W^2$ in Dirac-spinor field representation. 
The first one is simply $P^2 = - \mathds{1} \, \Box $, and the corresponding 
eigenvalue equation is the KG equation,
\begin{align} \label{eq:KGDiracspinor}
P^2 \psi  = - \Box \, \psi= m^2 \psi\,.
\end{align}
In order to determine $W^2$, Eq.\ (\ref{eq:W2}), in Dirac-spinor representation
it is useful to write down all spinor indices (Roman letters) explicitly:
\begin{align}
(W^2)_{ij} & = - \frac{1}{2} (I_{\alpha\beta})_{ik} (I^{\alpha\beta})_{kl} (P_\gamma)_{lm} 
(P^\gamma)_{mj} + (I_{\gamma \beta})_{ik} (I^{\alpha \beta})_{kl} 
(P_\alpha)_{lm} (P^\gamma)_{mj} \,.
\end{align}
Inserting the generator for translations (\ref{eq:spinorgen2}) this becomes
\begin{align}
(W^2)_{ij} & = \frac{1}{2} (I_{\alpha\beta})_{ik} (I^{\alpha\beta})_{kj} \Box - 
(I_{\gamma\beta})_{ik} (I^{\alpha\beta})_{kj} \partial_\alpha \partial^\gamma \,.
 \label{eq:spinorW2}
\end{align}
We first calculate the contraction of the generators in the 
second term, dropping spinor indices for the sake of clarity:
\begin{align}
I_{\gamma\beta}I^{\alpha\beta} & = \left[ \frac{1}{2} \, \sigma_{\gamma\beta} 
+ i\, \mathds{1}\, (x_\gamma \partial_\beta - x_\beta \partial_\gamma) \right] 
\left[ \frac{1}{2} \,\sigma^{\alpha\beta} 
+ i\, \mathds{1}\, (x^\alpha \partial^\beta - x^\beta \partial^\alpha) \right] \nonumber
\\
& = \frac{1}{4} \, \sigma_{\gamma\beta} \sigma^{\alpha\beta} 
+ \frac{i}{2}\, \sigma_{\gamma\beta} (x^\alpha \partial^\beta - x^\beta \partial^\alpha) 
+ \frac{i}{2}\, \sigma^{\alpha\beta} (x_\gamma \partial_\beta - x_\beta \partial_\gamma) 
\vphantom{\left[ \frac{1}{2} \, \sigma_{\gamma\beta} 
+ i\, \mathds{1}\, (x_\gamma \partial_\beta - x_\beta \partial_\gamma) \right] }
\nonumber
\\
& \quad - \mathds{1}\, (x_\gamma x^\alpha \Box - 2\, x_\gamma \partial^\alpha 
- x_\gamma x_\beta \partial^\beta \partial^\alpha
- g^\alpha_\gamma x_\beta\partial^\beta - x^\alpha x^\beta \partial_\beta \partial_\gamma 
 + x^2 \partial_\gamma \partial^\alpha )  
 \vphantom{\left[ \frac{1}{2} \, \sigma_{\gamma\beta} 
+ i\, \mathds{1}\, (x_\gamma \partial_\beta - x_\beta \partial_\gamma) \right] }
 \,. \label{eq:spinorcont1}
\end{align}
Contracting $\alpha$ and $\gamma$ indices, this becomes:
\begin{align}
I_{\alpha\beta} I^{\alpha\beta} & = 
\frac{1}{4} \, \sigma_{\alpha\beta} \sigma^{\alpha\beta} + i\, \sigma_{\alpha\beta} 
(x^\alpha \partial^\beta - x^\beta \partial^\alpha) - 2\, \mathds{1}\, ( x^2 \Box 
- 3\, x_\beta \partial^\beta -  x_\alpha x_\beta \partial^\alpha \partial^\beta ) \,. 
\label{eq:spinorcont2}
\end{align}
We now combine Eqs.\ (\ref{eq:spinorcont1}) and (\ref{eq:spinorcont2}) and
use the antisymmetry of $\sigma^{\alpha \beta}$ to
arrive at the following expression for the second Casimir operator (\ref{eq:spinorW2}):
\begin{align}
W^2 = \frac{1}{8} \sigma_{\alpha\beta} \sigma^{\alpha\beta} \Box  
- \frac{1}{4} \sigma_{\gamma\beta} \sigma^{\alpha\beta} \partial^\gamma \partial_\alpha 
\,.
\end{align}
This expression can be further simplified with the help of the anticommutation relations 
for the Dirac matrices:
\begin{align}
W^{2}= \frac{3}{2} \Box - \frac{1}{2} \slashed{\partial}\slashed{\partial} - \frac{1}{4} \Box
= \frac{5}{4} \Box - \frac{1}{2} \slashed{\partial}\slashed{\partial}\,.
\end{align}
The eigenvalue equation for the second Casimir operator (for $s=1/2$) then reads:
\begin{align}
W^2 \psi = \left( \frac{5}{4}\Box-\frac{1}{2}\slashed{\partial}\slashed{\partial}
\right) \psi = - m^2 s (s+1) \psi  \equiv - \frac{3}{4} m^2 \psi\,.
\end{align}
Using the KG equation (\ref{eq:KGDiracspinor}), this becomes:
\begin{align}
0  &  =\left(  -\slashed{\partial}\slashed{\partial}-m^{2}\right)  \psi\nonumber \\
&  =\left(  i\slashed{\partial}+m\right)  \left(i\slashed{\partial}-m\right)  \psi
\nonumber \\
\longrightarrow\ \  0 & = \left(i\slashed{\partial}-m\right)  \psi \,,
\end{align}
i.e., the Dirac equation.
\subsubsection{Spin one:}
With the results (\ref{eq:genvectrans}) and (\ref{eq:genveclor}) we now consider 
the eigenvalue equations for the Casimir operators in spin-one field representation.
Equation (\ref{eq:genvectrans}) yields
\begin{align}
\tensor{(P^\alpha)}{^\mu_\lambda} \tensor{(P_\alpha)}{^\lambda_\nu}
= - g^\mu_\lambda g^\lambda_\nu \partial^\alpha \partial_\alpha
= - g^\mu_\nu \Box\, ,
\end{align}
thus
\begin{align}
\tensor{(P^2)}{^\mu_\nu} A^\nu = - \Box\, g^\mu_\nu  A^\nu
= - \Box\,  A^\mu = m^2  A^\mu\, , \label{eq:KGvector}
\end{align}
i.e., the four components of the vector field fulfill the KG equation.

Equation (\ref{eq:W2}) with Eq.\ (\ref{eq:genveclor}) yields
\begin{align}
\tensor{(W^2)}{^\mu_\nu} & = - \frac{1}{2} \tensor{(I_{\alpha\beta})}{^\mu_\lambda} 
\tensor{(I^{\alpha\beta})}{^\lambda_\rho} \tensor{(P_\gamma)}{^\rho_\sigma} 
\tensor{(P^\gamma)}{^\sigma_\nu} + \tensor{(I_{\gamma\beta})}{^\mu_\lambda} 
\tensor{(I^{\alpha\beta})}{^\lambda_\rho} \tensor{(P_{\alpha})}{^\rho_\sigma} 
\tensor{(P^{\gamma})}{^\sigma_\nu} \,.
\end{align}
Using the generator for translations (\ref{eq:genvectrans}), this simplifies to:
\begin{align}
\tensor{(W^2)}{^\mu_\nu} & = \frac{1}{2} \tensor{(I_{\alpha\beta})}{^\mu_\lambda} 
\tensor{(I^{\alpha\beta})}{^\lambda_\nu} \Box - \tensor{(I_{\gamma\beta})}{^\mu_\lambda} 
\tensor{(I^{\alpha\beta})}{^\lambda_\nu} \partial^\gamma \partial_\alpha \,. \label{eq:vector5}
\end{align}
With Eq.\ (\ref{eq:genveclor}) the contraction of the generators in the second term is:
\begin{align}
\tensor{(I_{\gamma\beta})}{^\mu_\lambda} \tensor{(I^{\alpha\beta})}{^\lambda_\nu} 
& = i^2 \left[ g^\mu_\gamma g_{\beta\lambda} - g^\mu_\beta g_{\gamma\lambda} 
+ g^\mu_\lambda (x_\gamma \partial_\beta - x_\beta \partial_\gamma)  \right]
\vphantom{\partial^\delta} \nonumber
\\
& \quad \times \left[ g^{\alpha\lambda} g^\beta_\nu - g^{\beta\lambda} g^\alpha_\nu 
+ g^\lambda_\nu (x^\alpha \partial^\beta - x^\beta \partial^\alpha) 
\vphantom{g^\mu_\beta}\right] 
\vphantom{\partial^\delta} \nonumber
\\
& = 2\, g^\mu_\gamma g^\alpha_\nu + g^\mu_\nu g^\alpha_\gamma \nonumber
\\
& \quad -g^\mu_\gamma (x^\alpha \partial_\nu - x_\nu \partial^\alpha) 
+g_{\gamma\nu} (x^\alpha \partial^\mu-x^\mu \partial^\alpha ) 
\vphantom{\partial^\delta} \nonumber
\\
& \quad - g^{\mu\alpha} (x_\gamma \partial_\nu - x_\nu \partial_\gamma) 
+ g^\alpha_\nu (x_\gamma \partial^\mu- x^\mu\partial_\gamma ) 
\vphantom{\partial^\delta} \nonumber
\\
& \quad - g^\mu_\nu ( x_\gamma x^\alpha \Box - 2\, x_\gamma \partial^\alpha 
- x_\gamma x_\beta \partial^\beta \partial^\alpha - g^\alpha_\gamma x_\beta \partial^\beta
- x^\alpha  x_\beta  \partial^\beta \partial_\gamma+ x^2 \partial_\gamma \partial^\alpha )
 \vphantom{\partial^\delta} 
\,. \label{eq:vector6}
\end{align}
Contracting indices $\alpha$ and $\gamma$ results in:
\begin{align}
\tensor{(I_{\alpha\beta})}{^\mu_\lambda} \tensor{(I^{\alpha\beta})}{^\lambda_\nu} 
& = 6\, g^\mu_\nu + 4\, (x_\nu \partial^\mu - x^\mu \partial_\nu) \nonumber
\\
& \quad - 2\,g^\mu_\nu ( x^2 \Box  - 3\, x_\beta \partial^\beta
- x_\alpha x_\beta \partial^\alpha \partial^\beta) \, .\label{eq:vector7} 
\end{align}
Inserting Eqs.\ (\ref{eq:vector6}) and (\ref{eq:vector7}) into Eq.\ (\ref{eq:vector5}) 
the Casimir operator becomes:
\begin{align}
\tensor{(W^2)}{^\mu_\nu} & = 2\left( g^\mu_\nu \Box\, - \partial^\mu \partial_\nu 
\right)\,.
\end{align}
This result is well known, see Aurilia \& Umezawa \cite[p.1688]{Aurilia}.
Thus, using  $s=1$
the eigenvalue equation for the second Casimir operator is:
\begin{align}
2\, \Box\, A^{\mu}- 2\, \partial^{\mu}\partial_{\nu}A^{\nu}   &=
-m^{2}s(s+1)A^{\mu} \equiv -2\, m^2 A^\mu\nonumber 
\\ 
\longrightarrow \ \ 0 &= \left(\Box +m^2\right) A^\mu - \partial^{\mu}\partial_{\nu}A^{\nu}   
\,. \label{eq:vector8}
\end{align}
This is the Proca equation.
Using the KG equation (\ref{eq:KGvector}) and differentiating with respect to $x^\mu$ gives:
\begin{align}
0  &= \Box \,\partial_\nu A^\nu = - m^{2}\partial_{\nu}A^{\nu}\nonumber\\
\longrightarrow \ \ 0 &  =\partial_{\nu}A^{\nu}\,,
\end{align}
which is the Lorentz condition. 
\subsubsection{Spin $3/2$:}
With Eq.\ (\ref{eq:raritagen2}) the eigenvalue equation for the Casimir operator
$P^2$ reads:
\begin{align} \label{eq:KGrarita}
\tensor{(P^2)}{^\mu_\nu} \psi^\nu = - \Box\, \psi^\mu = m^2 \psi^\mu\, ,
\end{align}
i.e., all components of the Rarita-Schwinger field fulfill the KG equation.

The second Casimir operator (\ref{eq:W2}) is (again we explicitly denote spinor indices
as Roman letters):
\begin{align}
\tensor{{(W^2 )_{ij}} }{^\mu_\nu} &  
= - \frac{1}{2} \tensor{ {(I_{\alpha\beta})_{ik}} }{^\mu_\lambda}
\tensor{ {(I^{\alpha\beta})_{kl}} }{^\lambda_\rho} \tensor{ {(P_\gamma )_{lm}} }{^\rho_\sigma} 
\tensor{ {(P^\gamma )_{mj}} }{^\sigma_\nu} \nonumber
\\
& \quad + \tensor{ {(I_{\gamma\beta} )_{ik}} }{^\mu_\lambda} 
\tensor{ {(I^{\alpha\beta} )_{kl}} }{^\lambda_\rho} \tensor{ {(P_\alpha )_{lm}} }{^\rho_\sigma} 
\tensor{ {(P^\gamma )_{mj}} }{^\sigma_\nu}\,.
\end{align}
With the generators (\ref{eq:raritagen2}) this simplifies to
\begin{align}
\tensor{(W^2)}{^\mu_\nu} & = \frac{1}{2} \tensor{(I_{\alpha\beta})}{^\mu_\lambda}
\tensor{(I^{\alpha\beta})}{^\lambda_\nu} \Box - \tensor{(I_{\gamma\beta})}{^\mu_\lambda} 
\tensor{(I^{\alpha\beta})}{^\lambda_\nu} \partial_\alpha \partial^\gamma \,, \label{eq:raritaW2}
\end{align}
where spinor indices have been suppressed. We first compute the contraction of
the generators in the second term. With Eq.\ (\ref{eq:raritagen1}) we have
\begin{align}
\tensor{(I_{\gamma\beta})}{^\mu_\lambda} \tensor{(I^{\alpha\beta})}{^\lambda_\nu} 
& = \left[ i\, \mathds{1}\, ( g^\mu_\gamma g_{\beta\lambda} - g^\mu_\beta 
g_{\gamma\lambda} ) + \frac{1}{2} g^\mu_\lambda \sigma_{\gamma\beta} 
+ i\, \mathds{1}\, g^\mu_\lambda ( x_\gamma \partial_\beta- x_\beta \partial_\gamma ) \right]
\vphantom{\left[\frac{1}{2}\right]} \nonumber
\\
& \quad \times \left[ i\, \mathds{1}\, ( g^{\lambda\alpha} g^\beta_\nu - g^{\lambda\beta} 
g^\alpha_\nu ) + \frac{1}{2} g^\lambda_\nu \sigma^{\alpha\beta} 
+ i\, \mathds{1}\, g^\lambda_\nu ( x^\alpha \partial^\beta -  x^\beta \partial^\alpha ) \right] 
\vphantom{\left[\frac{1}{2}\right]} \nonumber
\\
& = \mathds{1}\left(  2\, g^\mu_\gamma g^\alpha_\nu + g^\mu_\nu g^\alpha_\gamma \right)
+ \frac{1}{4} \, g^\mu_\nu\, \sigma_{\gamma\beta} \sigma^{\alpha\beta} 
\vphantom{\left[\frac{1}{2}\right]} \nonumber
\\
& \quad - \mathds{1}\, g^\mu_\nu ( x_\gamma x^\alpha \Box - 2\, x_\gamma \partial^\alpha 
- x_\gamma x_\beta \partial^\beta \partial^\alpha - g^\alpha_\gamma x_\beta \partial^\beta
-  x^\alpha x_\beta \partial^\beta \partial_\gamma  + x^2 \partial_\gamma 
\partial^\alpha ) \vphantom{\left[\frac{1}{2}\right]}\nonumber 
\\
& \quad + \frac{i}{2} ( g^\mu_\gamma \tensor{\sigma}{^\alpha_\nu} - g_{\gamma\nu} 
\sigma^{\alpha\mu} + g^{\mu\alpha} \sigma_{\gamma\nu} - 
g^\alpha_\nu \tensor{\sigma}{_\gamma^\mu} ) 
\vphantom{\left[\frac{1}{2}\right]} \nonumber
\\
& \quad - \mathds{1}\, \left[ g^\mu_\gamma (x^\alpha \partial_\nu - x_\nu \partial^\alpha) 
- g_{\gamma\nu} (x^\alpha \partial^\mu - x^\mu \partial^\alpha) \right. 
\vphantom{\left[\frac{1}{2}\right]} \nonumber
\\
& \qquad \quad \left. + g^{\mu\alpha} (x_\gamma \partial_\nu - x_\nu \partial_\gamma) 
- g^\alpha_\nu (x_\gamma \partial^\mu - x^\mu \partial_\gamma) \right] 
\vphantom{\left[\frac{1}{2}\right]}\nonumber
\\
& \quad +\frac{i}{2} g^\mu_\nu \left[ \sigma_{\gamma\beta} ( x^\alpha \partial^\beta 
- x^\beta \partial^\alpha ) + \sigma^{\alpha\beta} ( x_\gamma \partial_\beta 
- x_\beta \partial_\gamma ) \right] \vphantom{\left[\frac{1}{2}\right]}   \,. \label{eq:rarita3}
\end{align}
Contracting the indices $\alpha$ and $\gamma$, we obtain the first part of the first
term in Eq.\ (\ref{eq:raritaW2}):
\begin{align}
\tensor{(I_{\alpha\beta})}{^\mu_\lambda} \tensor{(I^{\alpha\beta})}{^\lambda_\nu} 
& = 9\, \mathds{1}\, g^\mu_\nu 
- 2\, \mathds{1}\, g^\mu_\nu ( x^2 \Box - 3\, x_\alpha \partial^\alpha 
- x_\alpha x_\beta \partial^\alpha \partial^\beta ) 
+ 2\, i\, \tensor{\sigma}{^\mu_\nu} \nonumber
\\
& \quad  - 4\, \mathds{1}\, (x^\mu \partial_\nu - x_\nu \partial^\mu)
+i\, g^\mu_\nu \sigma_{\alpha\beta} 
(x^\alpha \partial^\beta - x^\beta \partial^\alpha) \vphantom{\frac{1}{4}} \,, \label{eq:rarita4}
\end{align}
where we used the antisymmetry of $\sigma^{\mu \nu}$ and $\sigma_{\alpha\beta}
\sigma^{\alpha \beta} = 12 \, \mathds{1}$.
Inserting Eqs.\ (\ref{eq:rarita3}) and (\ref{eq:rarita4}) into Eq.\ (\ref{eq:raritaW2}) one obtains:
\begin{align}
\tensor{(W^2)}{^\mu_\nu}  &  = \left( \frac{11}{4} \,\mathds{1}\, g_{\nu}^{\mu} + 
i \tensor{\sigma}{^\mu_\nu}\right) \Box - 2 \,\mathds{1}\,  \partial^\mu \partial_\nu  
+i \tensor{\sigma}{^\alpha^\mu} \partial_\alpha \partial_\nu -
i \tensor{\sigma}{^\alpha_\nu} \partial_\alpha \partial^\mu\nonumber 
\\ 
& = \mathds{1} \, \left( \frac{11}{4}\, g^\mu_\nu \, \Box - 2\, \partial^\mu \partial_\nu \right)
-\frac{1}{2} \left( \left[ \gamma^\mu, \gamma_\nu \right]_{-} \Box
 + \left[\slashed{\partial}, \gamma^\mu \right]_{-} \partial_\nu
 - \left[\slashed{\partial}, \gamma_\nu\right]_{-} \partial^\mu  \right)  \,.
\end{align}
Aurilia \& Umezawa presented this result (without deriving it) in Ref.\
\cite[p.1690]{Aurilia}. The eigenvalue equation for $s=\frac{3}{2}$ then reads
\begin{align}
\lefteqn{\hspace*{-4cm}
\left[ \mathds{1} \, \left(  \frac{11}{4}\, g^\mu_\nu \, \Box - 2\, \partial^\mu \partial_\nu \right)
-\frac{1}{2} \left( \left[ \gamma^\mu, \gamma_\nu \right]_{-} \Box
 + \left[\slashed{\partial}, \gamma^\mu \right]_{-} \partial_\nu
 - \left[\slashed{\partial}, \gamma_\nu\right]_{-} \partial^\mu  \right) \right] \psi^\nu}
 \nonumber \\
 & = -\frac{3}{2}\left(  \frac{3}{2}+1\right) m^{2}  \psi^{\mu}  \equiv - \frac{15}{4} m^2 \psi^\mu\,. 
 \label{eq:eigenvaluerarita}%
\end{align}
Differentiating with respect to $x^\mu$ and using Eq.\ (\ref{eq:KGrarita}) one finds
\begin{align}
0  &  =\left(  \Box + 5\, m^{2}\right) \partial_{\nu}\psi^{\nu}
\equiv 4 m^2  \partial_{\nu}\psi^{\nu} \nonumber \\
\longrightarrow \ \ 0  & = \partial_{\nu}\psi^{\nu} \,, \label{eq:raritaconstrai1}
\end{align}
which is the first constraint. Finally, multiplying Eq.\ (\ref{eq:eigenvaluerarita}) 
with $\gamma_{\mu}$ and using Eq.\ (\ref{eq:KGrarita}) we obtain:
\begin{align}
0  &  =\left(   \Box \gamma_\nu + 5 m^2 \gamma_\nu \right)
\psi^\nu \nonumber 
\\
& = 4 m^{2} \gamma_{\nu} \psi^{\nu}\nonumber
\\
\longrightarrow \ \ 0 &  = \gamma_{\nu}\psi^{\nu} \,,
\end{align}
which is the last constraint for the spin-3/2 field. The procedure can be generalized 
to arbitrary values of spin, see e.g.\ Ref.\ \cite{ks}.

\section{Legendre transformation}
\label{app:legendretransformation} 
In this appendix we derive the covariant Hamilton density via
the complete Legendre transformation (\ref{eq:covhamdens}) for the spin-two field.
The canonically conjugate field is given in Eq.\ (\ref{eq:momentumfield}).
In the course of the calculation, we shall also need the following contractions
of this field,
\begin{align}
\tensor{\Pi}{_\alpha^\alpha^\gamma}  &  =-2\,\partial^{\gamma}T+\partial
_{\alpha}T^{\alpha\gamma}+\partial_{\alpha}T^{\gamma\alpha}%
\vphantom{\frac{2}{3}}\,,\nonumber\\
\tensor{\Pi}{^\sigma_\alpha^\alpha}  &  =\frac{3}{2}\left(  \partial^{\sigma
}T-\partial_{\alpha}T^{\sigma\alpha}-\partial_{\alpha}T^{\alpha\sigma}\right)
\,,\nonumber\\
\tensor{\Pi}{_\alpha^\rho^\alpha}  &  =\frac{3}{2}\left(  \partial^{\rho
}T-\partial_{\alpha}T^{\rho\alpha}-\partial_{\alpha}T^{\alpha\rho}\right)  \,.
\label{eq:contractions}%
\end{align}
From its definition (\ref{eq:momentumfield}) it is obvious that the canonically
conjugate field is symmetric in the first two Lorentz indices. This naturally explains
why the latter two contractions are identical.

The Legendre transformation considerably simplifies by noting that
\begin{align}
\frac{1}{4}\Pi_{\alpha\beta\mu}\partial^{\mu} \left(T^{\alpha\beta}+T^{\beta\alpha}\right)
  & = \quad \frac{1}{8} \,  \left[ \partial_{\mu} \left(T_{\alpha\beta}
+T_{\beta \alpha}\right)\right] \partial^{\mu} \left(T^{\alpha\beta}
+T^{\beta \alpha}\right)  -\frac{1}{2} \, (\partial_{\mu} T) \partial^{\mu}T 
\nonumber \\
& \quad 
-  \frac{1}{4} \, \left[\partial_{\alpha}\left( T^{\alpha\beta} +T^{\beta \alpha} \right)\right]
\partial^{\mu} \left( T_{\mu\beta} + T_{\beta \mu} \right) \nonumber\\
&\quad + \vphantom{\partial_\mu T^\beta}\frac{1}{2} \, \left[\partial_{\mu}
\left( T^{\mu\nu}+ T^{\nu \mu} \right)\right] \partial_{\nu} T  \,.
\end{align}
is the kinetic part of the Lagrangian (\ref{eq:lagrangedichte}), so that
\begin{align}
\mathcal{L}  &  =\frac{1}{4}\Pi_{\sigma\rho\gamma}\partial^{\gamma}\left(
T^{\sigma\rho}+T^{\rho\sigma}\right) 
-\frac{m^{2}}{8}\left( T_{\mu\nu}+T_{\nu\mu} \right)\left(T^{\mu\nu}+T^{\nu\mu
}\right)  +\frac{m^{2}}{2}T^{2}%
\vphantom{\left(\frac{\partial^\gamma T^{\sigma\rho} + \partial^\gamma 
T^{\rho\sigma}}{2}\right)}\,.
\end{align}
The Legendre transformation now reads explicitly:
\begin{align}
\mathcal{H}_{\rm cov}  &  =\Pi_{\sigma\rho\gamma}\partial^{\gamma}T^{\sigma\rho
}-\mathcal{L}%
\vphantom{\left[\left(\frac{\partial^\gamma T^{\sigma\rho} - \partial^\gamma 
T^{\rho\sigma}}{2}\right) \right]}\nonumber\\
&  =\frac{1}{2}\, \Pi_{\sigma\rho\gamma}\left[  \partial^{\gamma} \left(  
T^{\sigma\rho}+T^{\rho\sigma}\right)  
+ \partial^{\gamma}\left(T^{\sigma\rho}-T^{\rho\sigma} \right)
\right]  -\mathcal{L}\nonumber\\
&  =\frac{1}{4}\Pi_{\sigma\rho\gamma}\partial^{\gamma}\left( T^{\sigma\rho}
+T^{\rho\sigma}\right)  +\frac{1}{2}\, \Pi_{\sigma\rho\gamma}\partial^{\gamma}
\left(T^{\sigma\rho}-T^{\rho\sigma}\right)
\vphantom{\left[\left(\frac{\partial^\gamma T^{\sigma\rho} - \partial^\gamma 
T^{\rho\sigma}}{2}\right) \right]}\nonumber\\
&  \quad +\frac{m^{2}}{8}\left( T_{\mu\nu}+T_{\nu\mu} \right)\left(T^{\mu\nu}+T^{\nu\mu
}\right)  -\frac{m^{2}}{2}T^{2}
\vphantom{\left[\left(\frac{\partial^\gamma T^{\sigma\rho} - \partial^\gamma 
T^{\rho\sigma}}{2}\right) \right]}\,.
\end{align}
The second term vanishes because $\Pi_{\sigma \rho \gamma}$ is symmetric
in the first two indices.
At this point one uses the expression for the momentum field
(\ref{eq:momentumfield}) to rewrite the first term:
\begin{align}
\mathcal{H}_{\rm cov}  &  = +\frac{1}{2}\,
\Pi_{\sigma\rho\gamma}\left[  \vphantom{\frac{1}{2}}\Pi^{\sigma\rho\gamma} 
+g^{\rho\sigma}\partial^{\gamma}T-\frac{1}{2}\left(  g^{\gamma\sigma}\partial^{\rho}T
+g^{\rho\gamma}\partial^{\sigma}T\right) \right. \nonumber\\
&  \quad+\left.\frac{1}{2}\, g^{\gamma\sigma}\partial_{\mu}\left(  T^{\mu\rho}+T^{\rho\mu}
\right) +\frac{1}{2}\, g^{\gamma\rho}\partial_{\mu}\left( T^{\sigma\mu}+T^{\mu\sigma}\right)
-\frac{1}{2}\,g^{\rho\sigma }\partial_{\mu}\left( T^{\mu\gamma}+T^{\gamma\mu}\right) 
\right]  \vphantom{\frac{1}{2}} \nonumber\\
&  \quad+  \frac{m^{2}}{8}\left( T_{\mu\nu}+T_{\nu\mu} \right)
\left(T^{\mu\nu}+T^{\nu\mu}\right)  -\frac{m^{2}}{2}T^{2}
\vphantom{\left[\left(\frac{\partial^\gamma T^{\sigma\rho} - \partial^\gamma 
T^{\rho\sigma}}{2}\right) \right]}\,.
\end{align}
Then, the derivatives of the tensor field are substituted by the contractions
(\ref{eq:contractions}) of the canonically conjugate field:
\begin{align}
\mathcal{H}_{\rm cov}  &  = + 
\frac{1}{2}\Pi_{\sigma\rho\gamma}\left(  \Pi^{\sigma\rho\gamma}%
-\frac{1}{2}\, g^{\sigma\rho}\tensor{\Pi}{_\alpha^\alpha^\gamma}-\frac{1}%
{3}\, g^{\rho\gamma}\tensor{\Pi}{^\sigma_\alpha^\alpha}-\frac{1}{3}\,g^{\sigma\gamma}
\tensor{\Pi}{_\alpha^\rho^\alpha}\right) \nonumber\\
&  \quad
+\frac{m^{2}}{8}\left( T_{\mu\nu}+T_{\nu\mu} \right)\left(T^{\mu\nu}+T^{\nu\mu
}\right)  -\frac{m^{2}}{2}T^{2}\vphantom{\left[\frac{1}{3}\right]}\,.
\end{align}
This then leads to Eq.\ (\ref{eq:hamiltonian}).
\section{Canonical equations}
\label{app:canonicalequations} 
The covariant Hamilton density contains all information about the system. 
In the present case these are the equations of
motion and the Fierz-Pauli constraints. 
In general one extracts them from the canonical equations
(\ref{eq:canonicalone}) and (\ref{eq:canonicaltwo}):
\begin{align}
\frac{\partial\mathcal{H}_{\text{cov}}}{\partial\Pi_{\alpha\beta\delta}}  &
=\partial^{\delta}T^{\alpha\beta}\nonumber\\
&  = \Pi^{\alpha\beta\delta}-\frac{1}{2}\,g^{\alpha\beta}
\tensor{\Pi}{_\mu^\mu^\delta}-\frac{1}{3}\,g^{\beta\delta}
\tensor{\Pi}{^\alpha_\mu^\mu}-\frac{1}{3}\,g^{\alpha\delta}
\tensor{\Pi}{_\mu^\beta^\mu}
\vphantom{\frac{\partial \mathcal{H}}{\partial \Pi_{\alpha\beta\delta}}}\,,
\label{eq:kanone}
\\
-\frac{\partial\mathcal{H}_{\text{cov}}}{\partial T_{\alpha\beta}}  &
=\partial_{\delta}\Pi^{\alpha\beta\delta}\nonumber\\
&  =-\frac{m^{2}}{2}\left(  T^{\alpha\beta}+T^{\beta\alpha}\right)
+m^{2}g^{\alpha\beta}T\,. \label{eq:kantwo}%
\end{align}
The right-hand side of the second equation (\ref{eq:kantwo}) 
is symmetric under the exchange of
$\alpha$ and $\beta$. Hence, the dynamical part of the momentum field is also
symmetric. We now prove that also $T^{\alpha\beta}$ is symmetric. To this end,
we exchange the indices $\alpha$ and $\beta$ in the first canonical equation
(\ref{eq:kanone}),
\begin{align}
\partial^{\delta}T^{\beta\alpha}
&  =\Pi^{\beta\alpha\delta}-\frac{1}{2}\,g^{\beta\alpha}%
\tensor{\Pi}{_\mu^\mu^\delta}-\frac{1}{3}\,g^{\alpha\delta}%
\tensor{\Pi}{^\beta_\mu^\mu}-\frac{1}{3}\,g^{\beta\delta}%
\tensor{\Pi}{_\mu^\alpha^\mu}\nonumber\\
&  =\Pi^{\alpha\beta\delta}-\frac{1}{2}g^{\alpha\beta}%
\tensor{\Pi}{_\mu^\mu^\delta}-\frac{1}{3}g^{\beta\delta}%
\tensor{\Pi}{^\alpha_\mu^\mu}-\frac{1}{3}g^{\alpha\delta}%
\tensor{\Pi}{_\mu^\beta^\mu}\,,
\end{align}
where we used the symmetry of the metric tensor and the canonically conjugate field and
exchanged the order of the last two terms.
Comparing this with Eq.\ (\ref{eq:kanone}), we observe that 
$T^{\alpha \beta} =T^{\beta \alpha}$.
This completes the proof the symmetry of the tensor field, which in turn simplifies the
second canonical equation (\ref{eq:kantwo}),
\begin{align}
\partial_{\delta}\Pi^{\alpha\beta\delta}  &  =-m^{2}T^{\alpha\beta}%
+m^{2}g^{\alpha\beta}T\vphantom{\frac{1}{3}}\,.\label{eq:canone}
\end{align}
For the remainder of the calculation it is useful to quote the following contractions of Eq.\
(\ref{eq:kanone}):
\begin{align}
\partial^{\delta}T  &  =-\tensor{\Pi}{_\mu^\mu^\delta}-\frac{2}{3}%
\tensor{\Pi}{_\mu^\delta^\mu}\,,\label{eq:ablkontraktion1}\\
\partial_{\mu}T^{\mu\beta}  &  =-\frac{2}{3}\tensor{\Pi}{_\mu^\beta^\mu}-\frac
{1}{2}\tensor{\Pi}{_\mu^\mu^\beta}\,. \label{eq:ablkontraktion2}%
\end{align}
Furthermore, by combining Eqs.\
(\ref{eq:ablkontraktion1}) and (\ref{eq:ablkontraktion2}), we obtain
\begin{equation}
\tensor{\Pi}{_\mu^\mu^\delta}=2\left( \partial_{\mu}T^{\mu\delta}-\partial^{\delta}T
\right)\,.
\label{eq:ablkontraktion3}%
\end{equation}
Taking the divergence of expression (\ref{eq:kanone}) and inserting
Eq.\ (\ref{eq:canone}) in the first term on the right-hand side yields
\begin{align}
\Box T^{\alpha\beta}&=\partial_{\delta}\Pi^{\alpha\beta\delta}-\frac{1}%
{2}g^{\alpha\beta}\partial_{\delta}\tensor{\Pi}{_\mu^\mu^\delta}-\frac{1}{3}\partial^{\beta
}\tensor{\Pi}{^\alpha_\mu^\mu}-\frac{1}%
{3}\partial^{\alpha}\tensor{\Pi}{^\beta_\mu^\mu}\nonumber \\
& = -m^{2}T^{\alpha\beta}+m^{2}g^{\alpha\beta}T
-\frac{1}{2}g^{\alpha\beta}\partial_{\delta}\tensor{\Pi}{_\mu^\mu^\delta}-\frac
{1}{3}\partial^{\alpha}\tensor{\Pi}{^\beta_\mu^\mu}-\frac{1}{3}\partial
^{\beta}\tensor{\Pi}{^\alpha_\mu^\mu}\,.
\end{align}
We now employ Eq.\ (\ref{eq:ablkontraktion2}) with the fact that $\Pi^{\alpha \beta \mu}$ is 
symmetric in the first two indices to rewrite the last two terms. We obtain
\begin{align}
\Box T^{\alpha\beta}+m^{2}T^{\alpha\beta}-m^{2}g^{\alpha\beta}T  &
=-\frac{1}{2}g^{\alpha\beta}\partial_{\delta}%
\tensor{\Pi}{_\mu^\mu^\delta}+\frac{1}{2}\partial^{\alpha}\partial_{\mu}%
T^{\mu\beta}+\frac{1}{4}\partial^{\alpha}%
\tensor{\Pi}{_\mu^\mu^\beta}\nonumber\\
&  \quad+\frac{1}{2}\partial^{\beta}\partial_{\mu}T^{\mu\alpha}+\frac{1}%
{4}\partial^{\beta}\tensor{\Pi}{_\mu^\mu^\alpha}\,.
\end{align}
Finally, Eq.\ (\ref{eq:ablkontraktion3}) is used to express the
remaining canonically conjugate fields by derivatives of tensor fields.
This gives the Fierz-Pauli equation (\ref{eq:symzwangsbew}).
All further steps, i.e., showing that $T=0$ and $\partial_{\alpha}T^{\alpha \beta}=0$ 
follow now as shown in Sec.\ \ref{sec:lagrangianhamiltonian}.
The covariant Hamilton approach is therefore completely equivalent to the
well-known Lagrangian formulation.

\section{Boosted polarization tensors}

\label{app:polarisationboost}

In this appendix, we compute the polarization tensors in a frame where the particle
is moving along the $z$-axis. To this end, we have to perform a Lorentz boost of
the tensors given in Eqs.\ (\ref{pol1}) -- (\ref{pol5}). 

In general, Lorentz boosts on vectors and tensors are given by:
\begin{align}
A^{\prime}{}^{\mu} &  =\tensor{L}{^\mu_\nu}A^{\nu}\,,\\
B^{\prime}{}^{\alpha\beta} &
=\tensor{L}{^\alpha_\mu}\tensor{L}{^\beta_\nu}B^{\mu\nu}%
\,,\label{eq:lorentzboost}%
\end{align}
where \[
\tensor{L}{^\mu_\nu}=%
\begin{pmatrix}
\gamma & 0 & 0 & -\beta\gamma\\
0 & 1 & 0 & 0\\
0 & 0 & 1 & 0\\
-\beta\gamma & 0 & 0 & \gamma
\end{pmatrix}
\,,
\]
with $\beta\equiv v$ being the three-velocity of the moving frame
(the velocity of the particle is equal to $-v$ in this frame) and
$\gamma=\frac{1}{\sqrt{1-\beta^{2}}}$.

Then, the polarization tensors transform as:
\begin{align}
(\epsilon^{\prime}{}^{\alpha\beta})& =%
\begin{pmatrix}
\gamma & 0 & 0 & -\beta\gamma\\
0 & 1 & 0 & 0\\
0 & 0 & 1 & 0\\
-\beta\gamma & 0 & 0 & \gamma
\end{pmatrix}
(\epsilon^{\mu\nu})%
\begin{pmatrix}
\gamma & 0 & 0 & -\beta\gamma\\
0 & 1 & 0 & 0\\
0 & 0 & 1 & 0\\
-\beta\gamma & 0 & 0 & \gamma
\end{pmatrix}
\,.
\end{align}
For $\vec{k}^{\, \prime}=(0,0,k_{z}^\prime)$ we obtain for $\lambda=1,\ldots,5$:
\begin{align*}
\epsilon^{\prime}{}^{\mu\nu}(\vec{k}^{\, \prime},1)& =\frac{1}{\sqrt{2}}%
\begin{pmatrix}
0 & 0 & 0 & 0\\
0 & 1 & 0 & 0\\
0 & 0 & -1 & 0\\
0 & 0 & 0 & 0
\end{pmatrix} \equiv \epsilon^{\mu\nu}(\vec{0},1)
\,, \\[0.1cm]
\epsilon^{\prime}{}^{\mu\nu}(\vec{k}^{\, \prime},2)& =\frac{1}{\sqrt{2}}%
\begin{pmatrix}
0 & 0 & 0 & 0\\
0 & 0 & 1 & 0\\
0 & 1 & 0 & 0\\
0 & 0 & 0 & 0
\end{pmatrix}\equiv \epsilon^{\mu\nu}(\vec{0},2)
\,, \\[0.1cm]
\epsilon^{\prime}{}^{\mu\nu}(\vec{k}^{\, \prime},3)& =\frac{1}{\sqrt{2}}%
\begin{pmatrix}
0 & -\beta\gamma & 0 & 0\\
-\beta\gamma & 0 & 0 & \gamma\\
0 & 0 & 0 & 0\\
0 & \gamma & 0 & 0
\end{pmatrix}
\,, \\[0.1cm]
\epsilon^{\prime}{}^{\mu\nu}(\vec{k}^{\, \prime},4)& =\frac{1}{\sqrt{2}}%
\begin{pmatrix}
0 & 0 & -\beta\gamma & 0\\
0 & 0 & 0 & 0\\
-\beta\gamma & 0 & 0 & \gamma\\
0 & 0 & \gamma & 0
\end{pmatrix}
\,, \\[0.1cm]
\epsilon^{\prime}{}^{\mu\nu}(\vec{k}^{\, \prime},5)& =\frac{1}{\sqrt{6}}%
\begin{pmatrix}
-2\beta^{2}\gamma^{2} & 0 & 0 & 2\beta\gamma^{2}\\
0 & 1 & 0 & 0\\
0 & 0 & 1 & 0\\
2\beta\gamma^{2} & 0 & 0 & -2\gamma^{2}%
\end{pmatrix}
\,.
\end{align*}
Note that the velocity of the particle is $k_z^\prime/\omega_k^\prime = -v$.

It is obvious that the boosted polarization tensors still obey the symmetry and
tracelessness conditions (\ref{eq:symmetrie}) and (\ref{eq:traceless}).
In order to check whether Eq.\ (\ref{eq:lorentzcond}) is fulfilled, we 
first compute the four-momentum of the particle in the moving frame,
\[%
\begin{pmatrix}
\gamma & 0 & 0 & -\beta\gamma\\
0 & 1 & 0 & 0\\
0 & 0 & 1 & 0\\
-\beta\gamma & 0 & 0 & \gamma
\end{pmatrix}%
\begin{pmatrix}
m\\
0\\
0\\
0
\end{pmatrix}
=%
\begin{pmatrix}
\gamma m\\
0\\
0\\
-\beta\gamma m
\end{pmatrix}
=k^{\prime}{}^{\mu}\equiv
\begin{pmatrix}
\omega_k^\prime\\
0\\
0\\
k_z^\prime
\end{pmatrix}\,,
\]
which confirms that $k_z^\prime = -v\, \omega_k^\prime$. Now one explicitly computes
$k^\prime_\mu \epsilon^{\prime}{}^{\mu\nu}(\vec{k}^{\, \prime},\lambda)$ for each
$\lambda = 1, \ldots , 5$ and confirms the validity of Eq.\ (\ref{eq:lorentzcond}).

\section{Completeness relation}

\label{app:completenessrelation}

In this appendix, we prove the completeness relation
(\ref{eq:vollrel}). To this end, we note that the left-hand side
of this equation depends on the three-momentum $\vec{k}$, i.e., in
a relativistic setting on the on-shell four-momentum $k^\mu = (\omega_k,\vec{k})^T$,
where $\omega_k = \sqrt{\vec{k}^{\,2} + m^2}$. Consequently, we may
tensor-decompose the left-hand side with respect to all possible rank-four tensor products 
formed from
$k^\mu$ and the rank-two tensor (\ref{Gproj}),
which projects onto the three-dimensional subspace
orthogonal to $k^\mu$.
Note that $G_{\mu \nu}$ is a projector only if $k^\mu$ is on-shell.
The tensor (\ref{Gproj}) fulfills the relations
\begin{align}
G^\mu_\alpha \tensor{G}{^\alpha^\nu} & = \tensor{G}{^\mu^\nu}\,, 
\vphantom{\left(\frac{k_\mu k_\nu}{m^2} \right)} \label{eq:GGproj} \\
G^\mu_\mu & = 3\, ,\vphantom{\left(\frac{k_\mu k_\nu}{m^2} \right)} 
\label{eq:Gtrace} \\
k^{\mu}G_{\mu\nu}& =k^{\mu}\left(  g_{\mu\nu}-\frac{k_{\mu}k_{\nu}}{m^{2}%
}\right)  =0\vphantom{\left(\frac{k_\mu k_\nu}{m^2} \right)}\,. \label{eq:orthokG}
\end{align}
The tensor decomposition reads (see also Ref.\ \cite{Zee}):
\begin{align}
\sum_{\lambda=1}^{5}\epsilon_{\mu\nu}(\vec{k},\lambda)\epsilon_{\alpha\beta
}(\vec{k},\lambda)  &  =A\, G_{\mu\nu}G_{\alpha\beta}+B_{1}\, G_{\mu\alpha}%
G_{\nu\beta}+B_{2}\, G_{\mu\beta}G_{\nu\alpha}\nonumber\\
&  \quad+C_{1}\, k_{\alpha}k_{\beta}G_{\mu\nu}+C_{2}\, k_{\mu}k_{\nu}G_{\alpha\beta
}+D_{1}\, k_{\mu}k_{\alpha}G_{\nu\beta}+D_{2}\, k_{\mu}k_{\beta}G_{\nu\alpha
}\vphantom{\sum_{\lambda = 1}^{5}}\nonumber\\
&  \quad+D_{3}\, k_{\nu}k_{\beta}G_{\mu\alpha}+D_{4}\ k_{\nu}k_{\alpha}G_{\mu\beta
}+E\ k_{\mu}k_{\nu}k_{\alpha}k_{\beta}\vphantom{\sum_{\lambda = 1}^{5}} \,.
\label{eq:ansatzcomplete}
\end{align}
The symmetry relation (\ref{eq:symm}) yields the conditions
\begin{align}
B_{1}  &  =B_{2}\equiv B \,,\nonumber\\
D_{1}  &  =D_{2}=D_{3}=D_{4}\equiv D\,,\nonumber
\end{align}
while the symmetry under simultaneous exchange
$(\mu \nu) \leftrightarrow (\alpha \beta)$ yields
\begin{align}
C_{1}  &  =C_{2} \equiv C\,.\nonumber
\end{align}
Then, Eq.\ (\ref{eq:ansatzcomplete}) can be simplified as:
\begin{align}
\sum_{\lambda=1}^{5}\epsilon_{\mu\nu}(\vec{k},\lambda)
\epsilon_{\alpha\beta}(\vec{k},\lambda)
&  =A\, G_{\mu\nu}G_{\alpha\beta}+B\, (G_{\mu\alpha}G_{\nu\beta}+G_{\mu\beta}%
G_{\nu\alpha})\nonumber\\
&  \quad+C\, (k_{\alpha}k_{\beta}G_{\mu\nu}+k_{\mu}k_{\nu}G_{\alpha\beta
})+E\, k_{\mu}k_{\nu}k_{\alpha}k_{\beta}%
\vphantom{\sum_{\lambda = 1}^{5}}\nonumber\\
&  \quad+D\, (k_{\mu}k_{\alpha}G_{\nu\beta}+k_{\mu}k_{\beta}G_{\nu\alpha}+k_{\nu
}k_{\beta}G_{\mu\alpha}+k_{\nu}k_{\alpha}G_{\mu\beta}%
)\vphantom{\sum_{\lambda = 1}^{5}}\,. \label{eq:ansatzcomplete_simp}
\end{align}
Using Eq.\ (\ref{eq:orthokG}) together with Eq.\ (\ref{eq:spatial}) leads to:
\begin{align}
0  &  =k^{\mu}\sum_{\lambda=1}^{5}\epsilon_{\mu\nu}(\vec{k},\lambda)\epsilon
_{\alpha\beta}(\vec{k},\lambda)\nonumber\\
&  =C\,  k^{2}k_{\nu}G_{\alpha\beta}  +D\, \left(  k^{2}k_{\alpha
}G_{\nu\beta}+k^{2}k_{\beta}G_{\nu\alpha}\right)  +E\,  k^{2}k_{\nu
}k_{\alpha}k_{\beta}  \vphantom{\sum_{\lambda = 1}^{5}}\,.
\end{align}
Contracting this result further with $k^\nu,\,k^\alpha$, and $k^\beta$, we
derive that the coefficients $C$, $D$, and $E$ vanish. This leaves only two
undetermined coefficients in Eq.\ (\ref{eq:ansatzcomplete_simp}),
\begin{align}
\sum_{\lambda=1}^{5}\epsilon_{\mu\nu}(\vec{k},\lambda)\epsilon_{\alpha\beta}%
(\vec{k},\lambda)& =A\,G_{\mu\nu}G_{\alpha\beta}+B\, (G_{\mu\alpha}G_{\nu\beta}+G_{\mu
\beta}G_{\nu\alpha})\,.\label{eq:ansatzcomplete_simp2}
\end{align}
The condition (\ref{eq:trace}) eliminates one of those coefficients,
\begin{align}
0  &  =g^{\mu\nu}\sum_{\lambda=1}^{5}\epsilon_{\mu\nu}(\vec{k},\lambda)\epsilon
_{\alpha\beta}(\vec{k},\lambda)\nonumber\\
&  =A\, G_\mu^\mu G_{\alpha\beta}+2B\, G_{\mu\alpha}%
G^\mu_\beta\vphantom{\sum_{\lambda = 1}^{5}}\nonumber\\
&  =(3\, A +2\, B)G_{\alpha\beta}\,. \vphantom{\sum_{\lambda = 1}^{5}}
\end{align}
Inserting the solution $A=-2B/3$ into Eq.\ (\ref{eq:ansatzcomplete_simp2}) we
find
\begin{equation}
\sum_{\lambda=1}^{5}\epsilon_{\mu\nu}(\vec{k},\lambda)\epsilon_{\alpha\beta}%
(\vec{k},\lambda) =2B \left[\frac{1}{2} (G_{\mu\alpha}G_{\nu\beta}
+G_{\mu \beta}G_{\nu\alpha}) -\frac{1}{3}\,G_{\mu\nu}G_{\alpha\beta}\right]\,.
\end{equation}
Finally, we determine the overall factor $B$. This will be done using the
orthonormality condition (\ref{eq:ortho}). Together with the properties
(\ref{eq:GGproj}), (\ref{eq:Gtrace}) we obtain
\[
5=\sum_{\lambda=1}^{5}\delta_{\lambda\lambda} =2\, B \left[ \frac{1}{2}(9 + 3) 
- \frac{1}{3} \, 3 \right] = 10\, B \ \longrightarrow \ B=\frac{1}{2}\,.
\]
As a final result, the completeness relation reads:
\begin{align}
\sum_{\lambda=1}^{5}\epsilon_{\mu\nu}(\vec{k},\lambda)
\epsilon_{\alpha\beta}(\vec{k},\lambda)
&  =\frac{1}{2}(G_{\mu\alpha}G_{\nu\beta}+G_{\mu\beta}G_{\nu\alpha})
-\frac{1}{3}G_{\mu\nu}G_{\alpha\beta}\, , \label{eq:completenessrel}%
\end{align}
cf.\ Eq.\ (\ref{eq:vollrel}).
This expression appears also in the numerator of the tree-level propagator discussed in
Sec.\ \ref{sec:propagator}. We remark that taking the value $B=1/2$ is consistent
with the normalization of the tree-level 
propagator. (For a different choice, see Ref.\ \cite[p.33]{Zee}.)

\section{Inversion of the differential operator}

\label{app:propagatorcalc}

In order to compute the tree-level propagator for spin-two fields
explicitly, one has to insert the differential operator
(\ref{eq:differentialoperator}) and the ansatz
(\ref{eq:approachprop}) into Eq.\ (\ref{eq:inversionprop}) and compare the
coefficients of the individual expressions:
\begin{align}
\frac{1}{2} \left( g_{\nu}^{\sigma} g_{\rho}^{\lambda} 
+ g_{\nu}^{\lambda} g_{\rho}^{\sigma}\right) 
& = \left[ \frac{1}{2} \left( k^{2} - m^{2} \right) \left( g_{\nu\alpha} g_{\rho\beta} 
+ g_{\nu\beta} g_{\rho\alpha}
- 2\, g_{\nu\rho} g_{\alpha\beta} \right) \right. \nonumber
\\
& \quad - \left. \frac{1}{2}\left(  g_{\rho\beta}k_{\nu}k_{\alpha}+g_{\rho\alpha
}k_{\nu}k_{\beta}+g_{\nu\alpha}k_{\rho}k_{\beta}+g_{\nu\beta}k_{\rho}%
k_{\alpha} -2\,  g_{\nu\rho}k_{\alpha}k_{\beta}-2\,g_{\alpha\beta}k_{\nu}k_{\rho}
\right)  \right] \nonumber
\\
& \vphantom{\left[\frac{1}{2}\right]} \quad\times\left[  A\, \left(  g^{\alpha\sigma}%
g^{\beta\lambda}+g^{\alpha\lambda}g^{\beta\sigma}\right) 
+B\,  g^{\alpha\beta}g^{\sigma\lambda} \right. \nonumber
\\
&  \qquad+C\, \left(  g^{\beta\lambda}k^{\alpha}k^{\sigma}+g^{\beta\sigma
}k^{\alpha}k^{\lambda}+g^{\alpha\sigma}k^{\beta}k^{\lambda}+g^{\alpha\lambda
}k^{\beta}k^{\sigma}\right) \vphantom{\left[\frac{1}{2}\right]} \nonumber
\\
&  \qquad+D\, \left(  g^{\sigma\lambda}k^{\alpha}k^{\beta}+g^{\alpha\beta
}k^{\sigma}k^{\lambda}\right) +\left.  E\,  k^{\alpha}k^{\beta}k^{\sigma}k^{\lambda}
\right] \vphantom{\left[\frac{1}{2}\right]} \nonumber
\\
&  = \vphantom{\left[\frac{1}{2}\right]} \frac{1}{2}\left(  k^{2}-m^{2}\right) 
\left[  2\, A\, \left(  g_{\nu}^{\sigma}g_{\rho}^{\lambda}+g_{\nu}^{\lambda}g_{\rho}^{\sigma}
-2\, g_{\nu\rho}g^{\sigma\lambda}\right)  -6\, B\,
g_{\nu\rho}g^{\sigma\lambda} \right. \nonumber
\\
& \vphantom{\left[\frac{1}{2}\right]} \qquad\qquad\qquad+2\, C\, \left(  
g_{\rho}^{\lambda}k_{\nu}k^{\sigma}+g_{\rho}^{\sigma}k_{\nu}k^{\lambda}
+g_{\nu}^{\sigma}k_{\rho}k^{\lambda}
+g_{\nu}^{\lambda}k_{\rho}k^{\sigma}
-4\, g_{\nu\rho}k^{\sigma}k^{\lambda}\right) \nonumber\\
& \vphantom{\left[\frac{1}{2}\right]} \qquad\qquad\qquad
+  2D\left(  g^{\sigma\lambda}k_{\nu}k_{\rho
}-3\, g_{\nu\rho}k^{\sigma}k^{\lambda}-g_{\nu\rho}g^{\sigma\lambda}%
k^{2}\right) \nonumber
\\
&  \qquad\qquad\qquad+\left. 2\, E\, \left(  k_{\nu}k_{\rho}k^{\sigma
}k^{\lambda}-  g_{\nu\rho}%
k^{\sigma}k^{\lambda}k^{2}\right)  \right] \vphantom{\left[\frac{1}{2}\right]} \nonumber
\\
& \vphantom{\left[\frac{1}{2}\right]} \quad-\frac{1}{2}\left[  
2\, A\, \left(  g_{\rho}^{\lambda}k_{\nu}k^{\sigma
}+g_{\rho}^{\sigma}k_{\nu}k^{\lambda}+g_{\nu}^{\sigma}k_{\rho}k^{\lambda
}+g_{\nu}^{\lambda}k_{\rho}k^{\sigma}-2\,g_{\nu\rho}k^{\sigma}k^{\lambda}
-2\,g^{\sigma\lambda }k_{\nu}k_{\rho}\right)  \right. \nonumber
\\
& \vphantom{\left[\frac{1}{2}\right]} \qquad\quad-2\, B\, \left( 
2\, g^{\sigma\lambda} k_{\nu}k_{\rho} 
+g_{\nu\rho}g^{\sigma\lambda}k^{2}\right) \nonumber
\\
& \vphantom{\left[\frac{1}{2}\right]} \qquad\quad+2\, C\, \left(  
g_{\rho}^{\lambda}k_{\nu}k^{\sigma}k^{2}+g_{\rho
}^{\sigma}k_{\nu}k^{\lambda}k^{2}+g_{\nu}^{\lambda}k_{\rho}k^{\sigma}%
k^{2}+g_{\nu}^{\sigma}k_{\rho}k^{\lambda}k^{2}
-4\,g_{\nu\rho}k^{\sigma}k^{\lambda}k^{2}\right) \nonumber
\\
& \vphantom{\left[\frac{1}{2}\right]} \qquad\quad
+2\, D\, \left(  g^{\sigma\lambda}k_{\nu}k_{\rho}k^{2}-g_{\nu\rho}k^{\sigma
}k^{\lambda}k^{2}-g^{\sigma\lambda}g_{\nu\rho}(k^{2})^2-2\,k_{\nu}k_{\rho}k^{\sigma
}k^{\lambda}\right) \nonumber
\\
&  \qquad\quad+\left.  2\, E\, \left(  k_{\nu}k_{\rho}k^{\sigma}k^{\lambda}%
k^{2}-g_{\nu\rho}k^{\sigma}k^{\lambda}(k^{2})^2\right)  \right] 
\vphantom{\left[\frac{1}{2}\right]}   \,.
\end{align}
Equating coefficients term by term one finds:\newline\newline\textbf{1.}
$g_{\nu}^{\sigma}g_{\rho}^{\lambda}$ and $g_{\nu}^{\lambda}g_{\rho}^{\sigma}%
$:
\begin{equation} \label{A}
\longrightarrow \ A=\frac{1}{2\left(  k^{2}-m^{2}\right)  }\,.
\end{equation}
\textbf{2.} $g_{\rho}^{\lambda}k_{\nu}k^{\sigma}$, $g_{\nu}^{\lambda}k_{\rho
}k^{\sigma}$, $g_{\rho}^{\sigma}k_{\nu}k^{\lambda}$ and $g_{\nu}^{\sigma
}k_{\rho}k^{\lambda}$:
\begin{align}
0  &  =\frac{1}{2}\left(  k^{2}-m^{2}\right)  2\, C-\frac{1}{2}\left(
2\, A+2\, Ck^{2}\right) \nonumber\\
&  =-m^{2}C-A\vphantom{\frac{1}{2}}\,,
\end{align}%
\begin{equation} \label{C}
\longrightarrow \ C=-\frac{1}{2\left(  k^{2}-m^{2}\right)  }\frac{1}{m^{2}}\,.
\end{equation}
\textbf{3.} $g_{\nu\rho}k^{\sigma}k^{\lambda}$:
\begin{align}
0  &  =-\frac{1}{2}\left(  k^{2}-m^{2}\right) \left(  8\, C+6\, D+2\, Ek^{2}\right)  +2\, A+4\, Ck^{2}
+Dk^{2}+E\, (k^{2})^2 \nonumber \\
&  =-2\, Dk^{2}+3\, Dm^{2}+4\, Cm^{2}+Em^{2}k^{2}+2\, A\,.\vphantom{\frac{1}{2}}
\end{align}%
\begin{align} \label{zwischen}
\longrightarrow \ -\frac{1}{ k^{2}-m^{2}  }=2\, Dk^{2}-3\, Dm^{2}-Em^{2}k^{2}\,.
\end{align}
\textbf{4.} $k_{\nu}k_{\rho}k^{\sigma}k^{\lambda}$:
\begin{align}
0  &  =\left(k^2-m^2\right) E+2\, D-Ek^{2}\nonumber\\
&  =-m^{2}E+2\, D\,.\vphantom{\frac{1}{2}}
\end{align}%
\begin{align} \label{zwischen2}
\longrightarrow \ E=\frac{2\, D}{m^{2}}\,.
\end{align}
Using this in Eq.\ (\ref{zwischen}) we obtain:
\begin{align} \label{D}
D=\frac{1}{3\left(  k^{2}-m^{2}\right)  }\frac{1}{m^{2}}\,.
\end{align}
Inserting this into Eq.\ (\ref{zwischen2}) results in
\begin{align} \label{E}
E=\frac{2}{3\left(  k^{2}-m^{2}\right)  }\frac{1}{m^{4}}\,.
\end{align}
\textbf{5.} $g^{\sigma\lambda}k_{\nu}k_{\rho}$:
\begin{align}
0  &  =\frac{1}{2}\left(  k^{2}-m^{2}\right)  2\, D +2\, A+2\, B-Dk^{2}\nonumber\\
&  =-Dm^{2}+2\, A+2\, B\,.\vphantom{\frac{1}{2}}
\end{align}
Inserting Eqs.\ (\ref{A}) and (\ref{D}) leads to:
\begin{align} \label{B}
B=-\frac{1}{3\left(  k^{2}-m^{2}\right)  }\,.
\end{align}
\textbf{6.} $g_{\nu\rho}g^{\sigma\lambda}$: 
\newline Although all coefficients
are already determined the coefficient of this tensor structure
leads to the condition
\begin{equation}
0=-\frac{1}{2}\left(  k^{2}-m^{2}\right)  \left(4A+6B+2Dk^{2}\right)  +Bk^{2}+D(k^{2})^2\,.
\end{equation}
Inserting the results (\ref{A}), (\ref{D}), and (\ref{B}), we observe that this expression
vanishes identically.

We now insert Eqs.\ (\ref{A}), (\ref{C}), (\ref{E}), (\ref{D}), and (\ref{B}) into 
Eq.\ (\ref{eq:approachprop}) and obtain the tree-level propagator as:
\begin{align}
P_{\alpha\beta\sigma\lambda}  &  =\frac{1}{2}\frac{1}{ k^{2}-m^{2} }
\left[    g_{\alpha\sigma}g_{\beta\lambda}%
+g_{\alpha\lambda}g_{\beta\sigma}  -\frac{2}{3}\,  g_{\alpha\beta}g_{\sigma\lambda}
  \right. \nonumber\\
&  \qquad\qquad\qquad-\frac{1}{m^{2}}\left(  g_{\beta\lambda}k_{\alpha
}k_{\sigma}+g_{\beta\sigma}k_{\alpha}k_{\lambda}+g_{\alpha\sigma}k_{\beta
}k_{\lambda}+g_{\alpha\lambda}k_{\beta}k_{\sigma}\right) \nonumber\\
&  \qquad\qquad\qquad+\frac{2}{3m^{2}}\left(  g_{\sigma\lambda}k_{\alpha
}k_{\beta}+g_{\alpha\beta}k_{\sigma}k_{\lambda}\right)
+\left.  \frac{4}{3m^{4}}\,  k_{\alpha}k_{\beta}k_{\sigma}k_{\lambda}  \right] \,.
\end{align}
The result can be expressed in terms of the projector (\ref{Gproj}) to assume the
form given in Eq.\ (\ref{eq:treelevelprop}).

\section{Commutation relations}

\label{app:commutators}

In this appendix we compute the commutators (\ref{eq:commutatorTT}), 
(\ref{eq:commutatorPP}), and (\ref{eq:commutatorTP}) for the quantized 
spin-two field (\ref{eq:loesung}) and its quantized canonically conjugate field
(\ref{eq:kankonjlosung}). We start with the derivation of the commutation 
relation (\ref{eq:commutatorTT}). With the help of the commutators (\ref{eq:antikomm}) 
we obtain
\begin{align}
\left[ \tensor{\hat{T}}{_\mu_\nu}(x) , \tensor{\hat{T}}{_\alpha_\beta}(y) \right]_{-}
& = \int \frac{\mathrm{d}^{3}k\, \mathrm{d}^{3}k^{\prime}}{\left(  2\pi\right)^{3}} 
\frac{1}{4\omega_{k}\omega_{k^{\prime}}} \sum_{\lambda,\lambda^{\prime}=1}^{5}
\epsilon_{\mu\nu}(\vec{k},\lambda) \epsilon_{\alpha\beta}(\vec{k}^{\prime},\lambda^{\prime}) 
\nonumber
\\
& \qquad \times \left\{ \left[ \hat{a}(\vec{k},\lambda) , \hat{a}(\vec{k}^{\prime},\lambda^{\prime})
\right]_- e^{-ikx-ik^{\prime}y} + \left[ \hat{a}(\vec{k},\lambda) , 
\hat{a}^{\dagger}(\vec{k}^{\prime},\lambda^{\prime}) \right]_-
e^{-ikx+ik^{\prime}y} \right. \vphantom{\sum_{\lambda,\lambda' = 1}^{5}} \nonumber
\\
& \qquad \left. + \left[ \hat{a}^{\dagger}(\vec{k},\lambda) , \hat{a}(\vec{k}^{\prime},
\lambda^{\prime}) \right]_- e^{+ikx-ik^{\prime}y} + \left[ \hat{a}^{\dagger}(\vec{k},\lambda) , 
\hat{a}^{\dagger}(\vec{k}^{\prime},\lambda^{\prime}) \right]_- 
e^{+ikx+ik^{\prime}y} \right\} \vphantom{\sum_{\lambda,\lambda' = 1}^{5}} \nonumber
\\
& = \int \frac{\mathrm{d}^{3}k\, \mathrm{d}^{3}k^{\prime}}{\left(2\pi\right)^{3}} 
\frac{1}{4\omega_{k}\omega_{k^{\prime}}} \sum_{\lambda,\lambda^{\prime}=1}^{5} 
\epsilon_{\mu\nu}(\vec{k},\lambda) \epsilon_{\alpha\beta}(\vec{k}^{\prime},\lambda^{\prime})
\nonumber
\\
& \qquad \times \left[ 2\, \omega_{k}\, \delta_{\lambda\lambda^{\prime}}\, 
\delta^{(3)}(\vec{k}-\vec{k}^{\prime}) e^{-ikx+ik^{\prime}y} 
- 2\, \omega_{k}\, \delta_{\lambda\lambda^{\prime}}\, \delta^{(3)}(\vec{k}-\vec{k}^{\prime})
e^{+ikx-ik^{\prime}y} \right] \vphantom{\sum_{\lambda,\lambda' = 1}^{5}} \nonumber
\\
& = \int \frac{\mathrm{d}^{3}k}{\sqrt{2\pi}^{6}}\frac{1}{2\omega_{k}} \sum_{\lambda=1}^{5} 
\epsilon_{\mu\nu}(\vec{k},\lambda) \epsilon_{\alpha\beta}(\vec{k},\lambda) \left[ e^{-ik(x-y)} 
- e^{+ik(x-y)} \right] \,.
\end{align}
With the completeness relation (\ref{eq:vollrel}) this becomes
\begin{align}
\left[ \tensor{\hat{T}}{_\mu_\nu}(x) , \tensor{\hat{T}}{_\alpha_\beta}(y) \right]_{-}
& = \int \frac{\mathrm{d}^{3}k}{\left(2\pi\right)^{3}} \frac{1}{2\omega_{k}} 
\left( - \frac{1}{3} g_{\mu\nu} g_{\alpha\beta} + \frac{1}{2} g_{\mu\alpha} g_{\nu\beta} 
+ \frac{1}{2} g_{\mu\beta} g_{\nu\alpha} + \frac{1}{3} g_{\mu\nu} 
\frac{k_{\alpha}k_{\beta}}{m^{2}} + \frac{1}{3} g_{\alpha\beta} 
\frac{k_{\mu}k_{\nu}}{m^{2}} \right. \vphantom{\int \frac{\mathrm{d}^3 k}{\sqrt{2\pi}^6}} 
\nonumber
\\
& \qquad \left. - \frac{1}{2} g_{\mu\alpha} \frac{k_{\nu}k_{\beta}}{m^{2}} 
- \frac{1}{2} g_{\nu\alpha} \frac{k_{\mu}k_{\beta}}{m^{2}} 
- \frac{1}{2} g_{\mu\beta} \frac{k_{\nu}k_{\alpha}}{m^{2}} 
- \frac{1}{2} g_{\nu\beta} \frac{k_{\mu}k_{\alpha}}{m^{2}} 
\vphantom{\int\frac{\mathrm{d}^3 k}{\sqrt{2\pi}^6}} 
+ \frac{2}{3} \frac{k_{\mu}k_{\nu}k_{\alpha}k_{\beta}}{m^{4}} \right) \nonumber
\\
& \qquad \times \left[ e^{-ik(x-y)} - e^{+ik(x-y)} \right] 
\vphantom{\int \frac{\mathrm{d}^3 k}{\sqrt{2\pi}^6}} \label{eq:nichtvertasuch} \,.
\end{align}
We restrict ourself to spatial components,
\begin{align}
\left[ \tensor{\hat{T}}{_i_j}(x) , \tensor{\hat{T}}{_k_l}(y) \right]_{-}
& = \int \frac{\mathrm{d}^{3}k}{\left(2\pi\right)^{3}} \frac{1}{i\, \omega_{k}} 
\left( - \frac{1}{3} \delta_{ij} \delta_{kl} + \frac{1}{2} \delta_{ik} \delta_{jl} 
+ \frac{1}{2} \delta_{il} \delta_{jk} - \frac{1}{3} \delta_{ij} \frac{k_{k}k_{l}}{m^{2}} 
- \frac{1}{3} \delta_{kl} \frac{k_{i}k_{j}}{m^{2}} \right. 
\vphantom{\int \frac{\mathrm{d}^3 k}{\sqrt{2\pi}^6}} \nonumber
\\
& \qquad \left. + \frac{1}{2} \delta_{ik} \frac{k_{j}k_{l}}{m^{2}} 
+ \frac{1}{2} \delta_{jk} \frac{k_{i}k_{l}}{m^{2}} + \frac{1}{2} \delta_{il} \frac{k_{j}k_{k}}{m^{2}} 
+ \frac{1}{2} \delta_{jl} \frac{k_{i}k_{k}}{m^{2}} + \frac{2}{3} \frac{k_{i}k_{j}k_{k}k_{l}}{m^{4}} 
\right) \sin[k(x-y)] \,.
\end{align}
We identify three different types of integrals:
\begin{align}
{\cal I}(x-y) & \equiv\int \frac{\mathrm{d}^{3}k}{\left(2\pi\right)^{3} \omega_{k}} \sin[k(x-y)] 
\nonumber \,,
\\
{\cal I}_{ij}(x-y)& \equiv \int \frac{\mathrm{d}^{3}k}{\left(2\pi\right)^{3} \omega_{k}} k_{i}k_{j} 
\sin[k(x-y)] \nonumber \,,
\\
{\cal I}_{ijkl}(x-y)& \equiv \int \frac{\mathrm{d}^{3}k}{\left(2\pi\right)^{3} \omega_{k}} 
k_{i}k_{j}k_{k}k_{l} \sin[k(x-y)] \nonumber \,.
\end{align}
We consider only equal-time commutation relation, and thus these integrals have to 
be evaluated at equal times $t_x = t_y \equiv t $, 
\begin{align}
{\cal I}(\vec{x}-\vec{y})& =-\int \frac{\mathrm{d}^{3}k}{\left(2\pi\right)^{3} \omega_{k}} 
\sin[\vec{k}\cdot (\vec{x}-\vec{y})] \nonumber \,,
\\
{\cal I}_{ij}(\vec{x}-\vec{y})& =-\int \frac{\mathrm{d}^{3}k}{\left(2\pi\right)^{3} \omega_{k}} 
k_{i}k_{j} \sin[\vec{k}\cdot (\vec{x}-\vec{y})] \nonumber \,,
\\
{\cal I}_{ijkl}(\vec{x}-\vec{y})& =-\int \frac{\mathrm{d}^{3}k}{\left(2\pi\right)^{3} 
\omega_{k}} k_{i}k_{j}k_{k}k_{l} \sin[\vec{k}\cdot (\vec{x}-\vec{y})] \nonumber \,.
\end{align}
Considering the symmetry of the integrands, one can convince oneself that all 
three integrals vanish identically and we obtain the 
commutation relation (\ref{eq:commutatorTT}).

The calculation of the second commutator proceeds analogously. 
We utilize the fact that the quantized canonically conjugate fields 
(\ref{eq:kankonjlosung}) can be expressed in terms of time derivatives of the 
quantized spin-two fields (\ref{eq:loesung}), cf.\ Eq.\ (\ref{eq:impulsfeld}):
\begin{align}
\left[ \tensor{\hat{\Pi}}{_\mu_\nu}(x) , \tensor{\hat{\Pi}}{_\alpha_\beta}(y) \right]_{-}
& = \partial_{0}^{x} \partial_{0}^{y} \left[ \tensor{\hat{T}}{_\mu_\nu}(x) , 
\tensor{\hat{T}}{_\alpha_\beta}(y) \right]_{-} \;,
\end{align}
where we used the fact that the derivative with respect to $t_y (t_x)$ commutes with the
field $\hat{T}_{\mu\nu}(x)$ ($\hat{T}_{\alpha \beta}(y)$). Using Eq.\ (\ref{eq:nichtvertasuch}) 
we observe that the two time derivatives yield a factor $\omega_k^2$ under the integral.
Since this does not change the symmetry of the integrand, we conclude that also
\begin{align}
\left[ \tensor{\hat{\Pi}}{_i_j}(t,\vec{x}) , \tensor{\hat{\Pi}}{_l_k}(t,\vec{y})\right]_{-} = 0 \,.
\end{align}

In order to derive the commutation relation (\ref{eq:commutatorTP}), 
we again use Eq.\ (\ref{eq:impulsfeld}) and take the time derivative out of the commutator
(\ref{eq:nichtvertasuch}),
\begin{align}
\left[ \tensor{\hat{T}}{_\mu_\nu}(x) , \tensor{\hat{\Pi}}{_\alpha_\beta}(y) \right]_{-}
& = \partial_{0}^{y} \left[ \tensor{\hat{T}}{_\mu_\nu}(x) , \tensor{\hat{T}}{_\alpha_\beta}(y) \right]_{-} \,.
\end{align} 
Note that the time derivative yields a factor $\pm i \omega_k$ in
front of the first (second) exponential in Eq.\ (\ref{eq:nichtvertasuch}). This then yields
\begin{align}
\left[ \tensor{\hat{T}}{_\mu_\nu}(x) , \tensor{\hat{\Pi}}{_\alpha_\beta}(y) \right]_{-}
& = \int \frac{\mathrm{d}^{3}k}{\left(2\pi\right)^{3}} \left( - \frac{1}{3} g_{\mu\nu} g_{\alpha\beta} 
+ \frac{1}{2} g_{\mu\alpha} g_{\nu\beta} + \frac{1}{2} g_{\mu\beta} g_{\nu\alpha} + \frac{1}{3} 
g_{\mu\nu} \frac{k_{\alpha}k_{\beta}}{m^{2}} + \frac{1}{3} g_{\alpha\beta} 
\frac{k_{\mu}k_{\nu}}{m^{2}} \right. \vphantom{\int \frac{\mathrm{d}^3 k}{\sqrt{2\pi}^6}} 
\nonumber
\\
& \qquad \left. - \frac{1}{2} g_{\mu\alpha} \frac{k_{\nu}k_{\beta}}{m^{2}} 
- \frac{1}{2} g_{\nu\alpha} \frac{k_{\mu}k_{\beta}}{m^{2}} - \frac{1}{2} g_{\mu\beta} 
\frac{k_{\nu}k_{\alpha}}{m^{2}} - \frac{1}{2} g_{\nu\beta} \frac{k_{\mu}k_{\alpha}}{m^{2}} 
+ \frac{2}{3} \frac{k_{\mu}k_{\nu}k_{\alpha}k_{\beta}}{m^{4}} \right) \nonumber
\\
& \qquad \times i\, \cos[k(x-y)] \vphantom{\int \frac{\mathrm{d}^3 k}{\sqrt{2\pi}^6}} \,.
\end{align}
Restricting ourself to spatial components we find:
\begin{align}
\left[ \tensor{\hat{T}}{_i_j}(x) , \tensor{\hat{\Pi}}{_k_l}(y) \right]_{-}
& = \int \frac{\mathrm{d}^{3}k}{\left(2\pi\right)^{3}} \left( - \frac{1}{3} \delta_{ij} \delta_{kl} 
+ \frac{1}{2} \delta_{ik} \delta_{jl} +\frac{1}{2} \delta_{il} \delta_{jk} - \frac{1}{3} \delta_{ij} 
\frac{k_{k}k_{l}}{m^{2}} - \frac{1}{3} \delta_{kl} \frac{k_{i}k_{j}}{m^{2}} \right. \vphantom{\int 
\frac{\mathrm{d}^3 k}{\sqrt{2\pi}^6}} \nonumber
\\
& \qquad \left. + \frac{1}{2} \delta_{ik} \frac{k_{j}k_{l}}{m^{2}} + \frac{1}{2} \delta_{jk} \frac{k_{i}
k_{l}}{m^{2}} + \frac{1}{2} \delta_{il} \frac{k_{j}k_{k}}{m^{2}} + \frac{1}{2} \delta_{jl} \frac{k_{i}
k_{k}}{m^{2}} + \frac{2}{3} \frac{k_{i}k_{j}k_{k}k_{l}}{m^{4}} \right) \nonumber
\\
& \qquad \times i\, \cos[k(x-y)] \vphantom{\int \frac{\mathrm{d}^3 k}{\sqrt{2\pi}^6}} \,.
\end{align}
There are again three different types of integrals which have to be discussed for 
equal times $t_x = t_y \equiv t $:
\begin{align}
{\cal J}(\vec{x}-\vec{y})& \equiv  \int \frac{\mathrm{d}^{3}k}{\left(2\pi\right)^{3}} 
\cos[\vec{k}\cdot (\vec{x}-\vec{y})]  
=\int \frac{\mathrm{d}^{3}k}{\left(2\pi\right)^{3}2} \left[ e^{+i\vec{k}\cdot (\vec{x}-\vec{y})} 
+ e^{-i\vec{k}\cdot (\vec{x}-\vec{y})} \right]  = \delta^{3}(\vec{x}-\vec{y}) \,, \\
{\cal J}_{ij}(\vec{x}-\vec{y})& \equiv  \int \frac{\mathrm{d}^{3}k}{\left(2\pi\right)^{3}} k_i k_j
\cos[\vec{k}\cdot (\vec{x}-\vec{y})]  
=-\partial_i \partial_j \delta^{3}(\vec{x}-\vec{y}) \,,  \\
{\cal J}_{ijkl}(\vec{x}-\vec{y})& \equiv  \int \frac{\mathrm{d}^{3}k}{\left(2\pi\right)^{3}} k_i k_j
k_k k_l \cos[\vec{k}\cdot (\vec{x}-\vec{y})]  
=\partial_i \partial_j \partial_k \partial_l \delta^{3}(\vec{x}-\vec{y}) \,.
\end{align}
Here we used the first integral to write the second 
and third in a compact form.

Combining all results the equal-time commutation relation reads
\begin{align}
\left[ \tensor{\hat{T}}{_i_j}(t,\vec{x}) , \tensor{\hat{\Pi}}{_l_k}(t,\vec{y})\right]_{-}
& = \left( - \frac{1}{3} \delta_{ij} \delta_{kl} 
+ \frac{1}{2} \delta_{ik} \delta_{jl} +\frac{1}{2} \delta_{il} \delta_{jk} 
+ \frac{1}{3} \delta_{ij} \frac{\partial_{k}\partial_{l}}{m^{2}} 
+ \frac{1}{3} \delta_{kl} \frac{\partial_{i}\partial_{j}}{m^{2}}
- \frac{1}{2} \delta_{ik} \frac{\partial_{j}\partial_{l}}{m^{2}}  \right. 
\vphantom{\int \frac{\mathrm{d}^3 k}{\sqrt{2\pi}^6}} \nonumber
\\
& \qquad \left. 
- \frac{1}{2} \delta_{jk} \frac{\partial_{i}\partial_{l}}{m^{2}} 
- \frac{1}{2} \delta_{il} \frac{\partial_{j}\partial_{k}}{m^{2}} 
- \frac{1}{2} \delta_{jl} \frac{\partial_{i}\partial_{k}}{m^{2}} 
+ \frac{2}{3} \frac{\partial_{i}\partial_{j}\partial_{k}\partial_{l}}{m^{4}} \right)
i \, \delta^{3}(\vec{x}-\vec{y}) \nonumber
\\
& = \left[ \frac{1}{2} \left( \tilde{G}_{ik}\tilde{G}_{jl} +\tilde{G}_{il}\tilde{G}_{jk} \right)
- \frac{1}{3}  \tilde{G}_{ij}\tilde{G}_{kl} \right] i \, \delta^{3}(\vec{x}-\vec{y})\,,
\end{align}
where we used $\tilde{G}_{ij}$ from Eq.\ (\ref{eq:tildeGproj}).

\section{Calculating the operators for conserved quantities}

\label{app:operators} 
In this appendix, we show the calculations leading to the 
operators described in Sec.\ \ref{sec:specialoperators}. 
We start with the Hamilton operator. One
inserts the quantized fields (\ref{eq:loesung}) and (\ref{eq:kankonjlosung})
into Eq.\ (\ref{eq:energiefunktion}) and arranges the resulting terms
according to products of annihilation and creation operators:
\begin{align}
\hat{H} & = \frac{1}{2} \int \mathrm{d}^{3}x \left( \hat{\Pi}_{\mu\nu} \hat{\Pi}^{\mu\nu} 
+ \vec{\nabla} \hat{T}_{\mu\nu} \cdot \vec{\nabla} \hat{T}^{\mu\nu} 
+ m^{2} \hat{T}_{\mu\nu} \hat{T}^{\mu\nu} \right) 
\vphantom{\sum_{\lambda,\lambda'=1}^{5}} \nonumber
\\
& = \frac{1}{2} \int \mathrm{d}^{3}x \int \frac{\mathrm{d}^{3}k\, 
\mathrm{d}^{3}k^{\prime}}{(2\pi)^{3}4\omega_{k}\omega_{k^{\prime}}} 
\sum_{\lambda,\lambda^{\prime}=1}^{5} \epsilon_{\mu\nu}(\vec{k},\lambda) 
\epsilon^{\mu\nu}(\vec{k}^{\prime},\lambda^{\prime}) \nonumber
\\
& \quad \times \left[ \hat{a}(\vec{k},\lambda) \hat{a}(\vec{k}^{\prime},\lambda^{\prime}) 
\left( - \omega_{k} \omega_{k^{\prime}} - \vec{k} \cdot \vec{k}^{\prime} + m^{2} \right) 
e^{-i(k+k^{\prime})x} \right. \vphantom{\sum_{\lambda,\lambda'=1}^{5}} \nonumber
\\
& \qquad + \hat{a}(\vec{k},\lambda) \hat{a}^{\dagger}(\vec{k}^{\prime},\lambda^{\prime}) 
\left( \omega_{k} \omega_{k^{\prime}} + \vec{k} \cdot\vec{k}^{\prime} 
+ m^{2} \right) e^{-i(k-k^{\prime})x} \vphantom{\sum_{\lambda,\lambda'=1}^{5}} \nonumber
\\
& \qquad + \hat{a}^{\dagger}(\vec{k},\lambda) \hat{a}(\vec{k}^{\prime},\lambda^{\prime}) 
\left( \omega_{k} \omega_{k^{\prime}} + \vec{k} \cdot\vec{k}^{\prime} 
+ m^{2} \right) e^{+i(k-k^{\prime})x} \vphantom{\sum_{\lambda,\lambda'=1}^{5}} \nonumber
\\
& \qquad + \left. \hat{a}^{\dagger}(\vec{k},\lambda) 
\hat{a}^{\dagger}(\vec{k}^{\prime},\lambda^{\prime}) 
\left( - \omega_{k} \omega_{k^{\prime}} - \vec{k}\cdot \vec{k}^{\prime} 
+ m^{2} \right) e^{+i(k+k^{\prime})x} \right] \vphantom{\sum_{\lambda,\lambda'=1}^{5}} \,.
\end{align}%
Using the completeness relation for the plane waves, 
$\int {\rm d}^3x e^{i (\vec{k} \mp \vec{k}^{\, \prime})\cdot \vec{x} }
= (2\pi)^3 \delta^{3}(\vec{k} \mp \vec{k}^{\, \prime})$ we obtain
\begin{align}
\hat{H} & = \frac{1}{2} \int \frac{\mathrm{d}^{3}k\, 
\mathrm{d}^{3}k^{\prime}}{(2\pi)^{3}4\omega_{k}\omega_{k^{\prime}}} 
\sum_{\lambda,\lambda^{\prime}=1}^{5} \epsilon_{\mu\nu}(\vec{k},\lambda) 
\epsilon^{\mu\nu}(\vec{k}^{\prime},\lambda^{\prime}) 
\vphantom{\sum_{\lambda,\lambda'=1}^{5}} \nonumber
\\
& \quad \times \left[ \hat{a}(\vec{k},\lambda) \hat{a}(\vec{k}^{\prime},\lambda^{\prime}) 
\left( - \omega_{k} \omega_{k^{\prime}} - \vec{k} \cdot \vec{k}^{\prime} 
+ m^{2} \right) e^{-i(\omega_{k} + \omega_{k^{\prime}})t}\,  
(2\pi)^{3}\,  \delta^{3}(\vec{k}+\vec{k}^{\prime}) \right. 
\vphantom{\sum_{\lambda,\lambda'=1}^{5}}\nonumber
\\
& \qquad + \hat{a}(\vec{k},\lambda) \hat{a}^{\dagger}(\vec{k}^{\prime},\lambda^{\prime}) 
\left( \omega_{k} \omega_{k^{\prime}} + \vec{k}\cdot  \vec{k}^{\prime} 
+ m^{2} \right) e^{-i(\omega_{k}-\omega_{k^{\prime}})t}\, 
(2\pi)^{3}\, \delta^{3} (\vec{k}-\vec{k}^{\prime}) 
\vphantom{\sum_{\lambda,\lambda'=1}^{5}} \nonumber
\\
&  \qquad+\hat{a}^{\dagger}(\vec{k},\lambda)\hat{a}(\vec{k}^{\prime},\lambda^{\prime})
\left(  \omega_{k}\omega_{k^{\prime}}+\vec{k}\cdot \vec{k}^{\prime}+m^{2}\right)  
e^{+i(\omega_{k}-\omega_{k^{\prime}})t}\, 
(2\pi)^{3}\, \delta^{3}(\vec{k}-\vec{k}^{\prime})
\vphantom{\sum_{\lambda,\lambda'=1}^{5}}\nonumber
\\
&  \qquad+\left.  \hat{a}^{\dagger}(\vec{k},\lambda)
\hat{a}^{\dagger}(\vec{k}^{\prime},\lambda^{\prime})
\left(  -\omega_{k}\omega_{k^{\prime}}-\vec{k} \cdot \vec{k}^{\prime}
+m^{2}\right)  e^{+i(\omega_{k}+\omega_{k^{\prime}})t}\, 
(2\pi)^{3}\, \delta^{3}(\vec{k}+\vec{k}^{\prime})\right]
\vphantom{\sum_{\lambda,\lambda'=1}^{5}}\nonumber
\\
&  =\frac{1}{2}\int\frac{\mathrm{d}^{3}k\, 2(\omega_{k})^{2}}{4(\omega_{k})^{2}}
\sum_{\lambda,\lambda^{\prime}=1}^{5}\epsilon_{\mu\nu}(\vec{k},\lambda)
\epsilon^{\mu\nu}(\vec{k},\lambda^{\prime})
\left[  \hat{a}(\vec{k},\lambda)\hat{a}^{\dagger}(\vec{k},\lambda^{\prime})
+\hat{a}^{\dagger}(\vec{k},\lambda)\hat{a}(\vec{k},\lambda^{\prime})\right] \;.\nonumber
\end{align}
Employing the orthogonality relation (\ref{eq:ortho}) and the commutation 
relation (\ref{eq:antikomm}), we arrive at
\begin{align}
\hat{H} &  =\frac{1}{2}\int \mathrm{d}^{3}k \sum_{\lambda=1}^{5}
\left[  \hat{a}^{\dagger}(\vec{k},\lambda)\hat{a}(\vec{k},\lambda)
+ \omega_{k}\, \delta^{3}(0)\right]\;.
\end{align}
The definition (\ref{eq:numberop}) of the number operator then yields
Eq.\ (\ref{eq:HOp}).

In complete analogy one determines the momentum operator as the
quantized version of Eq.\ (\ref{eq:impulsfunktion}):
\begin{align}
\hat{\vec{P}}  &  =-\int\left(  \hat{\Pi}_{\alpha\beta}\vec{\nabla}\hat
{T}^{\alpha\beta}\right)  \mathrm{d}^{3}%
x\vphantom{\sum_{\lambda,\lambda'=1}^{5}} \nonumber
\\
&  =-\int\frac{\mathrm{d}^{3}k\, \mathrm{d}^{3}k^{\prime}}{(2\pi)^{3}4\omega
_{k}\omega_{k^{\prime}}}\sum_{\lambda,\lambda^{\prime}=1}^{5}\epsilon_{\mu\nu
}(\vec{k},\lambda)\epsilon^{\mu\nu}(\vec{k}^{\prime},\lambda^{\prime
})\nonumber
\\
&  \quad\times\left[  \hat{a}(\vec{k},\lambda)\hat{a}(\vec{k}^{\prime}%
,\lambda^{\prime})\left(  \omega_{k}\vec{k}^{\prime}\right)  e^{-i(\omega
_{k}+\omega_{k^{\prime}})t}\, (2\pi)^{3}\, \delta^{3}(\vec{k}+\vec{k}^{\prime
})\right.  \vphantom{\sum_{\lambda,\lambda'=1}^{5}}\nonumber
\\
&  \qquad+\hat{a}(\vec{k},\lambda)\hat{a}^{\dagger}(\vec{k}^{\prime}%
,\lambda^{\prime})\left(  -\omega_{k}\vec{k}^{\prime}\right)  e^{-i(\omega
_{k}-\omega_{k^{\prime}})t}\, (2\pi)^{3}\, \delta^{3}(\vec{k}-\vec{k}^{\prime
})\vphantom{\sum_{\lambda,\lambda'=1}^{5}}\nonumber
\\
&  \qquad+\hat{a}^{\dagger}(\vec{k},\lambda)\hat{a}(\vec{k}^{\prime}%
,\lambda^{\prime})\left(  -\omega_{k}\vec{k}^{\prime}\right)  e^{+i(\omega
_{k}-\omega_{k^{\prime}})t}\, (2\pi)^{3}\, \delta^{3}(\vec{k}-\vec{k}^{\prime
})\vphantom{\sum_{\lambda,\lambda'=1}^{5}}\nonumber
\\
&  \qquad+\left.  \hat{a}^{\dagger}(\vec{k},\lambda)\hat{a}^{\dagger}(\vec
{k}^{\prime},\lambda^{\prime})\left(  \omega_{k}\vec{k}^{\prime}\right)
e^{+i(\omega_{k}+\omega_{k^{\prime}})t}\, (2\pi)^{3}\, \delta^{3}(\vec{k}+\vec
{k}^{\prime})\right]  \vphantom{\sum_{\lambda,\lambda'=1}^{5}} \nonumber 
\\
&  = \int\frac{\mathrm{d}^{3}k\ \vec{k}}{4\omega_{k}} \left\{
\sum_{\lambda,\lambda^{\prime}=1}^{5}\epsilon_{\mu\nu}(\vec{k},\lambda)
\epsilon^{\mu\nu}(\vec{k},\lambda^{\prime})
\left[\hat{a}^{\dagger}(\vec{k},\lambda')\hat{a}(\vec{k},\lambda)
+ \hat{a}^{\dagger}(\vec{k},\lambda)\hat{a}(\vec{k},\lambda^{\prime})
+ 2 \omega_k \delta_{\lambda \lambda'} \delta^3(0) \right] \right.
\nonumber
\\
&  \qquad \qquad \quad + \left. 
 \sum_{\lambda,\lambda^{\prime}=1}^{5}\epsilon_{\mu\nu}(\vec{k},\lambda)
\epsilon^{\mu\nu}(-\vec{k},\lambda^{\prime}) 
\left[\hat{a}(\vec{k},\lambda)\hat{a}(-\vec{k},\lambda^{\prime})e^{-2i\omega_{k}t}
+ \hat{a}^{\dagger}(\vec{k},\lambda)\hat{a}^{\dagger}(-\vec{k},\lambda^{\prime})
e^{+2i\omega_{k}t}\right] \right\} \,,
\vphantom{\sum_{\lambda,\lambda'=1}^{5}}\nonumber
\end{align}
where we used the commutation relation (\ref{eq:antikomm}). We now
observe that last term in the first line and all terms in the second line vanish, since
the integrand is an odd function of $\vec{k}$. Using the orthogonality relation
(\ref{eq:ortho}) and the definition (\ref{eq:numberop}) we then obtain Eq.\ (\ref{eq:POp}).

Next we calculate the spin operator. We first
quantize Eq.\ (\ref{eq:spinfunktion}), where we select the $z$--direction as quantization axis:
\begin{align}
\hat{S}_{z} &
= - 2 \int \mathrm{d}^{3}x  \left( \hat{\Pi}_{1\rho} \hat{T}^{2\rho} - \hat{\Pi}_{2\rho} 
\hat{T}^{1\rho} \right) \vphantom{\sum_{\lambda,\lambda'=1}^{5}}\nonumber
\\
& = - 2\, i \int \frac{\mathrm{d}^{3}k\, \mathrm{d}^{3}k^{\prime}}{(2\pi)^{3}4\omega_{k^{\prime}}}
\sum_{\lambda,\lambda^{\prime}=1}^{5} \left[ \epsilon_{1\rho}(\vec{k},\lambda) 
\epsilon^{2\rho}(\vec{k}^{\prime},\lambda^{\prime}) - \epsilon_{2\rho}(\vec{k},\lambda) 
\epsilon^{1\rho}(\vec{k}^{\prime},\lambda^{\prime}) \right] \nonumber
\\
& \quad \times \left[ - \hat{a}(\vec{k},\lambda) \hat{a}(\vec{k}^{\prime},\lambda^{\prime}) 
e^{-i(\omega_{k}+\omega_{k^{\prime}})t}\, (2\pi)^{3}\, \delta^{3}(\vec{k}+\vec{k}^{\prime}) \right.
\vphantom{\sum_{\lambda,\lambda'=1}^{5}} \nonumber
\\
& \qquad - \hat{a}(\vec{k},\lambda) \hat{a}^{\dagger}(\vec{k}^{\prime},\lambda^{\prime}) 
e^{-i(\omega_{k}-\omega_{k^{\prime}})t}\, (2\pi)^{3}\, \delta^{3}(\vec{k}-\vec{k}^{\prime}) 
\vphantom{\sum_{\lambda,\lambda'=1}^{5}} \nonumber
\\
& \qquad + \hat{a}^{\dagger}(\vec{k},\lambda) \hat{a}(\vec{k}^{\prime},\lambda^{\prime}) 
e^{+i(\omega_{k}-\omega_{k^{\prime}})t}\, (2\pi)^{3}\, \delta^{3}(\vec{k}-\vec{k}^{\prime}) 
\vphantom{\sum_{\lambda,\lambda'=1}^{5}} \nonumber
\\
& \qquad + \left. \hat{a}^{\dagger}(\vec{k},\lambda) 
\hat{a}^{\dagger}(\vec{k}^{\prime},\lambda^{\prime}) 
e^{+i(\omega_{k}+\omega_{k^{\prime}})t}\, (2\pi)^{3}\, \delta^{3}(\vec{k}+\vec{k}^{\prime}) \right] 
\vphantom{\sum_{\lambda,\lambda'=1}^{5}} \nonumber \\
& =- i\int\frac{\mathrm{d}^{3}k}{2\omega_{k}%
}\left\{  \sum_{\lambda,\lambda^{\prime}=1}^{5}\left[  \epsilon_{1\rho}(\vec
{k},\lambda)\epsilon^{2\rho}(-\vec{k},\lambda^{\prime})-\epsilon_{2\rho}%
(\vec{k},\lambda)\epsilon^{1\rho}(-\vec{k},\lambda^{\prime})\right]  \right.
\nonumber\\
&  \quad\times\left[  -\hat{a}(\vec{k},\lambda)\hat{a}(-\vec{k},\lambda
^{\prime})e^{-2i\omega_{k}t}+\hat{a}^{\dagger}(\vec{k},\lambda)\hat
{a}^{\dagger}(-\vec{k},\lambda^{\prime})e^{2i\omega_{k}t}\right]
\vphantom{\left[\sum_{\lambda,\lambda'=1}^{5}\right]}\nonumber\\
&  \qquad \qquad \;\; +\sum_{\lambda,\lambda^{\prime}=1}^{5}
\left[  \epsilon_{1\rho}(\vec{k},\lambda)\epsilon^{2\rho}(\vec{k},\lambda^{\prime})
-\epsilon_{2\rho}(\vec{k},\lambda)\epsilon^{1\rho}(\vec{k},\lambda^{\prime})\right]
\vphantom{\left[\sum_{\lambda,\lambda'=1}^{5}\right]}\nonumber\\
&  \quad\times\left.  \left[  -\hat{a}(\vec{k},\lambda)
\hat{a}^{\dagger}(\vec{k},\lambda^{\prime})+\hat{a}^{\dagger}(\vec{k},\lambda)
\hat{a}(\vec{k},\lambda^{\prime})\right]  \vphantom{\sum_{\lambda,\lambda'=1}^{5}}\right\}
\,.\nonumber
\end{align}
Taking a closer look at the first sum, we see that it vanishes by symmetry
when we substitute $\vec{k}\rightarrow-\vec{k}$ and exchange $\lambda \leftrightarrow
\lambda^{\prime}$. We thus arrive at
\begin{align}
\hat{S}_z &  =- i\int\frac{\mathrm{d}^{3}k}{2\omega_{k}}
\sum_{\lambda,\lambda^{\prime}=1}^{5}\left[  \epsilon_{1\rho}(\vec{k},\lambda)
\epsilon^{2\rho}(\vec{k},\lambda^{\prime})-\epsilon_{2\rho}(\vec{k},\lambda)
\epsilon^{1\rho}(\vec{k},\lambda^{\prime})\right] 
\left[  \hat{a}^{\dagger}(\vec{k},\lambda)\hat{a}(\vec{k},\lambda^{\prime})
-\hat{a}(\vec{k},\lambda)\hat{a}^{\dagger}(\vec{k},\lambda^{\prime})\right] \,.\nonumber
\end{align}
Further progress is made by employing the explicit form
of the polarization tensors. Here it does not matter whether we use the
polarization tensors in the rest frame, Eqs.\ (\ref{pol1}) -- (\ref{pol5}), or the boosted ones from
App.\ \ref{app:polarisationboost}. It turns out that only the terms
$(\lambda, \lambda') = (1,2),\, (2,1),\, (3,4), \, (4,3)$ give a nonvanishing contribution:
\begin{align}
\hat{S}_z &  =-i\int\frac{\mathrm{d}^{3}k}{4\omega_{k}%
}\left\{ 2\left[  \hat{a}^{\dagger}(\vec{k},1)\hat{a}(\vec{k},2)-\hat{a}%
(\vec{k},1)\hat{a}^{\dagger}(\vec{k},2)\right] -2\left[  \hat{a}^{\dagger}(\vec{k},2)\hat{a}(\vec
{k},1)-\hat{a}(\vec{k},2)\hat{a}^{\dagger}(\vec{k},1)\right]
\vphantom{\int \frac{\mathrm{d}^3 k}{4 \omega_k}} \right.\nonumber
\\
&  \qquad\qquad\quad \left. +\left[  \hat{a}^{\dagger}(\vec{k},3)\hat{a}(\vec
{k},4)-\hat{a}(\vec{k},3)\hat{a}^{\dagger}(\vec{k},4)\right]
\vphantom{\int \frac{\mathrm{d}^3 k}{4 \omega_k}} - \left[  \hat{a}^{\dagger}(\vec{k},4)\hat{a}%
(\vec{k},3)-\hat{a}(\vec{k},4)\hat{a}^{\dagger}(\vec{k},3)\right]  \right\}
\vphantom{\int \frac{\mathrm{d}^3 k}{4 \omega_k}} \nonumber
\\
&  =-i\int\frac{\mathrm{d}^{3}k}{2\omega_{k}}\left[  -2\, \hat{a}^{\dagger}%
(\vec{k},2)\hat{a}(\vec{k},1)+2\, \hat{a}^{\dagger}(\vec{k},1)\hat{a}(\vec
{k},2) - \hat{a}^{\dagger}(\vec{k},4)\hat{a}(\vec
{k},3)+\hat{a}^{\dagger}(\vec{k},3)\hat{a}(\vec{k},4)\right]
\vphantom{\int \frac{\mathrm{d}^3 k}{4 \omega_k}}\,.
\end{align}
This operator is not diagonal and cannot be directly expressed via number
operators. This is, however, possible by switching the basis of polarization
tensors to a circularly polarized one:
\begin{align}
\hat{a}(\vec{k},+)  &  =\frac{1}{\sqrt{2}}\left[  \hat{a}(\vec{k},1)-i\hat{a}(\vec{k},2)\right]  \,,
\nonumber\\
\hat{a}(\vec{k},-)  &  =\frac{1}{\sqrt{2}}\left[  \hat{a}(\vec{k},1)+i\hat{a}(\vec{k},2)\right]  \,,
\nonumber\\
\hat{a}(\vec{k},\Delta)  &  =\frac{1}{\sqrt{2}}\left[  \hat{a}(\vec{k},3)-i\hat{a}(\vec{k},4)\right]  \,,
\nonumber\\
\hat{a}(\vec{k},\Box)  &  =\frac{1}{\sqrt{2}}\left[  \hat{a}(\vec{k},3)+i\hat{a}(\vec{k},4)\right]  \,,
\nonumber\\
\hat{a}(\vec{k},0)  &  =\hat{a}(\vec{k},5)\,.\vphantom{\frac{1}{\sqrt{2}}} \label{eq:circpol}
\end{align}
The inverse transformations are given by:
\begin{align*}
\hat{a}(\vec{k},1)  &  =\frac{1}{\sqrt{2}}\left[  \hat{a}(\vec{k},+)+\hat{a}(\vec{k},-)\right]  \,,\\
\hat{a}(\vec{k},2)  &  =\frac{i}{\sqrt{2}}\left[  \hat{a}(\vec{k},+)-\hat{a}(\vec{k},-)\right]  \,,\\
\hat{a}(\vec{k},3)  &  =\frac{1}{\sqrt{2}}\left[  \hat{a}(\vec{k},\Delta)
+\hat{a}(\vec{k},\Box)\right]  \,,\\
\hat{a}(\vec{k},4)  &  =\frac{i}{\sqrt{2}}\left[  \hat{a}(\vec{k},\Delta)-\hat{a}(\vec{k},\Box)\right]  \,,\\
\hat{a}(\vec{k},5)  &  =\hat{a}(\vec{k},0)\,.\vphantom{\frac{1}{\sqrt{2}}}
\end{align*}
Finally, using the operators (\ref{eq:circpol}) one finds an elegant expression 
for the $z$-component of the spin operator:
\begin{align}
\hat{S}_z &  =\int\frac{\mathrm{d}^{3}k}{2\omega_{k}%
}\left\{  2\left[  \hat{a}^{\dagger}(\vec{k},+)\hat{a}(\vec{k},+)-\hat
{a}^{\dagger}(\vec{k},-)\hat{a}(\vec{k},-)\right] + \left[  \hat{a}^{\dagger}(\vec{k},\Delta)\hat
{a}(\vec{k},\Delta)-\hat{a}^{\dagger}(\vec{k},\Box)\hat{a}(\vec{k}%
,\Box)\right] \right\}\, .
\end{align}
Defining appropriate number operators for the individual polarization directions
in analogy to Eq.\ (\ref{eq:numberop}), we arrive at expression (\ref{eq:spinoperator}).


\newpage

\begin{thebibliography}{99}                                                                                      


\bibitem{proca}
A.~Proca, ``Sur la theorie ondulatoire des electrons positifs et negatifs'', J.\ Phys.\ Radium \textbf{7} (1936) 347.


\bibitem{Fierz}
M.~Fierz and W.~Pauli, ``On relativistic wave equations for particles of arbitrary spin in an electromagnetic field'', Proc.\ Roy.\ Soc.\ Lond.\ A \textbf{173} (1939) 211.


\bibitem{Fierzalt}
M.~Fierz, ``\"{U}ber die relativistische Theorie kr\"{a}ftefreier Teilchen mit beliebigem Spin'', Helvetica Physica Acta \textbf{12} (1939).


\bibitem{Ryder}
L.~H.~Ryder, ``Quantum Field Theory'', Cambridge, UK: Univ.\ Pr.\ (1985).


\bibitem{Reinhardt}
W.~Greiner and J.~Reinhardt, ``Field quantization'', Berlin, Germany: Springer (1996), (Theoretical physics, 7a).


\bibitem{Greiner}
W.~Greiner, ``Relativistic quantum mechanics: Wave equations'', Berlin, Germany: Springer (1990), (Theoretical physics, 3).


\bibitem{Macfarlane}
A.~J.~Macfarlane and W.~Tait, ``Tensor formulation of spin-1 and spin-2 fields'', Commun.\ Math.\ Phys.\ \textbf{24} (1972) 211.


\bibitem{Bargmann:1948ck}
V.~Bargmann and E.~P.~Wigner, ``Group Theoretical Discussion of Relativistic Wave Equations'', Proc.\ Nat.\ Acad.\ Sci.\  {\bf 34} (1948) 211.


\bibitem{Huang}
S.~Z.~Huang, T.~N.~Ruan, N.~Wu and Z.~P.~Zheng, ``Wavefunctions for particles with arbitrary spin'', Commun.\ Theor.\ Phys.\ \textbf{37} (2002) 63.


\bibitem{Aurilia}
A.~Aurilia and H.~Umezawa, ``Theory of high-spin fields'', Phys.\ Rev.\ \textbf{182} (1969) 1682.


\bibitem{Bouatta}
N.~Bouatta, G.~Compere and A.~Sagnotti, ``An Introduction to free higher-spin fields'', hep-th/0409068.


\bibitem{Chang2}
S.~J.~Chang, ``Lagrange Formulation for Systems with Higher Spin'', Phys.\ Rev.\ \textbf{161} (1967) 1308.


\bibitem{Barut}
A.~O.~Barut and R.~Raczka, ``Theory Of Group Representations And Applications'', Singapore, Singapore: World Scientific (1986).


\bibitem{Dalmazi}
D.~Dalmazi, ``A note on the nonuniqueness of the massive Fierz-Pauli theory and spectator fields'', Phys.\ Rev.\ D \textbf{88} (2013) 4, 045003 [arXiv:1307.1109 [hep-th]].


\bibitem{Hinterbichler}
K.~Hinterbichler, ``Theoretical Aspects of Massive Gravity'', Rev.\ Mod.\ Phys.\ \textbf{84} (2012) 671 [arXiv:1105.3735[hep-th]].


\bibitem{Baaklini}
N.~S.~Baaklini and M.~Tuite, ``Dirac Quantization of Spin-2 Field'', J.\ Phys.\ A \textbf{12} (1979) L13. 


\bibitem{Wagenaar}
J.~W.~Wagenaar and T.~A.~Rijken, ``On the Quantization of the Higher Spin Fields'', 
Phys.\ Rev.\ D \textbf{80} (2009) 104027 [arXiv:0905.3893 [hep-th]].


\bibitem{Rivers}
R.~Rivers, ``Lagrangian Theory for Neutral Massive Spin-2 Fields'',
Nuovo Cim.\ \textbf{34} (1964) 2, 386.


\bibitem{Bhargava}
S.~C.~Bhargava and H.~Watanabe, ``The Lagrangian formalism of the theory of spin two fields'', Nuclear Physics, Volume \textbf{87}, Issue 4 (1966).


\bibitem{Dalmazi2}
D.~Dalmazi, ``Massive spin-2 particle from a rank-2 tensor'', Phys.\ Rev.\ D \textbf{87} (2013) 12, 125027 [arXiv:1305.1513[hep-th]].


\bibitem{StruckmeierUn}
J.~Struckmeier and H.~Reichau, ``General $U(N)$ gauge transformations in the realm of covariant Hamiltonian field theory'', [arXiv:1205.5754 [hep-th]].


\bibitem{StruckmeierUn2}
J.~Struckmeier, ``Generalized U(N) gauge transformations in the realm of the extended covariant Hamilton formalism of field theory'', J.\ Phys.\ G \textbf{40} (2013) 015007 [arXiv:1206.4452[nucl-th]].


\bibitem{Chang1}
S.~J.~Chang, ``Quantization of Spin-2 Fields'', Phys.\ Rev.\ \textbf{148} (1966) 1259.


\bibitem{Carroll}
S.~M.~Carroll, ``Lecture notes on general relativity'', gr-qc/9712019.


\bibitem{Zee}
A.~Zee, ``Quantum field theory in a nutshell'', Princeton, UK: Princeton Univ. Pr. (2010).


\bibitem{Deser}
S.~Deser and A.~Waldron, ``Inconsistencies of massive charged gravitating higher spins'', Nucl.\ Phys.\ B \textbf{631} (2002) 369 [hep-th/0112182].


\bibitem{arkani}
N.~Arkani-Hamed, H.~Georgi and M.~D.~Schwartz, ``Effective field theory for massive gravitons and gravity in theory space'', Annals Phys.\ \textbf{305} (2003) 96 [hep-th/0210184].


\bibitem {rubakov}
V.~A.~Rubakov and P.~G.~Tinyakov, ``Infrared-modified gravities and massive gravitons'', Phys.\ Usp.\ \textbf{51} (2008) 759 [arXiv:0802.4379 [hep-th]].


\bibitem {colladay}
D.~Colladay and V.~A.~Kostelecky, ``Lorentz violating extension of the standard model'', Phys.\ Rev.\ D \textbf{58} (1998) 116002 [hep-ph/9809521].


\bibitem{Suzuki}
M.~Suzuki, ``Tensor meson dominance: Phenomenology of the f2 meson'', Phys.\ Rev.\ D \textbf{47} (1993) 1043.


\bibitem{Ye}
Z.~C.~Ye, X.~Wang, X.~Liu and Q.~Zhao, ``The mass spectrum and strong decays of isoscalar tensor mesons'', Phys.\ Rev.\ D \textbf{86} (2012) 054025 [arXiv:1206.0097 [hep-ph]].


\bibitem{Anisovich}
V.~V.~Anisovich, ``Systematization of tensor mesons and the determination of the 2++ glueball'', JETP Lett.\ \textbf{80} (2004) 715 [Pisma Zh.\ Eksp.\ Teor.\ Fiz.\ \textbf{80} (2004) 845] [hep-ph/0412093].


\bibitem{Burakovsky}
L.~Burakovsky and J.~T.~Goldman, ``Towards resolution of the enigmas of P wave meson
spectroscopy'', \ Phys.\ Rev.\ D \textbf{57} (1998) 2879 [hep-ph/9703271].


\bibitem{Cotanch}
S.~R.~Cotanch and R.~A.~Williams, ``Tensor glueball photoproduction and decay'', Phys.\ Lett.\ B \textbf{621} (2005) 269 [nucl-th/0505074].


\bibitem{Giacosa}
F.~Giacosa, T.~Gutsche, V.~E.~Lyubovitskij and A.~Faessler, ``Decays of tensor mesons and the tensor glueball in an effective field approach'', \ Phys.\ Rev.\ D \textbf{72} (2005) 114021 [hep-ph/0511171].


\bibitem{olive}
K.~A.~Olive \emph{et al.} (Particle Data Group), \emph{Chin. Phys.} \textbf{C38}, 090001 (2014).


\bibitem{Wang}
Z.~G.~Wang and Z.~Y.~Di, ``Masses and decay constants of the heavy tensor mesons with QCD sum rules'', Eur.\ Phys.\ J.\ A \textbf{50} (2014) 143 [arXiv:1405.5092 [hep-ph]].


\bibitem{chenzhu}
W.~Chen, Z.~X.~Cai and S.~L.~Zhu, ``Masses of the tensor mesons with $J^{P}=2^{-}$'', Nucl.\ Phys.\ B \textbf{887} (2014) 201 [arXiv:1107.4949 [hep-ph]].


\bibitem{ks}
S.~L.~Lyakhovich, A.~A.~Sharapov and K.~M.~Shekhter, ``Massive spinning particle in any dimension. 1. Integer spins'', Nucl.\ Phys.\ B \textbf{537} (1999) 640 [hep-th/9805020].


\end{thebibliography}
\end{document}